\documentclass[twocolumn]{aastex63}

\received{2 May, 2023}
\revised{11 August, 2023}
\accepted{19 September, 2023}
\submitjournal{AJ}

\newcommand{\periodplanetb}{\ensuremath{104.854_{-0.002}^{+0.001}}}   
\newcommand{\Kplanetb}{\ensuremath{7.7\pm1.1}}   
\newcommand{\eccplanetb}{\ensuremath{0.09_{-0.02}^{+0.01}}}   
\newcommand{\omegaplanetb}{\ensuremath{350_{-4}^{+2}}}   
\newcommand{\Maplanetb}{\ensuremath{111\pm2}}   
\newcommand{\massplanetb}{\ensuremath{0.17\pm0.02}}   
\newcommand{\axisplanetb}{\ensuremath{0.4254 \pm 0.002}}   
\newcommand{\periodplanetc}{\ensuremath{273.69_{-0.22}^{+0.26}}}   
\newcommand{\Kplanetc}{\ensuremath{9.1_{-0.4}^{+0.5}}}   
\newcommand{\eccplanetc}{\ensuremath{0.096_{-0.009}^{+0.008}}}   
\newcommand{\omegaplanetc}{\ensuremath{8_{-7}^{+4}}}   
\newcommand{\Maplanetc}{\ensuremath{192_{-3}^{+4}}}   
\newcommand{\massplanetc}{\ensuremath{0.28_{-0.01}^{+0.02}}}   
\newcommand{\axisplanetc}{\ensuremath{0.807 \pm 0.003}}   
\newcommand{\RVoffFEROS}{\ensuremath{51326.7_{-3.1}^{+2.9}}}   
\newcommand{\RVoffHARPS}{\ensuremath{51350.4_{-0.9}^{+1.0}}}   
\newcommand{\RVoffCHIRON}{\ensuremath{-0.5_{-3.9}^{+4.2}}}   
\newcommand{\RVoffCORALIE}{\ensuremath{51331.0_{-3.3}^{+3.1}}} 
\newcommand{\RVjittFEROS}{\ensuremath{19.9_{-2.1}^{+2.6}}}   
\newcommand{\RVjittHARPS}{\ensuremath{5.5_{-0.6}^{+0.8}}}   
\newcommand{\RVjittCHIRON}{\ensuremath{11.9_{-3.7}^{+4.8}}}   
\newcommand{\RVjittCORALIE}{\ensuremath{5.4_{-3.5}^{+4.4}}}   
\newcommand{\periodplanetbmaxLn}{\ensuremath{104.855} }  
\newcommand{\KplanetbmaxLn}{\ensuremath{7.7}}   
\newcommand{\eccplanetbmaxLn}{\ensuremath{0.102}}   
\newcommand{\omegaplanetbmaxLn}{\ensuremath{351}}   
\newcommand{\MaplanetbmaxLn}{\ensuremath{110}}   
\newcommand{\massplanetbmaxLn}{\ensuremath{0.17}}   
\newcommand{\axisplanetbmaxLn}{\ensuremath{0.4254}}   
\newcommand{\periodplanetcmaxLn}{\ensuremath{273.55}}   
\newcommand{\KplanetcmaxLn}{\ensuremath{8.8}}   
\newcommand{\eccplanetcmaxLn}{\ensuremath{0.107}}   
\newcommand{\omegaplanetcmaxLn}{\ensuremath{11}}   
\newcommand{\MaplanetcmaxLn}{\ensuremath{190}}   
\newcommand{\massplanetcmaxLn}{\ensuremath{0.27}}   
\newcommand{\axisplanetcmaxLn}{\ensuremath{0.806}}   
\newcommand{\RVoffFEROSmaxLn}{\ensuremath{51324.5}}   
\newcommand{\RVoffHARPSmaxLn}{\ensuremath{51349.4}}   
\newcommand{\RVoffCHIRONmaxLn}{\ensuremath{-0.2}}   
\newcommand{\RVoffCORALIEmaxLn}{\ensuremath{51327.1}} 
\newcommand{\RVjittFEROSmaxLn}{\ensuremath{21.1}}   
\newcommand{\RVjittHARPSmaxLn}{\ensuremath{5.1}}   
\newcommand{\RVjittCHIRONmaxLn}{\ensuremath{14.9}}   
\newcommand{\RVjittCORALIEmaxLn}{\ensuremath{4.3}}   

\newcommand{\julietpb}{\ensuremath{0.1015\pm{0.0005}}}
\newcommand{\julietbb}{\ensuremath{0.45_{-0.03}^{+0.02}}}
\newcommand{\julietrho}{\ensuremath{3431_{-140}^{+160}}}
\newcommand{\julietqoneTESS}{\ensuremath{0.30\pm0.03}}
\newcommand{\julietqtwoTESS}{\ensuremath{0.36_{-0.06}^{+0.05}}}
\newcommand{\julietqoneASTEP}{\ensuremath{0.62\pm0.06}}
\newcommand{\julietqtwoASTEP}{\ensuremath{0.07\pm0.04}}
\newcommand{\julietqoneLCOone}{\ensuremath{0.32_{-0.07}^{+0.08}}}
\newcommand{\julietqtwoLCOone}{\ensuremath{0.30\pm0.07}}
\newcommand{\julietqoneLCOtwo}{\ensuremath{0.61_{-0.06}^{+0.08}}}
\newcommand{\julietqtwoLCOtwo}{\ensuremath{0.43_{-0.06}^{+0.07}}}
\newcommand{\julietqoneLCOthr}{\ensuremath{0.29_{-0.06}^{+0.07}}}
\newcommand{\julietqtwoLCOthr}{\ensuremath{0.22_{-0.06}^{+0.07}}}
\newcommand{\julietqoneLCOfour}{\ensuremath{0.33_{-0.06}^{+0.07}}}
\newcommand{\julietqtwoLCOfour}{\ensuremath{0.25_{-0.06}^{+0.07}}}
\newcommand{\julietqoneLCOfive}{\ensuremath{0.71\pm0.07}}
\newcommand{\julietqtwoLCOfive}{\ensuremath{0.43\pm0.07}}
\newcommand{\julietqoneLCOsix}{\ensuremath{0.29_{-0.08}^{+0.07}}}
\newcommand{\julietqtwoLCOsix}{\ensuremath{0.25_{-0.05}^{+0.06}}}
\newcommand{\julietqonePEST}{\ensuremath{0.83\pm0.07}}
\newcommand{\julietqtwoPEST}{\ensuremath{0.49_{-0.07}^{+0.06}}}
\newcommand{\julietqoneNEOS}{\ensuremath{0.67_{-0.07}^{+0.06}}}
\newcommand{\julietqtwoNEOS}{\ensuremath{0.41\pm0.07}}
\newcommand{\julietqoneHwd}{\ensuremath{0.50_{-0.07}^{+0.08}}}
\newcommand{\julietqtwoHwd}{\ensuremath{0.35_{-0.07}^{+0.08}}}
\newcommand{\julietmfluxTESStwo}{\ensuremath{0.0002_{-0.0013}^{+0.0012}}}
\newcommand{\julietsigmaTESStwo}{\ensuremath{402\pm8}}
\newcommand{\julietmfluxTESSten}{\ensuremath{0.003\pm0.002}}
\newcommand{\julietsigmawTESSten}{\ensuremath{327\pm13}}
\newcommand{\julietmfluxTESSthir}{\ensuremath{-0.0007_{-0.0005}^{+0.0006}}}
\newcommand{\julietsigmawTESSthir}{\ensuremath{371_{-11}^{+10}}}
\newcommand{\julietmfluxTESStn}{\ensuremath{0.0010_{-0.0005}^{+0.0006}}}
\newcommand{\julietsigmawTESStn}{\ensuremath{641\pm9}}
\newcommand{\julietmfluxTESStt}{\ensuremath{0.006_{-0.006}^{+0.005}}}
\newcommand{\julietsigmawTESStt}{\ensuremath{419_{-7}^{+8}}}
\newcommand{\julietmfluxTESSts}{\ensuremath{0.0012\pm0.0008}}
\newcommand{\julietsigmawTESSts}{\ensuremath{604_{-9}^{+10}}}
\newcommand{\julietmfluxTESSst}{\ensuremath{-0.0003_{-0.0028}^{+0.0030}}}
\newcommand{\julietsigmawTESSst}{\ensuremath{493\pm8}}
\newcommand{\julietmfluxASTEPone}{\ensuremath{0.0004\pm0.0001}}
\newcommand{\julietsigmawASTEPone}{\ensuremath{960_{-25}^{+34}}}
\newcommand{\julietmfluxASTEPtwo}{\ensuremath{0.00027\pm0.00007}}
\newcommand{\julietsigmawASTEPtwo}{\ensuremath{999.1_{-0.6}^{+0.9}}}
\newcommand{\julietmfluxASTEPthr}{\ensuremath{-0.00088\pm0.00007}}
\newcommand{\julietsigmawASTEPthr}{\ensuremath{998_{-1}^{+2}}}
\newcommand{\julietmfluxPEST}{\ensuremath{0.00004_{-0.00009}^{+0.00008}}}
\newcommand{\julietsigmawPEST}{\ensuremath{996\pm3}}
\newcommand{\julietmfluxNEOS}{\ensuremath{-0.00002\pm0.00003}}
\newcommand{\julietsigmawNEOS}{\ensuremath{999.92_{-0.05}^{+0.08}}}
\newcommand{\julietmfluxHwd}{\ensuremath{-0.050\pm0.0001}}
\newcommand{\julietsigmawHwd}{\ensuremath{995_{-3}^{+5}}}
\newcommand{\julietmfluxLCOone}{\ensuremath{-0.000003_{-0.000071}^{+0.000075}}}
\newcommand{\julietsigmawLCOone}{\ensuremath{779_{-95}^{+91}}}
\newcommand{\julietmfluxLCOtwo}{\ensuremath{-0.0003\pm0.0002}}
\newcommand{\julietsigmawLCOtwo}{\ensuremath{989_{-7}^{+9}}}
\newcommand{\julietmfluxLCOthr}{\ensuremath{0.0003\pm0.0001}}
\newcommand{\julietsigmawLCOthr}{\ensuremath{994_{-4}^{+7}}}
\newcommand{\julietmfluxLCOfo}{\ensuremath{-0.0001\pm0.0001}}
\newcommand{\julietsigmawLCOfo}{\ensuremath{986_{-9}^{+14}}}
\newcommand{\julietmfluxLCOfi}{\ensuremath{-0.00003\pm0.00014}}
\newcommand{\julietsigmawLCOfi}{\ensuremath{990_{-7}^{+12}}}
\newcommand{\julietmfluxLCOsix}{\ensuremath{-0.0082\pm0.00006}}
\newcommand{\julietsigmawLCOsix}{\ensuremath{998_{-1}^{+2}}}
\newcommand{\julietGPsigmaTESStwo}{\ensuremath{0.0031_{-0.0007}^{+0.0005}}}
\newcommand{\julietGPrhoTESStwo}{\ensuremath{4.0_{-0.7}^{+0.5}}}
\newcommand{\julietGPsigmaTESSten}{\ensuremath{0.0086_{-0.0006}^{+0.0005}}}
\newcommand{\julietGPrhoTESSten}{\ensuremath{0.70_{-0.04}^{+0.03}}}
\newcommand{\julietGPsigmaTESSthir}{\ensuremath{0.0031\pm0.0002}}
\newcommand{\julietGPrhoTESSthir}{\ensuremath{0.71_{-0.04}^{+0.03}}}
\newcommand{\julietGPsigmaTESStn}{\ensuremath{0.0035\pm0.0002}}
\newcommand{\julietGPrhoTESStn}{\ensuremath{0.42\pm0.02}}
\newcommand{\julietGPsigmaTESStt}{\ensuremath{0.009_{-0.004}^{+0.002}}}
\newcommand{\julietGPrhoTESStt}{\ensuremath{9_{-3}^{+2}}}
\newcommand{\julietGPsigmaTESSts}{\ensuremath{0.0054\pm0.0003}}
\newcommand{\julietGPrhoTESSts}{\ensuremath{0.42\pm0.02}}
\newcommand{\julietGPsigmaTESSst}{\ensuremath{0.007_{-0.002}^{+0.001}}}
\newcommand{\julietGPrhoTESSst}{\ensuremath{4.2_{-0.8}^{+0.6}}}
\newcommand{\julietTTESStwo}{\ensuremath{2458361.0283\pm0.0008}}
\newcommand{\julietTTESSten}{\ensuremath{2458570.732\pm0.001}}
\newcommand{\julietTTESSthir}{\ensuremath{2458675.6182\pm0.0009}}
\newcommand{\julietTTESStn}{\ensuremath{2459095.1087_{-0.0009}^{+0.0010}}}
\newcommand{\julietTTESStt}{\ensuremath{2459200.0050\pm0.0006}}
\newcommand{\julietTTESSts}{\ensuremath{2459304.8770_{-0.0010}^{+0.0009}}}
\newcommand{\julietTTESSst}{\ensuremath{2460038.9735\pm0.0004}}
\newcommand{\julietTASTEPone}{\ensuremath{2459304.8774\pm0.0008}}
\newcommand{\julietTASTEPtwo}{\ensuremath{2459409.7229\pm0.0007}}
\newcommand{\julietTASTEPthr}{\ensuremath{2459829.2283\pm0.0005}}
\newcommand{\julietTLCOone}{\ensuremath{2459200.04\pm0.03}}
\newcommand{\julietTLCOtwo}{\ensuremath{2459619.482\pm0.002}}
\newcommand{\julietTLCOthr}{\ensuremath{2459619.480\pm0.001}}
\newcommand{\julietTLCOfo}{\ensuremath{2459619.486\pm0.001}}
\newcommand{\julietTLCOfi}{\ensuremath{2459619.485\pm0.001}}
\newcommand{\julietTLCOsix}{\ensuremath{2459934.0818\pm0.0003}}
\newcommand{\julietTPEST}{\ensuremath{2459200.019\pm0.001}}
\newcommand{\julietTNEOS}{\ensuremath{2459304.900_{-0.003}^{+0.002}}}
\newcommand{\julietTHwd}{\ensuremath{2459304.759\pm0.002}}
\newcommand{\julietP}{\ensuremath{104.87236\pm0.00005}}

\newcommand{\juliettzero}{\ensuremath{2458256.1286\pm0.0006}}
\newcommand{\julieta}{\ensuremath{0.480_{-0.008}^{+0.007}}}
\newcommand{\julietRp}{\ensuremath{0.810\pm0.005}}

\shorttitle{TOI-199 b: A transiting warm giant with TTVs seen from Antarctica}
\shortauthors{Hobson et al.}
\graphicspath{{./}{figures/}}

\begin{document}

\title{TOI-199 b: A well-characterized 100-day transiting warm giant planet with TTVs seen from Antarctica}

\correspondingauthor{Melissa J. Hobson}
\email{mhobson@astro.puc.cl}

\author[0000-0002-5945-7975]{Melissa J. Hobson}
\affiliation{Max-Planck-Institut für Astronomie, Königstuhl 17, 69117 Heidelberg, Germany}
\affiliation{Millennium Institute for Astrophysics, Chile}

\author{Trifon Trifonov}
\affiliation{Max-Planck-Institut für Astronomie, Königstuhl 17, 69117 Heidelberg, Germany}
 \affiliation{Department
 of Astronomy, Sofia University ``St Kliment Ohridski'', 5 James Bourchier Blvd, BG-1164 Sofia, Bulgaria}
  \affiliation{Zentrum f\"ur Astronomie der Universt\"at Heidelberg, Landessternwarte,
              K\"onigstuhl 12, 69117 Heidelberg, Germany}
\author[0000-0002-1493-300X]{Thomas Henning}
\affiliation{Max-Planck-Institut für Astronomie, Königstuhl 17, 69117 Heidelberg, Germany}

\author[0000-0002-5389-3944]{Andrés Jordán}
\affiliation{Millennium Institute for Astrophysics, Chile}
\affiliation{Facultad de Ingeniería y Ciencias, Universidad Adolfo Ibáñez, Av. Diagonal las Torres 2640, Peñalolén, Santiago, Chile}

\author{Felipe Rojas}
\affiliation{Millennium Institute for Astrophysics, Chile}
\affiliation{Instituto de Astrofísica, Facultad de Física, Pontificia Universidad Católica de Chile, Av. Vicuña Mackenna 4860, 782-0436 Macul, Santiago, Chile}

\author{Nestor Espinoza}
\affiliation{Space Telescope Science Institute, 3700 San Martin Drive, Baltimore, MD 21218, USA}

\author{Rafael Brahm}
\affiliation{Millennium Institute for Astrophysics, Chile}
\affiliation{Facultad de Ingeniería y Ciencias, Universidad Adolfo Ibáñez, Av. Diagonal las Torres 2640, Peñalolén, Santiago, Chile}

\author[0000-0003-3130-2768]{Jan Eberhardt}
\affiliation{Max-Planck-Institut für Astronomie, Königstuhl 17, 69117 Heidelberg, Germany}

\author{Mat\'ias I. Jones}
\affiliation{European Southern Observatory, Alonso de Córdova 3107, Vitacura, Casilla 19001, Santiago, Chile}

\author{Djamel Mekarnia} 
\affiliation{Université Côte d'Azur, Observatoire de la Côte d'Azur, CNRS, Laboratoire Lagrange, Bd de l'Observatoire, CS 34229, F-06304 Nice cedex 4, France}

\author{Diana Kossakowski}
\affiliation{Max-Planck-Institut für Astronomie, Königstuhl 17, 69117 Heidelberg, Germany}

\author[0000-0001-8355-2107]{Martin Schlecker}
\affiliation{Steward Observatory and Department of Astronomy, The University of Arizona, Tucson, AZ 85721, USA}

\author{Marcelo Tala Pinto}
\affiliation{Millennium Institute for Astrophysics, Chile}
\affiliation{Facultad de Ingeniería y Ciencias, Universidad Adolfo Ibáñez, Av. Diagonal las Torres 2640, Peñalolén, Santiago, Chile}

\author[0000-0003-0974-210X]{Pascal José Torres Miranda}
\affiliation{Millennium Institute for Astrophysics, Chile}
\affiliation{Instituto de Astrofísica, Facultad de Física, Pontificia Universidad Católica de Chile, Av. Vicuña Mackenna 4860, 782-0436 Macul, Santiago, Chile}


\author{Lyu Abe} 
\affiliation{Université Côte d'Azur, Observatoire de la Côte d'Azur, CNRS, Laboratoire Lagrange, Bd de l'Observatoire, CS 34229, F-06304 Nice cedex 4, France}

\author[0000-0003-1464-9276]{Khalid Barkaoui}
\affiliation{Astrobiology Research Unit, Universit\'e de Li\`ege, 19C All\'ee du 6 Ao\^ut, 4000 Li\`ege, Belgium}
\affiliation{Department of Earth, Atmospheric and Planetary Science, Massachusetts Institute of Technology, 77 Massachusetts Avenue, Cambridge, MA 02139, USA}
\affiliation{Instituto de Astrof\'isica de Canarias (IAC), Calle V\'ia L\'actea s/n, 38200, La Laguna, Tenerife, Spain}

\author{Philippe Bendjoya} 
\affiliation{Université Côte d'Azur, Observatoire de la Côte d'Azur, CNRS, Laboratoire Lagrange, Bd de l'Observatoire, CS 34229, F-06304 Nice cedex 4, France}

\author{François Bouchy} 
\affiliation{Observatoire de Genève, Département d'Astronomie, Université de Genève, Chemin Pegasi 51b, 1290 Versoix, Switzerland}

\author{Marco Buttu} 
\affiliation{INAF Osservatorio Astronomico di Cagliari, Via della Scienza 5 - 09047 Selargius CA, Italy}

\author[0000-0002-0810-3747]{Ilaria~Carleo} 
\affiliation{Instituto de Astrof\'isica de Canarias (IAC), Calle V\'ia L\'actea s/n, 38200, La Laguna, Tenerife, Spain}

\author[0000-0001-6588-9574]{Karen~A.~Collins} 
\affiliation{Center for Astrophysics \textbar \ Harvard \& Smithsonian, 60 Garden Street, Cambridge, MA 02138, USA}

\author[0000-0001-8020-7121]{Knicole D. Col\'{o}n} 
\affiliation{NASA Goddard Space Flight Center, 8800 Greenbelt Rd, Greenbelt, MD 20771, USA}

\author[0000-0001-7866-8738]{Nicolas Crouzet} 
\affiliation{Leiden Observatory, Leiden University, P.O. Box 9513, 2300 RA Leiden, The Netherlands}

\author[0000-0003-2313-467X]{Diana~Dragomir} 
\affiliation{Department of Physics and Astronomy, University of New Mexico, 210 Yale Blvd NE, Albuquerque, NM 87106, USA}

\author[0000-0002-3937-630X]{Georgina~Dransfield} 
\affiliation{School of Physics \& Astronomy, University of Birmingham, Edgbaston, Birmingham B15 2TT, UK}

\author{Thomas Gasparetto} 
\affiliation{Institute of Polar Sciences - CNR, via Torino, 155 - 30172 Venice-Mestre, Italy}

\author{Robert~F.~Goeke} 
\affiliation{Department of Physics and Kavli Institute for Astrophysics and Space Research, Massachusetts Institute of Technology, Cambridge, MA 02139, USA}

\author[0000-0002-7188-8428]{Tristan Guillot} 
\affiliation{Université Côte d'Azur, Observatoire de la Côte d'Azur, CNRS, Laboratoire Lagrange, Bd de l'Observatoire, CS 34229, F-06304 Nice cedex 4, France}

\author[0000-0002-3164-9086]{Maximilian N. Günther} 
\affiliation{European Space Agency (ESA), European Space Research and Technology Centre (ESTEC), Keplerlaan 1, 2201 AZ Noordwijk, The Netherlands}

\author[0000-0003-4894-7271]{Saburo Howard}
\affiliation{Université Côte d'Azur, Observatoire de la Côte d'Azur, CNRS, Laboratoire Lagrange, Bd de l'Observatoire, CS 34229, F-06304 Nice cedex 4, France}

\author[0000-0002-4715-9460]{Jon~M.~Jenkins} 
\affiliation{NASA Ames Research Center, Moffett Field, CA 94035, USA}

\author[0000-0002-0076-6239]{Judith Korth}
\affiliation{Lund Observatory, Division of Astrophysics, Department of Physics, Lund University, Box 43, 22100 Lund, Sweden}

\author[0000-0001-9911-7388]{David~W.~Latham} 
\affiliation{Center for Astrophysics \textbar \ Harvard \& Smithsonian, 60 Garden Street, Cambridge, MA 02138, USA}

\author[0000-0001-9699-1459]{Monika Lendl} 
\affiliation{Observatoire de Genève, Département d'Astronomie, Université de Genève, Chemin Pegasi 51b, 1290 Versoix, Switzerland}

\author[0000-0001-6513-1659]{Jack J. Lissauer}
\affiliation{Space Science \& Astrobiology Division, MS 245-3, NASA Ames Research Center, Moffett Field, CA 94035, USA}

\author[0000-0002-9312-0073]{Christopher R. Mann} 
\affiliation{Département de Physique, Université de Montréal, Montréal, QC, Canada}
\affiliation{Trottier Institute for Research on Exoplanets (\emph{iREx})}

\author[0000-0002-4510-2268]{Ismael~Mireles} 
\affiliation{Department of Physics and Astronomy, University of New Mexico, 210 Yale Blvd NE, Albuquerque, NM 87106, USA}

\author[0000-0003-2058-6662]{George~R.~Ricker} 
\affiliation{Department of Physics and Kavli Institute for Astrophysics and Space Research, Massachusetts Institute of Technology, Cambridge, MA 02139, USA}

\author{Sophie Saesen} 
\affiliation{Observatoire de Genève, Département d'Astronomie, Université de Genève, Chemin Pegasi 51b, 1290 Versoix, Switzerland}

\author[0000-0001-8227-1020]{Richard P. Schwarz}
\affiliation{Center for Astrophysics \textbar \ Harvard \& Smithsonian, 60 Garden Street, Cambridge, MA 02138, USA}

\author[0000-0002-6892-6948]{S.~Seager} 
\affiliation{Department of Physics and Kavli Institute for Astrophysics and Space Research, Massachusetts Institute of Technology, Cambridge, MA 02139, USA}
\affiliation{Department of Earth, Atmospheric and Planetary Sciences, Massachusetts Institute of Technology, Cambridge, MA 02139, USA}
\affiliation{Department of Aeronautics and Astronautics, MIT, 77 Massachusetts Avenue, Cambridge, MA 02139, USA}

\author[0000-0003-3904-6754]{Ramotholo Sefako} 
\affiliation{South African Astronomical Observatory, P.O. Box 9, Observatory, Cape Town 7935, South Africa}

\author[0000-0002-1836-3120]{Avi Shporer}
\affiliation{Department of Physics and Kavli Institute for Astrophysics and Space Research, Massachusetts Institute of Technology, Cambridge, MA 02139, USA}

\author[0000-0003-2163-1437]{Chris Stockdale}
\affiliation{Hazelwood Observatory, Australia}

\author[0000-0002-3503-3617]{Olga Suarez} 
\affiliation{Université Côte d'Azur, Observatoire de la Côte d'Azur, CNRS, Laboratoire Lagrange, Bd de l'Observatoire, CS 34229, F-06304 Nice cedex 4, France}

\author[0000-0001-5603-6895]{Thiam-Guan Tan} 
\affiliation{Perth Exoplanet Survey Telescope, Perth, Western Australia} 

\author[0000-0002-5510-8751]{Amaury H. M. J. Triaud} 
\affiliation{School of Physics \& Astronomy, University of Birmingham, Edgbaston, Birmingham B15 2TT, UK}

\author{Solène Ulmer-Moll} 
\affiliation{Observatoire de Genève, Département d'Astronomie, Université de Genève, Chemin Pegasi 51b, 1290 Versoix, Switzerland}
\affiliation{Physikalisches Institut, University of Bern, Gesellschaftsstrasse 6, 3012 Bern, Switzerland}

\author[0000-0001-6763-6562]{Roland~Vanderspek} 
\affiliation{Department of Physics and Kavli Institute for Astrophysics and Space Research, Massachusetts Institute of Technology, Cambridge, MA 02139, USA}

\author[0000-0002-4265-047X]{Joshua~N.~Winn} 
\affiliation{Department of Astrophysical Sciences, Princeton University, 4 Ivy Lane, Princeton, NJ 08544, USA}

\author[0000-0002-5402-9613]{Bill~Wohler} 
\affiliation{NASA Ames Research Center, Moffett Field, CA 94035, USA}
\affiliation{SETI Institute, Mountain View, CA 94043, USA}

\author{George~Zhou} 
\affiliation{University of Southern Queensland, Centre for Astrophysics, West Street, Toowoomba, QLD 4350, Australia}




\begin{abstract}

We present the spectroscopic confirmation and precise mass measurement of the warm giant planet TOI-199 b. This planet was first identified in \textit{TESS} photometry and confirmed using ground-based photometry from ASTEP in Antarctica including a full 6.5\,h long transit, PEST, Hazelwood, and LCO; space photometry from NEOSSat; and radial velocities (RVs) from FEROS, HARPS, CORALIE, and CHIRON. Orbiting a late G-type star, TOI-199\,b has a $\mathrm{\periodplanetb \, d}$ period, a mass of $\mathrm{\massplanetb \, M_J}$, and a radius of $\mathrm{\julietRp \, R_J}$. It is the first warm exo-Saturn with a precisely determined mass and radius. The \textit{TESS} and ASTEP transits show strong transit timing variations, pointing to the existence of a second planet in the system. The joint analysis of the RVs and TTVs provides a unique solution for the non-transiting companion TOI-199 c, which has a period of $\mathrm{\periodplanetc \, d}$ and an estimated mass of $\mathrm{\massplanetc \, M_J}$. This period places it within the conservative Habitable Zone.

\end{abstract}



\section{Introduction} \label{sec:intro}

In the quest to understand how planets form and evolve, warm giant planets are a vital piece of the puzzle. Warm giants are generally defined as planets with sizes $\mathrm{R_p \geq 4 R_\oplus}$ and periods $\mathrm{10 \,d \lesssim P \lesssim 300 \, d}$. Unlike their closer-in hot Jupiter cousins, they are far enough from their host stars to not be strongly irradiated, and thus are not expected to be affected by radius inflation \citep[see, e.g.][]{Sarkis20}. Likewise, tidal circularisation is not expected to impact their orbital parameters, meaning these can potentially be used to disentangle their migration history \citep[see][for a review]{Dawson18}. 

Identifying and characterizing a large sample of warm giants is the necessary first step towards understanding this population. Transiting warm giants around bright stars are especially valuable, as both their radii (via transits) and their masses (via radial velocities) can be measured, and thus their density computed and their internal structure modelled. The \textit{Transiting Exoplanet Survey Satellite} mission \citep[\textit{TESS},][]{Ricker15} is proving invaluable for the detection of these warm giants, with some 60 already published (according to the table of \textit{TESS} planets on the NASA Exoplanet Archive\footnote{located at \url{https://exoplanetarchive.ipac.caltech.edu}}, as of 6 February 2023) and hundreds more expected from yield simulations of the prime and extended missions \citep{Sullivan15, Barclay18, Kunimoto22}. 
Thanks primarily to \textit{TESS} and its predecessor Kepler \citep{Keplermission}, some 80 warm giants have had their radii and masses characterized to better than 25\%, according to the TEPCAT catalogue \citep{Southworth11}\footnote{Available at \url{https://www.astro.keele.ac.uk/jkt/tepcat/}, accessed 6th October 2022.}. The majority of these are on the shorter end of the 10-300 d period range, with half having periods shorter than 20 days. Longer-period transiting warm giants remain rare. 
The Warm gIaNts with tEss collaboration \citep[WINE, e.g.][]{Brahm19PARSEC, Jordan20, Schlecker20, Hobson21, Trifonov21} seeks to confirm and characterize warm giant candidates from \textit{TESS}, using a variety of photometric and spectroscopic facilities.

While the transit technique has been immensely fruitful, leading the list of planet discoveries with over 3900 of the over 5200 known exoplanets\footnote{According to the NASA Exoplanet Archive, last accessed 6th February 2023}, it is of necessity limited by orbital geometry – the planet must cross between the star and the observer – and biased towards short-period planets whose transits are more likely to be observed. However, for multi-planetary systems where the planets exert strong gravitational influences on each other, the transit timing variations \citep[TTVs, ][]{Agol05, Holman05} technique can help to overcome these limitations in some cases: by measuring the TTVs for a transiting planet, we can detect non-transiting planets, and measure the masses of all bodies involved. The first detection of the TTV effect was for Kepler-19 b by \cite{Ballard11}, but they could not unambiguously determine the mass and period of the non-transiting companion. The first planet with a unique solution determined from TTVs was KOI-872 c \citep{Nesvorny12}. Since then, this technique has enabled the detection of 24 planets\footnote{As of 6th February 2023, according to the NASA Exoplanet Archive}, many of which are warm giants \citep[e.g.][a pair of warm giants where only the inner one transits]{Trifonov21}.

In this paper, we present the TOI-199 system, consisting of two giant planets. The inner planet is transiting, with a period of \periodplanetb d, and was found in the \textit{TESS} photometry. The outer planet does not transit, but was revealed by the TTVs it induces on the transits of the inner planet. The paper is organized as follows. We present the observational data in Section \ref{sec:obs}. In Section \ref{sec:analysis} we analyse the data, and characterize the star and planet. Our results are discussed in Section \ref{sec:disc}, and finally, we provide our conclusions in Section \ref{sec:conc}.

\section{Observations} \label{sec:obs}

\subsection{\textit{TESS} photometry}

TOI-199 was observed by the \textit{TESS} prime and first and second extended missions between 25th July 2018 and 6th April 2023, in Sectors 1-13, 27, 29-37, 39, and 61-63, using camera 4. Sector 1 was observed with CCD 4; Sectors 2, 3, and 4 with CCD 1; Sectors 5, 6, and 7 with CCD 2; Sectors 8, 9, 10, and 11 with CCD 3; Sectors 12, 13, and 27 with CCD 4; Sectors 29, 30, and 31 with CCD 1; Sectors 32, 33, and 34 with CCD 2; Sectors 35, 36, and 37 with CCD 3; Sector 39 with CCD 4; Sector 61 with CCD 2; and Sectors 62 and 63 with CCD 3. 

The 2-minute cadence data for TOI-199 were processed by the \textit{TESS} Science Processing Operation Center \citep[SPOC,][]{Jenkins16} at NASA Ames Research Center. A potential transit signal was identified in the SPOC transit search \citep{Jenkins02, Jenkins10, Jenkins20} of the light curve, and designated as a \textit{TESS} Object of Interest (TOI) by the \textit{TESS} Science Office based on SPOC data validation results \citep{Twicken18, Li19} indicating that it was consistent with a transiting planet. The SPOC difference image centroiding analysis locates the source of the transit signal to within $1.0 \pm 2.5"$ of the target star's catalog position \citep{Twicken18}. The planetary candidate TOI-199.01, so named because it is the first transit signal detected in this light curve, is listed in the ExoFOP-TESS archive\footnote{Located at \url{https://exofop.ipac.caltech.edu/tess/target.php?id=309792357}}, with a period of $\mathrm{P_{01} = 104.87075\, d}$. Prior to its release as a TOI, the WINE team had independently identified the initial single transit event in Sector 2 and begun follow-up.

Seven transits of the planetary candidate TOI-199.01 were observable for \textit{TESS}, in sectors 2, 10, 13, 29, 32, and 35 respectively. Numbering the first observed transit in sector 2 as transit 1, these correspond to transits 1, 3, 4, 8, 9, 10, and 17 (note that transit 2 fell in the data gap between sectors 5 and 6). 

\begin{figure*}[htb!]
    \centering
    \includegraphics[width=1\hsize]{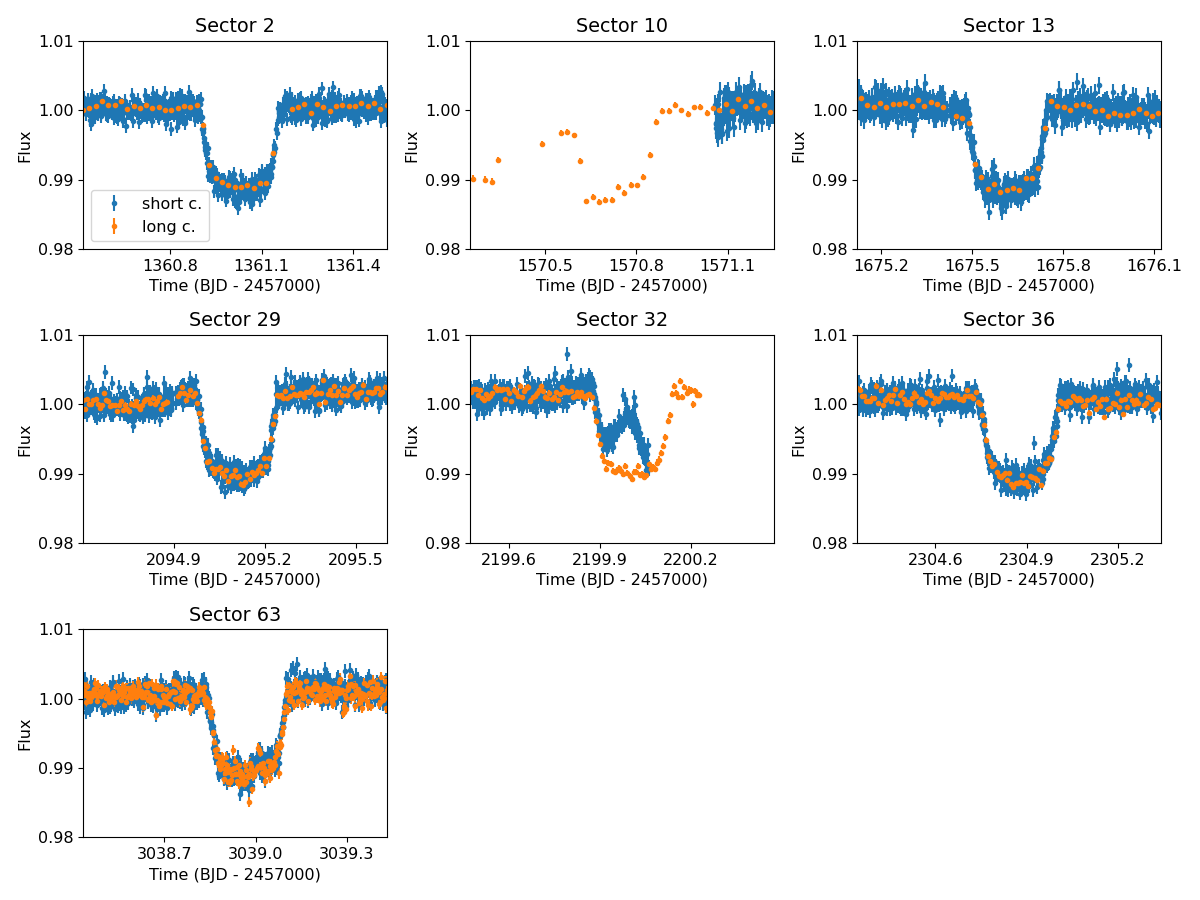}  
    \caption{\textit{TESS} light curves for TOI-199, for the seven sectors with transits. The short-cadence PDCSAP light curves are shown in blue, and the FFI light curves extracted with \texttt{tesseract} in orange. For Sectors 10 and 32 the transits could not be correctly recovered from the PDCSAP light curves.}
    \label{fig:lightcurve}
\end{figure*}

Unfortunately, one of the transits, close to the start of sector 10, is cut off by quality checks in the 2-minute cadence Presearch Data Conditioning \citep[PDCSAP,][]{Stumpe12, Stumpe14, Smith12} light curve, and another at the end of sector 32 is distorted. However, both are clearly seen in the light curves obtained from the full frame images (FFIs), which we extracted with our own \texttt{tesseract} pipeline (Rojas et al. in prep. \footnote{Publicly available at \url{https://github.com/astrofelipe/tesseract}}). We attempted to re-reduce the 2-minute cadence data to recover them, but were unable to retrieve the transit at the edge of sector 10, where the light curve drops off sharply. 
Therefore, we chose to use the FFI light curves from \texttt{tesseract} for the whole of our analysis. The PDCSAP and FFI light curves for the seven sectors with transits are shown in Fig. \ref{fig:lightcurve}; note that the cadence for the FFIs changed from 30 minutes for the prime mission (top row) to 10 minutes for the first extended mission (middle row), and to 200s for the second extended mission (bottom row).

\subsection{Follow-up photometry}

At $21\arcsec$ per pixel, the \textit{TESS} pixels are relatively large, meaning that nearby companions can contaminate its photometry. Follow-up photometry is used to identify such cases of false positives, and to monitor transits not observed by TESS. TOI-199 was observed by five facilities, described in the following sub-sections.

\subsubsection{\textit{ASTEP}}

Antarctica Search for Transiting ExoPlanets \citep[ASTEP,][]{Guillot15, Mekarnia16} is a 0.4\,m telescope equipped with a Wynne Newtonian coma corrector, located on the East Antarctic Plateau. Until December 2021 it was equipped with a 4k $\times$ 4k front-illuminated FLI Proline KAF-16801E CCD with an image scale of 0.''93 pixel$^{-1}$ resulting in a 1$^ {o} \times 1^ {o}$ corrected field of view. The focal instrument dichroic plate split the beam into a blue wavelength channel for guiding, and a non-filtered red science channel roughly matching an Rc transmission curve \citep{Abe13}.

In January 2022, the focal box was replaced with a new one with two high sensitivity cameras including an Andor iKon$-$L 936 at red wavelengths. The image scale is 1.39'' pixel$^{-1}$ with a transmission curve centred on 850 $\pm$138~nm \citep{Schmider22}.

The telescope is automated or remotely operated when needed. Due to the extremely low data transmission rate at the Concordia Station, the data are processed on-site using IDL \citep{Mekarnia16} and Python \citep{Dransfield22} aperture photometry pipelines.

Three observations of TOI-199 were made with ASTEP, all of them with clear sky, winds between 3 and 5 m~s$^{-1}$ and temperature ranging between $-62$ and $-69^o$C. An egress was observed on 31st March 2021 with a FWHM of 4.3'', corresponding to the transit observed by \textit{TESS} in Sector 36. Full transits were observed on 13th July 2021 and 6th September 2022 with a FWHM of 4.3 and 5.9'' respectively, which were not observed by \textit{TESS}. The Moon, 80\% illuminated, was present during the 31st March 2021 and 6th September 2022 observations. The light curves are shown in Fig. \ref{fig:followup-lc} (top row).

\subsubsection{\textit{PEST}}

The Perth Exoplanet Survey Telescope (PEST) is a backyard observatory in Perth, Australia, operated by Thiam-Guan Tan. PEST was used to observe a transit egress of TOI-199.01 on 16th December 2021, corresponding to the transit observed by \textit{TESS} in Sector 32. The observation was conducted with an Rc filter and 30s integration times. At the time, the 0.3 m telescope was equipped with a $1530 \times 1020$ SBIG ST-8XME camera with an image scale of $1.2\arcsec\mathrm{/pixel}$ resulting in a $31\arcmin \times 21\arcmin$ field of view.  A custom pipeline based on C-Munipack was used to calibrate the images and extract the differential photometry using an aperture with radius $8.6\arcsec$. The light curve is shown in Fig. \ref{fig:followup-lc} (middle row, left panel).

\subsubsection{\textit{NEOSSat}}

The Near Earth Object Surveillance Satellite (NEOSSat, \citealt{NEOSSat}) is a Canadian microsatellite with a 0.15 m F/6 telescope which has a $0.86 \degr \times 0.86 \degr$ field of view. Although designed for detecting and tracking near-Sun asteroids, it is also suitable for transit observations of bright stars \citep{Fox22}. NEOSSat observed a full transit of TOI-199.01 on 31st March 2021 with no filter, corresponding to the one observed by \textit{TESS} in Sector 36. The light curve is shown in Fig. \ref{fig:followup-lc} (middle row, centre panel).

\subsubsection{\textit{Hazelwood}}

The Hazelwood Observatory is a backyard observatory operated by Chris Stockdale in Victoria, Australia, with a 0.32 m Planewave CDK telescope working at f/8, a SBIG STT3200 $\mathrm{2.2k \times 1.5k}$ CCD, giving a $20\arcmin \times 13\arcmin$ field of view and $0.55\arcsec$ per pixel. The camera is equipped with B, V, Rc, Ic, g’, r’, i’ and z’ filters (Astrodon Interference). Typical FWHM is $2.2\arcsec$ to $3.0\arcsec$. The Hazelwood Observatory observed an egress of TOI-199.01 in Rc on 31st March 2021, corresponding to the transit observed by \textit{TESS} in Sector 36. Time series images were collected, bias, dark and flat fielded and the transit analysed with {\tt AstroImageJ} \citep{Collins:2017}. The reduced data was then uploaded to ExoFOP. The light curve is shown in Fig. \ref{fig:followup-lc} (middle row, right panel).

\subsubsection{\textit{LCO}}

\begin{figure*}[htb!]
    \centering
    \includegraphics[width=1\hsize]{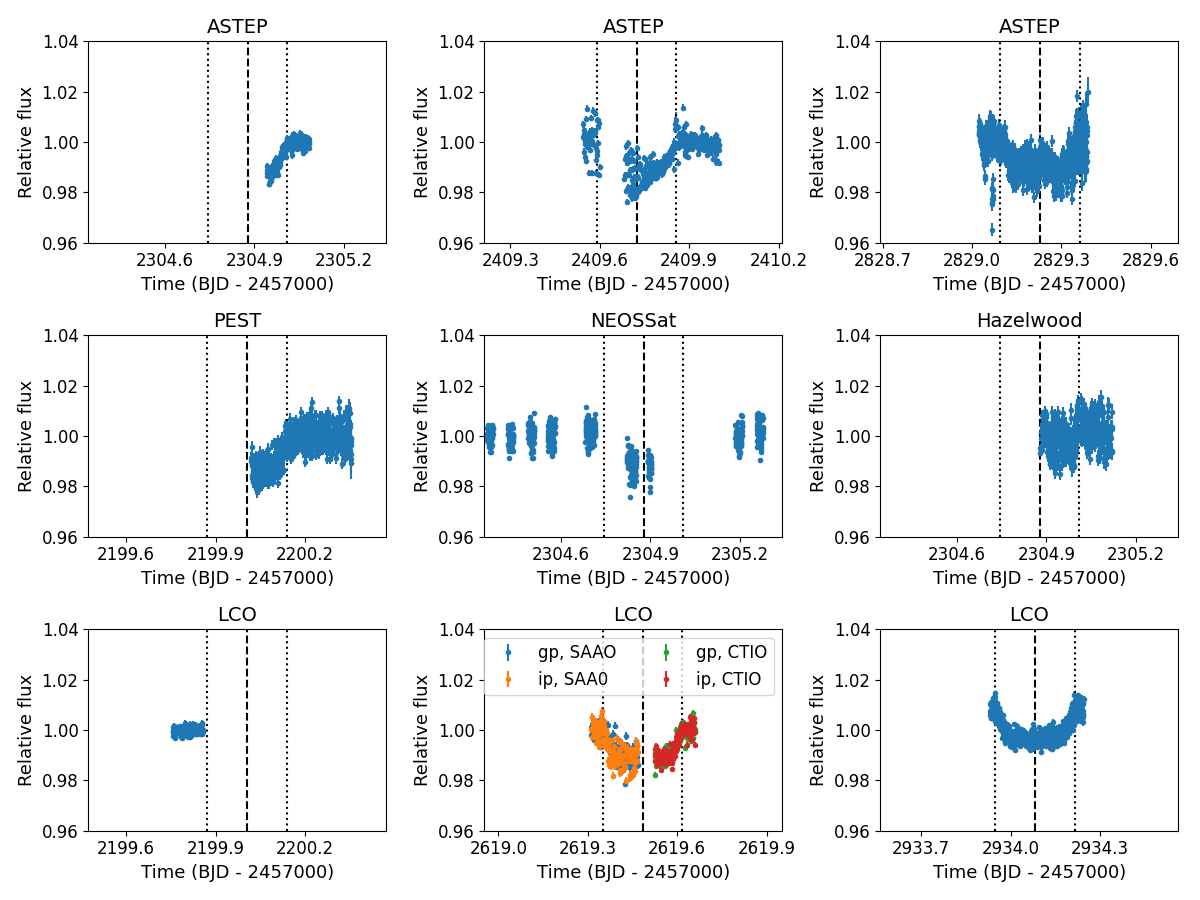}  
    \caption{Follow-up photometry for TOI-199. Top row: Light curves for ASTEP observations - an egress on 31st March 2021 (left), a full transit on 13th July 2021 (centre), and a full transit on 6th September 2022 (right). Middle Row: Light curves for PEST (left, egress on 16th December 2020), NEOSSat (centre, full transit on 31st March 2021), and Hazelwood (right, egress on 31st March 2021) observations. Bottom row: light curves for LCO observations - a pre-ingress flat curve on 16th December 2020 (left), two ingresses and egresses on 8th February 2022 (centre), for which the points are colour-coded by filter and site, and a full transit on 20th December 2022 (right). In all panels, the dashed vertical line indicates the transit midpoint, and the dotted vertical lines the egress and ingress.}
    \label{fig:followup-lc}
\end{figure*}

The Las Cumbres Observatory global telescope network \citep[LCOGT,][]{Brown13} is a globally distributed network of telescopes. The 1\,m telescopes are equipped with $4096\times4096$ SINISTRO cameras having an image scale of $0.389\farcs$ per pixel, resulting in a $26\arcmin\times26\arcmin$ field of view. LCO observed TOI-199 several times with the Sinistro instrument: a pre-ingress flat curve on 16th December 2021, corresponding to the transit observed by \textit{TESS} in Sector 32, from the Cerro Tololo Inter-American Observatory (CTIO) site in zs filter; an ingress on 8th February 2022, from the South African Astronomical Observatory (SAAO) site, in alternating gp and ip filters; the egress of the same transit, from the CTIO site, also in alternating gp and ip filters; and a full transit on 20th Decemeber 2022, from the Siding Springs Observatory (SSO) site. The images were calibrated by the standard LCOGT {\tt BANZAI} pipeline \citep{McCully:2018}, and photometric data were extracted using {\tt AstroImageJ}. The light curves are shown in Fig. \ref{fig:followup-lc} (bottom row).

\subsection{High-resolution imaging}

\begin{figure}[htb]
    \centering
    \includegraphics[width=1\hsize]{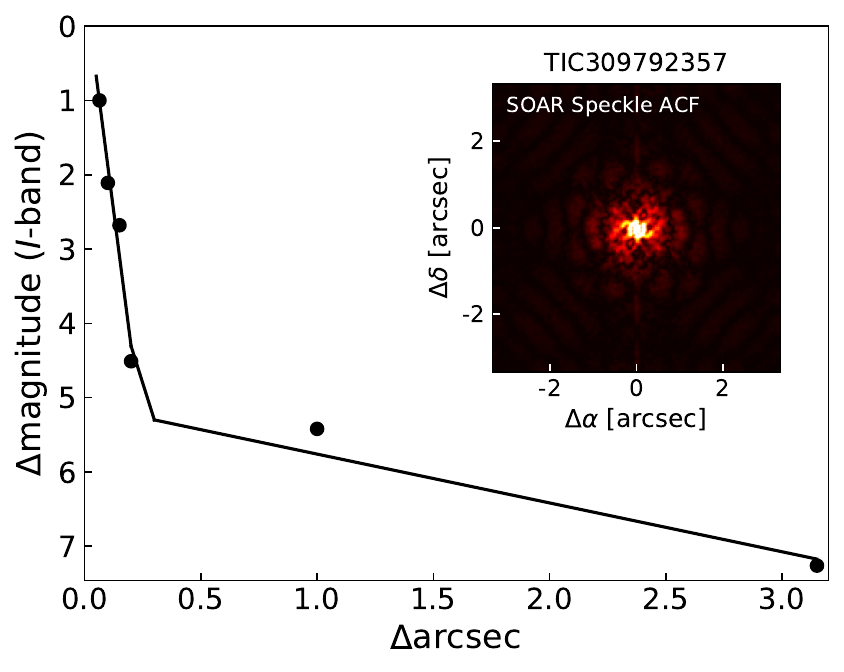}
    \caption{Contrast curve and speckle auto-correlation function from the HRCam at SOAR for TOI-199. The black points and solid line indicate the $5\sigma$ contrast curve; the inset shows the speckle auto-correlation function. No nearby sources are detected.}
    \label{fig:speckle-imaging}
\end{figure}

Given the relatively large pixel size of \textit{TESS}, high-resolution imaging from the ground is a valuable tool to assess possible blends and contamination from nearby sources. The SOAR \textit{TESS} survey \citep{Ziegler20} observes \textit{TESS} exoplanet candidate hosts with speckle imaging using the high-resolution camera (HRCam) imager on the 4.1-m Southern Astrophysical Research (SOAR) telescope at Cerro Pachón, Chile \citep{Tokovinin18}, in order to detect such nearby sources. TOI-199 was observed on the night of 18 February 2019, with no nearby sources detected within $3\arcsec$. The contrast curve and speckle auto-correlation function are shown in Fig. \ref{fig:speckle-imaging}.

\subsection{Spectroscopic data}

\begin{figure}[htb]
    \centering
    \includegraphics[width=1\hsize]{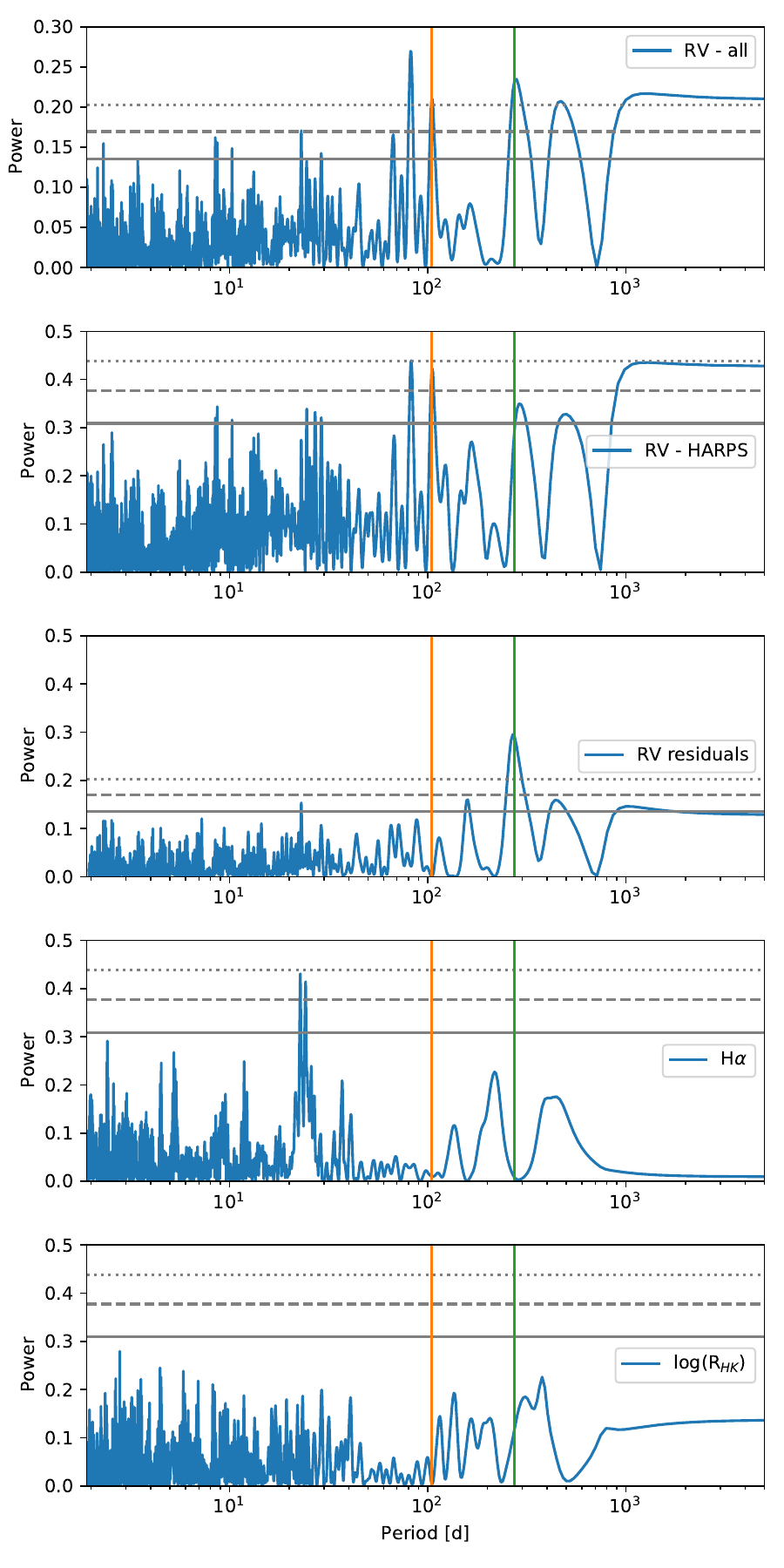}
    \caption{Periodograms of: the joint RVs (top panel); the HARPS RVs alone (second panel); the joint RV residuals to a one-planet fit for TOI-199 b (third panel); the HARPS H$_\alpha$ indicator (fourth panel); and the HARPS log($R^\prime_{HK}$) activity indicator (bottom panel). The grey solid, dashed, and dotted horizontal lines indicate false alarm probability levels of 10\%, 1\% and 0.1\% respectively. The periods of the two planets are indicated with vertical orange and green lines. We note that the peak at $\sim$81d is a one-year alias of the $\sim$104d planet.}
    \label{fig:periodograms}
\end{figure}

The WINE consortium carried out spectroscopic follow-up of TOI-199 with the HARPS, FEROS, CORALIE, and CHIRON spectrographs. We also received additional CORALIE and CHIRON RVs from other teams, which we incorporated into the analysis. All the resulting radial velocities (RVs) are shown in Fig. \ref{fig:RVs_TTVs} (top panel). Figure \ref{fig:periodograms} shows the GLS periodograms of the joint RVs from all four spectrographs (top panel); of the HARPS RVs alone (second panel); of the joint RV residuals to a fit of the inner planet alone (third panel); and of the H$_\alpha$ and log($R^\prime_{HK}$) activity indicators computed from the HARPS spectra (fourth and fifth panels).

\subsubsection{HARPS}

We observed TOI-199 with the HARPS spectrograph \citep{Mayor03} at the 3.6m telescope at La Silla, resolving power $\mathrm{R=120\,000}$, between 12th December 2018 and 9th March 2021. We obtained 46 spectra, under Program IDs 0101.C-0510, 0102.C-0451, 0104.C-0413, and 106.21ER.001. For the first three programs (December 2018 through February 2020), we operated in simultaneous sky mode, obtaining 26 spectra with a 900s exposure duration. For the last (December 2020 through March 2021), we switched to simultaneous Fabry-Perot mode as part of an overall revision of our observing program, obtaining 20 spectra with an increased 1200s exposure duration. The spectra have a median SNR of 62. We processed the spectra using the \texttt{CERES} pipeline \citep{Brahm17CERES}, which provides both RVs and several activity indicators: the CCF bisector \citep[BIS, e.g.][]{Queloz01}, and the H$_\alpha$ \citep{Boisse09}, log($R^\prime_{HK}$) \citep{Duncan91,Noyes84}, Na~II, and He~I \citep{Gomes11} activity indices. The resulting RVs and activity indicators are listed in Table \ref{tab:harps-data}; the RVs have a median error of 2.4 m/s.

\subsubsection{FEROS}

We observed TOI-199 with the FEROS spectrograph \citep{Kaufer99} at the MPG 2.2m telescope at La Silla, resolving power $\mathrm{R=50\,000}$, between 27th November 2018 and 4th March 2020. We obtained 48 spectra, under Program IDs 0102.A-9003, 0102.A-9006, 0102.A-9011, 0102.A-9029, 0103.A-9008, and 0104.A-9007 in Object-Calibration mode, with a 500s exposure duration (save three spectra with increased exposure time due to poor observing conditions, two of 700s and one of 1000s). The spectra have a median SNR of 90. As with the HARPS data, we processed the spectra using the \texttt{CERES} pipeline. The resulting RVs and activity indicators are listed in Table \ref{tab:feros-data}; the RVs have a median error of 6.3 m/s.

\subsubsection{CORALIE}

TOI-199 was observed with the CORALIE spectrograph \citep{CORALIE} at the Swiss 1.2m Euler telescope at La Silla by both the WINE and Swiss teams. In total, 19 spectra were obtained between 27th December 2018 and 2nd December 2019, with a median exposure duration of 1200s. CORALIE is a fibre-fed echelle spectrograph with a $2\arcsec$ science fibre, and a secondary fibre with a Fabry-Perot for simultaneous wavelength calibration. It has a spectral resolution of $\mathrm{R=60\,000}$. RVs are extracted using the standard CORALIE DRS by cross-correlating the spectra with a binary G2V template \citep{1996A&AS..119..373B,2002Msngr.110....9P}. The BIS, FWHM, and other line-profile diagnostics were also computed via the CORALIE DRS, as was the $\mathrm{H_\alpha}$ index for each spectrum to check for possible variation in stellar activity. The resulting RVs and activity indicators are listed in Table \ref{tab:coralie-data}; the RVs have a median error of 13.7 m/s.

\subsubsection{CHIRON}

TOI-199 was observed with the CHIRON spectrograph \citep{CHIRON} at the SMARTS 1.5-meter telescope at CTIO by both the WINE team and Dr Carleo. In total, 26 spectra were obtained between 11th March 2019 and 13th December 2020; the WINE team observed with single 1500s exposures, while Dr Carleo observed sets of three 600s exposures. RVs were extracted following \cite{Jones19}. The resulting RVs and activity indicators are listed in table \ref{tab:chiron-data}; the RVs have a median error of 11.3 m/s.

\section{Analysis} \label{sec:analysis}

\subsection{Stellar parameters}\label{sec:starparam}

We use a two-part process to characterise the host star. First, we employed the co-added FEROS spectra to determine the atmospheric parameters of TOI-199. To do this, we used the \texttt{ZASPE} code \citep{Brahm17ZASPE}, which compares the observed spectrum to a grid of synthetic models (generated from the ATLAS9 model atmospheres, \citealt{Castelli03}) in order to determine the effective temperature $\mathrm{T_{eff}}$, surface gravity $\mathrm{\log{g}}$, metallicity [Fe/H], and projected rotational velocity $\mathrm{v \sin{i}}$. 

Second, we followed the second procedure described in \cite{Brahm19PARSEC} to determine the physical parameters of the host star. To summarize, we compare the broadband photometric measurements (converted to absolute magnitudes via the \textit{Gaia} DR3 \citep{GAIA2016, GaiaDR3} parallax) with the stellar evolutionary models of \cite{Bressan12}. We use the \texttt{emcee} package \citep{Foreman-Mackey13} to sample the posterior distributions. Through this procedure, we determine the age, mass, radius, luminosity, density, and extinction. This also enables us to compute new values for $\mathrm{T_{eff}}$ and $\mathrm{\log{g}}$, the latter of which is more precise than the one previously determined through \texttt{ZASPE}. Therefore, we iterate the entire procedure, using the new $\mathrm{\log{g}}$ as an additional input parameter for \texttt{ZASPE}. One iteration was sufficient for the $\mathrm{\log{g}}$ value to converge. 

The final stellar parameters computed, and other observational properties of TOI-199 are presented in Table \ref{tab:starparams}.  We find that TOI-199 is a late G-type star. The quoted error bars for the stellar parameters are internal errors, computed as the $\pm 1 \sigma$ interval around the median of the posterior for each parameter; they do not take into account systematic errors in the stellar models. 

\begin{table}[htb]
\begin{center}
\caption{Stellar parameters of TOI-199.}
\label{tab:starparams}
\centering
\begin{tabular}{lcr}
\hline \hline
Parameter & Value                        & Reference \\
\hline
Names     &      CD-60 1144               &           \\
          & TIC309792357, TOI-199       & \textit{TESS}      \\
          &  J05202531-5953443    & 2MASS     \\
          & 4762582895440787712 & \textit{Gaia} DR3  \\
RA \dotfill (J2000) &  $05^{\mathrm h} 20^{\mathrm m} 25.3151633088^{\mathrm s}$ & \textit{Gaia} DR3  \\
DEC \dotfill (J2000) & $-59{\arcdeg} 53{\arcmin} 44.463213656{\arcsec}$  & \textit{Gaia} DR3  \\
pm$^{\rm RA}$ \hfill [mas yr$^{-1}$] & 45.864 & \textit{Gaia} DR3  \\
pm$^{\rm DEC}$ \hfill [mas yr$^{-1}$] & 58.445 & \textit{Gaia} DR3  \\
$\pi$ \dotfill [mas] & 9.8296  & \textit{Gaia} DR3  \\
\hline
T \dotfill [mag] & 10.0391 & \textit{TESS}      \\
B \dotfill [mag] &  11.59 & Tycho-2     \\
V \dotfill [mag] & 10.70 & Tycho-2     \\
J \dotfill [mag] & 9.329 & 2MASS     \\
H \dotfill [mag] & 8.952 & 2MASS     \\
K \dotfill [mag] & 8.814 & 2MASS     \\
\hline
$T_{\rm eff}$ \dotfill [K] & $5255_{-10}^{+12}$ (126)  & this work \\
Spectral type  \dotfill & G9V  & PM13 \\
Fe/H \dotfill [dex] & $0.22 \pm 0.03$ & this work \\
$\log{g}$ \dotfill [dex] & $4.582_{-0.006}^{0.003}$  & this work \\
$v \sin{i}$ \dotfill [$\mathrm{kms^{-1}}$] & $0.5 \pm 0.6$ & this work \\
R$_\star$ \dotfill [R$_\odot$] & $0.820 \pm 0.003$ (0.03) & this work \\
M$_\star$ \dotfill [M$_\odot$] & $0.936_{-0.005}^{+0.003}$ (0.009) & this work \\
L$_\star$ \dotfill [L$_\odot$] & $0.459_{-0.003}^{+0.005}$ (0.009) & this work \\
$\rho_\star$ \dotfill [g cm$^{-3}$] & $2.39_{-0.04}^{+0.02}$ & this work \\
Age \dotfill [Gyr] & $0.8_{-0.6}^{+1.2}$ (0.72) & this work \\
A$_V$ \dotfill [mag] & $0.01_{-0.01}^{+0.02}$ & this work \\
$\log R'_{\rm hk}$ \dotfill & $-4.77 \pm 0.13$ & this work \\
\hline
\end{tabular}
\end{center}
    \textit{TESS}: \textit{TESS} Input Catalog \citep{Stassun2019}; 2MASS: Two-micron All Sky Survey \citep{2MASS}; \textit{Gaia} DR3: \textit{\textit{Gaia}} Data Release 3 \citep{GAIA2016, GaiaDR3}; Tycho-2: the Tycho-2 Catalogue \citep{Tycho-2}; PM13: using the tables of \cite{Pecaut13}. Values in parentheses correspond to the fundamental uncertainty floors following \cite{Tayar22}.
\end{table}

Recently, \cite{Tayar22} examined the uncertainty floors on fundamental stellar parameters. Specifically, they find fundamental floors of $2.4\%$ for $T_{\rm eff}$, $2\%$ for L$_\star$, and $4.2\%$ for R$_\star$ due to the uncertainties on input observables such as the bolometric flux and angular diameter. We also report these values in Table \ref{tab:starparams}. For M$_\star$ and the stellar age, they perform a comparison between four model grids. We use the online interface provided to carry out the same comparison for TOI-199, and report the standard deviation of the resulting values in Table \ref{tab:starparams}.

\subsection{TTV extraction}

\begin{figure}[htb]
    \centering
    \includegraphics[width=1\hsize]{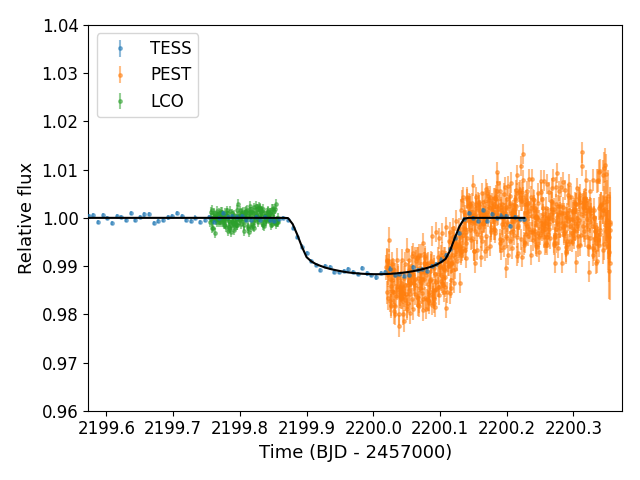}
    \caption{\textit{TESS} FFI light curve for sector 32 extracted with \texttt{tesseract} (blue points), PEST light curve (orange points), LCO light curve (green points) and fitted model to the \textit{TESS} data (black line) for transit 9.}
    \label{fig:transit9}
\end{figure}

\begin{figure}[htb]
    \centering
    \includegraphics[width=1\hsize]{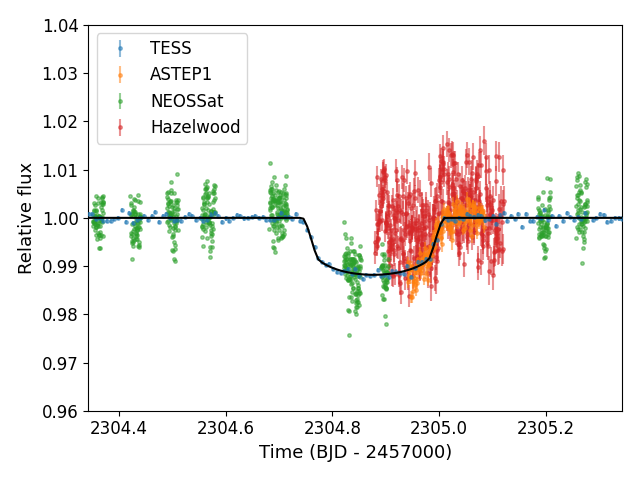}
    \caption{\textit{TESS} FFI light curve for sector 36 extracted with \texttt{tesseract} (blue points), ASTEP light curve (orange points), NEOSSat light curve (green points), Hazelwood light curve (red points), and fitted model to the \textit{TESS} data (black line) for transit 10.}
    \label{fig:transit10}
\end{figure}

We used the \texttt{juliet}\footnote{Available at \url{https://github.com/nespinoza/juliet}} software \citep[][]{Espinoza19juliet} to extract the TTVs. We focussed on the \textit{TESS}, ASTEP, and LCO transits, as the second and third ASTEP transits and second and third LCO transits are the only ones not also observed by \textit{TESS}. The other light curves are consistent with the \textit{TESS} observations, as shown in Figs. \ref{fig:transit9} and \ref{fig:transit10}.

\begin{figure*}[htb]
    \centering
    \includegraphics[width=1\hsize]{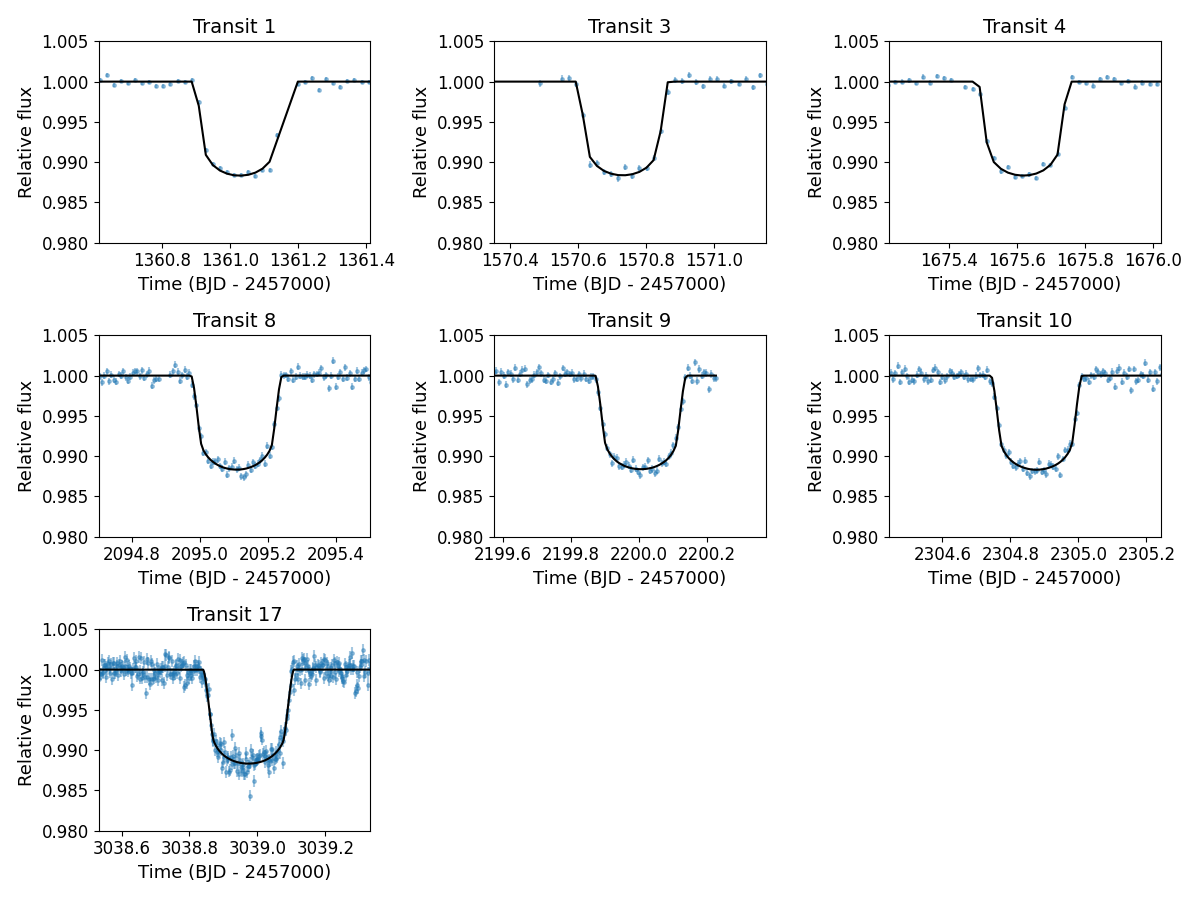}
    \caption{\textit{TESS} FFI light curves, extracted with \texttt{tesseract} (blue points), and fitted models (black lines) for the seven transits observed with \textit{TESS}. The GP components of each model have been subtracted.}
    \label{fig:tess-juliet}
\end{figure*}

\begin{figure*}[htb]
    \centering
     \includegraphics[width=1\hsize]{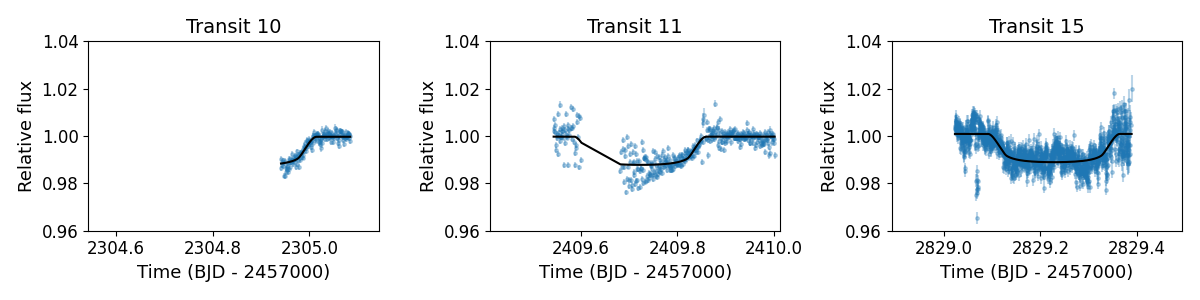}
    \caption{ASTEP light curves (blue points) and fitted models (black lines) for the three transits observed with ASTEP.}
    \label{fig:astep-juliet}
\end{figure*}

\begin{figure*}[htb]
    \centering
     \includegraphics[width=1\hsize]{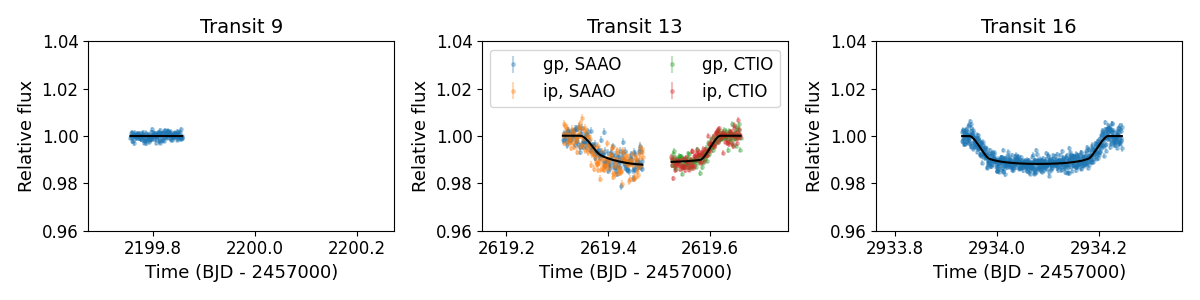}
    \caption{LCO light curves (coloured points) and fitted models (black lines) for the three transits observed with LCO. The data for transit 13 are coloured by filter and site, as in Fig. \ref{fig:followup-lc}.}
    \label{fig:lco-juliet}
\end{figure*}

\begin{figure*}[htb]
    \centering
    \includegraphics[width=1\hsize]{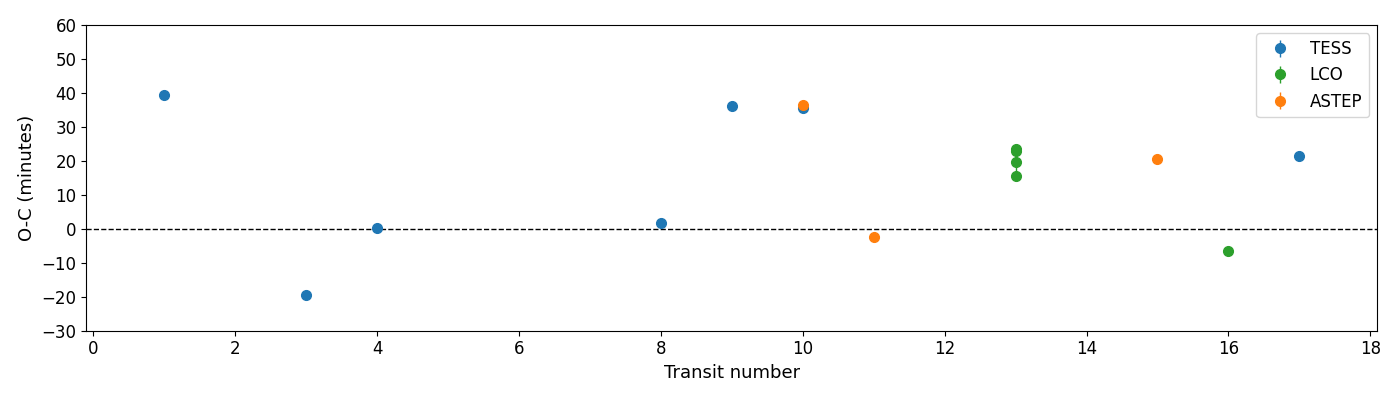}
    \caption{O-C plot (difference between predicted and observed transit midpoint times) for the \textit{TESS}, ASTEP, and LCO transits. The error bars are generally smaller than the marker size.}
    \label{fig:O-C}
\end{figure*}

The \texttt{tesseract} FFI light curves are uncorrected for systematics and other effects. We used a two-step method to fit a Gaussian Process (GP) to each \textit{TESS} sector, first optimizing the hyperparameters on the out-of transit data, with broad priors of $\mathcal{J}(10^{-6},10^{6})$ for $\sigma_\mathrm{{GP, TESS, i}}$, where $\mathcal{J}(a,b)$ is a Jeffreys or log-uniform distribution between $a$ and $b$, and $\mathcal{J}(10^{-3},10^{3})$ for $\rho_\mathrm{{GP, TESS, i}}$ for each \textit{TESS} sector $\mathrm{i}$. We then used the resulting posterior parameters as priors on the GP hyperparameters in the full fit including the transits, adopting the upper and lower $1\sigma$ bounds from the posteriors as limits on new Jeffreys distributions. 
Each transit is fit independently, with individual priors for the flux offsets $m_\mathrm{{flux,instrument,transit}}$ and jitters $\sigma_{\mathrm{instrument,transit}}$.  \texttt{juliet} also requires priors for the dilution factors $m_\mathrm{{d,instrument,transit}}$, which we have fixed to 1 for all instruments and transits.
The limb-darkening parameters q$_\mathrm{{1,instrument}}$ and q$_\mathrm{{2,instrument}}$ are common to all transits observed with each instrument (save for LCO, where the observations were made at different sites or/and with different filters). For each q$_\mathrm{{i,instrument}}$ we adopted truncated normal priors, where the mean of each prior was based on the derivation of limiting quadratic coefficients from the nonlinear coefficients described by \cite{Espinoza15}\footnote{as implemented in the public repository https://github.com/nespinoza/limb-darkening}. We used ATLAS model atmospheres interpolated over 100 $\mu$-points, as in \citep{Claret11}, and instrument response functions from the instrument website (for LCO, which provides profiles for all available filters) or from the SVO Filter Profile Service (\citealt{SVO2012, SVO2020}; for TESS there is a specific profile; we adopted a generic Johnson V filter profile for NEOSSat, which observed without filter, and a Cousins R filter for the rest of the instruments, which observed with R filters. For the transit times $\mathrm{T_{b,instrument,transit}}$, we adopt uniform priors with a width of 6h, and midpoints determined from the orbital period listed in ExoFOP. The ExoFOP values for the planet-to-star radius ratio p$_\mathrm{b}$ and the impact parameter b$_\mathrm{b}$ were also used as priors. The eccentricity $\mathrm{e_b}$ and $\omega_\mathrm{b}$ were fixed to 0 and $90\arcdeg$, respectively. While we may expect planets at $\mathrm{\sim 100\,d}$ to have non-circular orbits, the shape of the transit will be noticeably altered only for large eccentricities, so a fixed eccentricity suffices for the TTV extraction (we note the eccentricity is a free parameter in the full RV+TTV model, see Section 3.3). The priors and posteriors for the full fit are listed in Table \ref{tab:juliet-planet-pp} for the planetary parameters, and in Table \ref{tab:juliet-inst-pp} for the instrumental and GP parameters. The fitted \textit{TESS}, ASTEP, and LCO transits are shown in Figs. \ref{fig:tess-juliet}, \ref{fig:astep-juliet}, and \ref{fig:lco-juliet} respectively, and the O-C plot is shown in Fig. \ref{fig:O-C}.

\startlongtable
\begin{deluxetable*}{lll}
\tablecaption{Prior and posterior planetary parameter distributions for the TTV extraction with \texttt{juliet}.\textit{Top}: Fitted parameters. \textit{Bottom}: derived orbital parameters and physical parameters. \label{tab:juliet-planet-pp}}
\tablehead{\colhead{Parameter} & \colhead{Prior\tablenotemark{a}} & \colhead{Posterior}}
\startdata
p$_\mathrm{b}$ \dotfill &$\mathcal{N}(0.1023,0.00915)$  & \julietpb \\
b$_\mathrm{b}$ \dotfill &$\mathcal{N}(0.47,0.555)$  & \julietbb \\
e$_\mathrm{b}$ & 0.0 (fixed) & 0.0 (fixed) \\
$\omega_\mathrm{b}$ & 90.0 (fixed)  & 90.0 (fixed) \\                
$\rho$ \dotfill & $\mathcal{J}(100.0,10000.0)$ & \julietrho \\       
$\mathrm{T_{b,TESS, 2}}$   & $\mathcal{U}(2458360.89,2458361.14)$ & \julietTTESStwo \\
$\mathrm{T_{b,TESS, 10}}$  \dotfill & $\mathcal{U}(2458570.63,2458570.88)$ & \julietTTESSten \\
$\mathrm{T_{b,TESS, 13}}$  \dotfill & $\mathcal{U}(2458675.50,2458675.75)$ & \julietTTESSthir \\
$\mathrm{T_{b,TESS, 29}}$  \dotfill & $\mathcal{U}(2459094.98,2459095.23)$ & \julietTTESStn \\
$\mathrm{T_{b,TESS, 32}}$  \dotfill & $\mathcal{U}(2459199.85,2459200.10)$ & \julietTTESStt \\
$\mathrm{T_{b,TESS, 36}}$  \dotfill & $\mathcal{U}(2459304.72,2459304.97)$ & \julietTTESSts \\
$\mathrm{T_{b,TESS, 63}}$  \dotfill & $\mathcal{U}(2460038.81,2460039.06)$ & \julietTTESSst \\
$\mathrm{T_{b,ASTEP, 1}}$  \dotfill & $\mathcal{U}(2459304.72,2459304.97)$  & \julietTASTEPone \\
$\mathrm{T_{b,ASTEP,2}}$  \dotfill & $\mathcal{U}(2459409.59,2459409.84)$ &\julietTASTEPtwo \\
$\mathrm{T_{b,ASTEP,3}}$  \dotfill & $\mathcal{U}(2459829.07,2459829.32)$ & \julietTASTEPthr \\
$\mathrm{T_{b,PEST}}$  \dotfill & $\mathcal{U}(2459199.85,2459200.10)$ & \julietTPEST \\
$\mathrm{T_{b,NEOSSat}}$  \dotfill & $\mathcal{U}(2459304.72,2459304.97)$   & \julietTNEOS \\
$\mathrm{T_{b,Hwd}}$  \dotfill & $\mathcal{U}(2459304.72,2459304.97)$ & \julietTHwd \\
$\mathrm{T_{b,LCO,1}}$  \dotfill & $\mathcal{U}(2459199.85,2459200.10)$ & \julietTLCOone \\
$\mathrm{T_{b,LCO,2}}$  \dotfill & $\mathcal{U}(2459619.33,2459619.58)$ & \julietTLCOtwo \\
$\mathrm{T_{b,LCO,3}}$  \dotfill & $\mathcal{U}(2459619.33,2459619.58)$ & \julietTLCOthr \\
$\mathrm{T_{b,LCO,4}}$  \dotfill & $\mathcal{U}(2459619.33,2459619.58)$ & \julietTLCOfo \\
$\mathrm{T_{b,LCO,5}}$  \dotfill & $\mathcal{U}(2459619.33,2459619.58)$ & \julietTLCOfi \\
$\mathrm{T_{b,LCO,6}}$  \dotfill & $\mathcal{U}(2459933.94,2459934.19)$ & \julietTLCOsix \\
\hline
$\mathrm{P_b}$\tablenotemark{b} \dotfill [d]  & \multicolumn{1}{c}{--} & \julietP \\
$\mathrm{a_b}$ \dotfill [au]  & \multicolumn{1}{c}{--} &\julieta \\
T0$_\mathrm{b}$\tablenotemark{b} \dotfill [BJD]  & \multicolumn{1}{c}{--} & \juliettzero \\
$\mathrm{R_b}$ \dotfill [$\mathrm{R_J}$]  & \multicolumn{1}{c}{--} & \julietRp \\
$\mathrm{i_b}$ \dotfill [deg]  & \multicolumn{1}{c}{--} & $89.81 \pm 0.02$ \\
\enddata
\tablenotetext{a}{$\mathcal{U}(a,b)$ indicates a uniform distribution between $a$ and $b$; $\mathcal{N}(a, b)$ a normal distribution with mean $a$ and standard deviation $b$; $\mathcal{J}(a,b)$ a Jeffreys or log-uniform distribution between $a$ and $b$.}
\tablenotetext{b}{\texttt{juliet} computes the P and T0 in a TTV fit as the slope and intercept, respectively, of a least-squares fit to the transit times. We adopt the value from the full TTV+RV model (Sect. \ref{sec3.3}) as our final period.}
\end{deluxetable*}

\subsection{Joint TTV and RV model}
\label{sec3.3}

We used the \texttt{Exo-Striker}\footnote{Available at \url{https://github.com/3fon3fonov/exostriker}} software \citep{TrifonovExoStriker} to jointly fit the RV and the TTV data of TOI-199. We use a very similar approach to the TTV+RV N-body modelling scheme analysis done for the TOI-2202 system in \citet{Trifonov21}; we refer to this paper for more details on the TTV+RV modeling scheme used here. Briefly, the N-body RV modelling used is native to \texttt{Exo-Striker}, whereas the TTV modelling uses the \texttt{TTVFast} package \citep{Deck14}. The fitted parameters for each planet in our joint model are the semi-amplitude $K$, the orbital period~$P$, eccentricity $e$, argument of periastron $\omega$, and mean-anomaly $M_0$. For our fitting analysis, we assumed coplanar, edge-on and prograde 
two-planet system (i.e., $i_b$,$i_c$ = 90$^\circ$ and $\Delta i$ = 0$^\circ$), and for the stellar mass of TOI-199 we adopt our best estimate of $0.936_{-0.005}^{+0.003}$ $M_\odot$ (note that the stellar mass uncertainties are not included in the modelling, but are incorporated into the final uncertainties of the derived planetary parameters).
The time step in the dynamical model was set to 1\,day,  assuring adequate model precision. Additionally, we optimise two parameters for each Doppler data set, the  offset and the RV ``jitter". This latter parameter is added in quadrature to the nominal uncertainties of the RV data \citep{Baluev2009}. 

We constructed parameter posteriors by the nested sampling algorithm \citep{Skilling2004}, implemented via the \texttt{dynesty} package \citep{Speagle20}.
For TOI-199 b, we used the previous light curve characterization with \texttt{juliet} and the RV period search results to constrain the priors on the parameters. For the period of the perturber, we first ran initial joint fits with broad priors in $P_c$ between 150 and 330 days, $e_c$ between 0 and 0.4, and $\omega_c$ and M0$_c$ between 0$^\circ$ and 360$^\circ$ covering first- and second-order mean-motion resonance (MMR) period ratios. From these, we could constrain the orbital period to a range of 265-290 d for the final fits.

We tested three models: two planets, two planets plus a linear trend, and two planets plus a quadratic trend. A visual hint of a long-term RV variation prompted the addition of trends.
The comparison of the Bayesian likelihood factors favours the model with two planets and no trend ($\Delta\ln{\mathcal{Z}} = 15$ between the no-trend and quadratic-trend models). Therefore, we adopted the model with two planets and no additional trends, although we intend to continue monitoring this system for potential long-term RV variations that could point to the existence of more distant companions. The arguments of pericenter $\omega_b$ and $\omega_c$ required careful handling of the priors to avoid circular multimodality and ensure convergence within a reasonable time frame; we constrained them through preliminary fits. The final priors adopted are listed in Table \ref{tab:exos-priors-posteriors}.

\begin{figure}[htb]
    \centering
    \includegraphics[width=1\hsize]{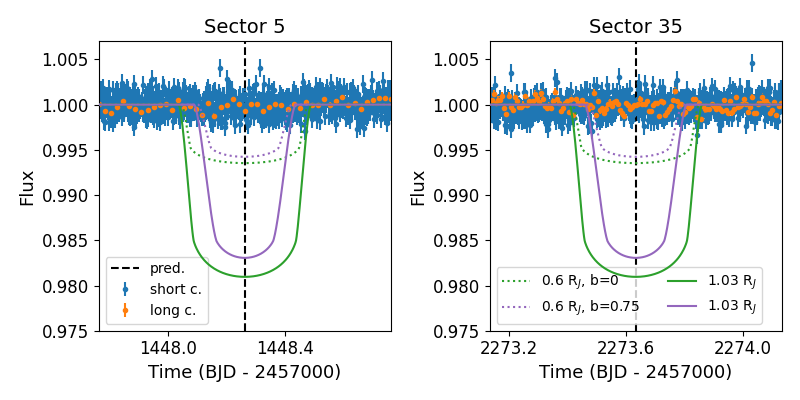}
    \caption{Predicted transits for TOI-199 c, which does not transit. Blue and orange points indicate the PDCSAP and \texttt{tesseract} light curves respectively, as in Fig. \ref{fig:lightcurve}. The vertical dashed line indicates the expected time of mid-transit. Model transits from \texttt{batman} are shown for two radii, $R_c=1.03\,R_J$ (solid lines) and $R_c=0.6\,R_J$ (dotted lines), and for two impact parameters, $\mathrm{b=0}$ (green) and $\mathrm{b=0.75}$ (purple).}
    \label{fig:transits_c}
\end{figure}

The joint TTV and RV model results in TOI-199 b having a period of $\mathrm{\periodplanetb \, d}$, an eccentricity of $\eccplanetb$ , a radius of $\mathrm{0.888 \pm 0.006 \, R_J}$, and a mass of $\mathrm{\massplanetb \, M_J}$. Meanwhile, TOI-199 c has a period of $\mathrm{\periodplanetc \, d}$, an eccentricity of $\eccplanetc$, and a minimum dynamical mass (since in the dynamical model we fix $i_c$ = 90$^\circ$) of $m_c$ = $\mathrm{\massplanetc \, M_J}$. If it were a transiting planet, \textit{TESS} should have seen transits in Sectors 5 and 35, but the light curves are flat around the predicted times of transit. In Fig. \ref{fig:transits_c} we show the light curves together with model transits for the expected range of radii from interior models (Sec. \ref{interior-models}) and impact parameters of $\mathrm{b=0}$ and $\mathrm{b=0.75}$, any of which should have been easily detectable in the \textit{TESS} light curves. Given the stellar radius, the planet's semi-major axis, and its predicted radii, this sets an upper limit on its inclination of $i_c\lessapprox 89.7$. Our model assumption that the system is edge-on and coplanar (i.e., $i_b,i_c$ = 90$^{\circ}$, and $\Delta i$ = 0$^{\circ}$) is still reasonable, since large mutual inclinations are unlikely, whereas small deviations from an edge-on configuration will not lead to a significant discrepancy from the derived (minimum) dynamical masses.

\begin{figure*}
    \centering
    \includegraphics[width=18cm]{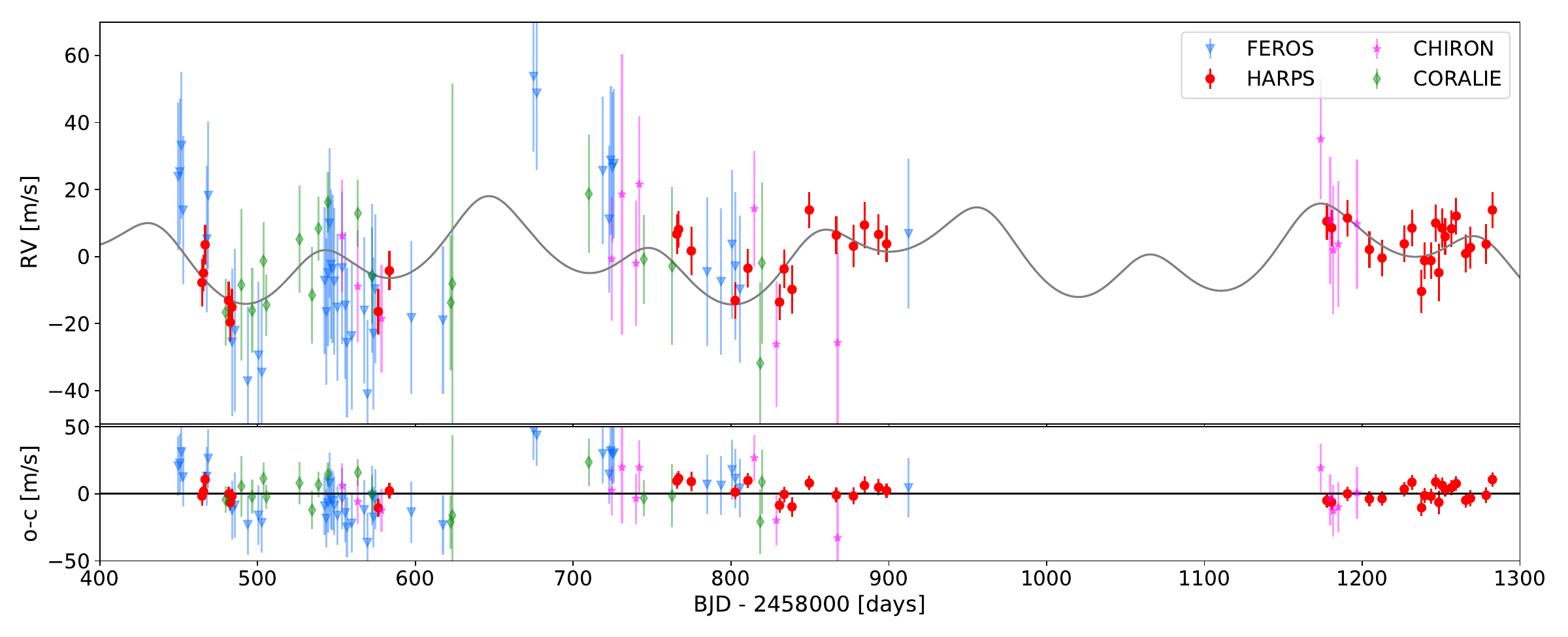} \\ \vspace{0.2cm}
       \includegraphics[width=5.9cm]{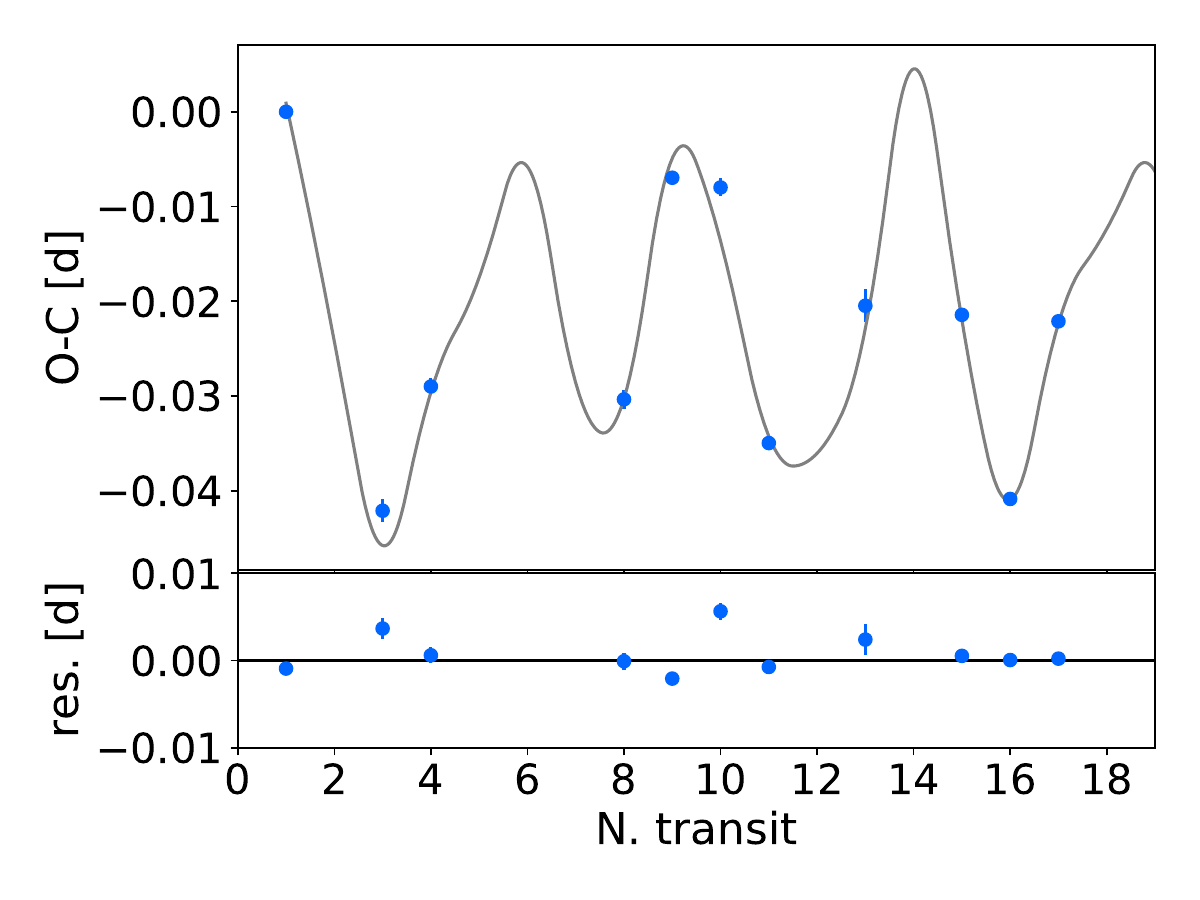} 
    \includegraphics[width=5.9cm]{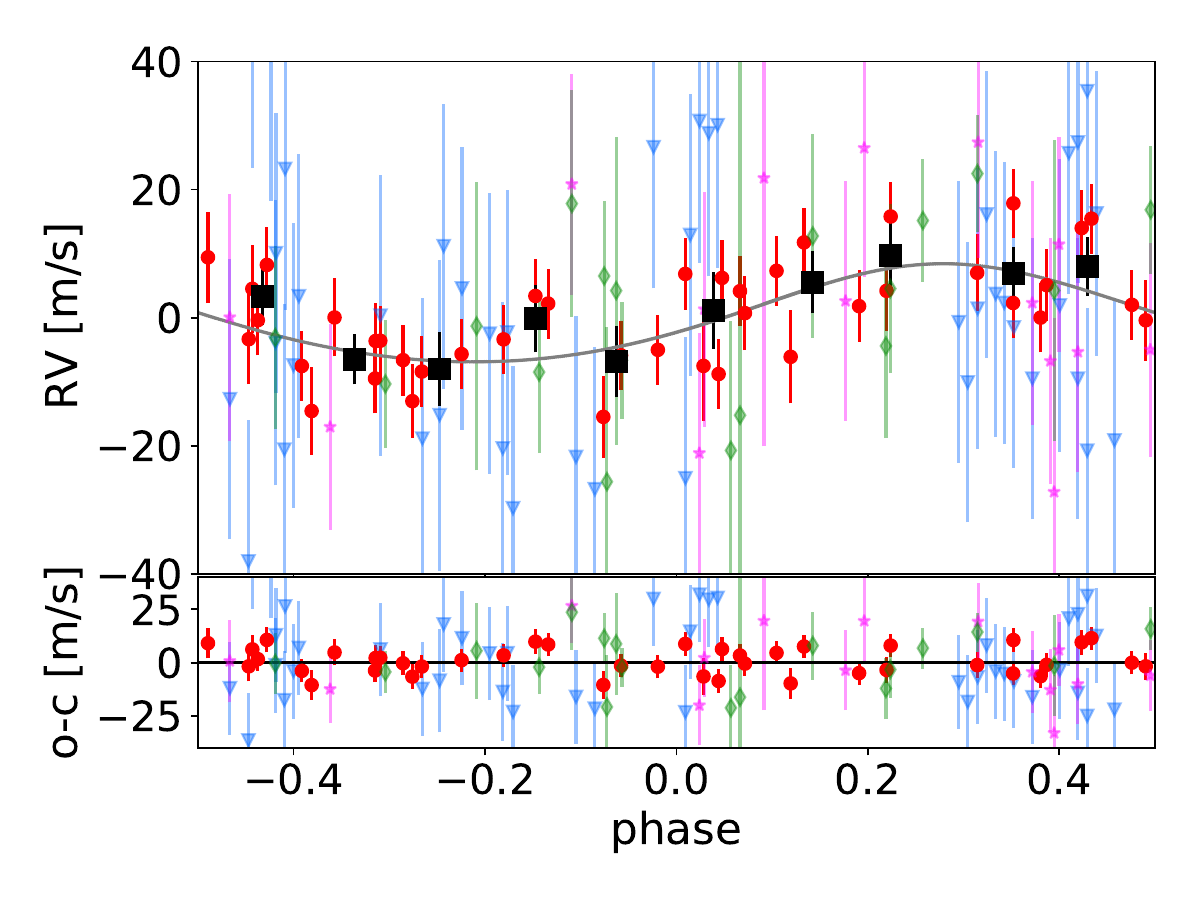} \put(-140,50){TOI\,199\,b} 
    \includegraphics[width=5.9cm]{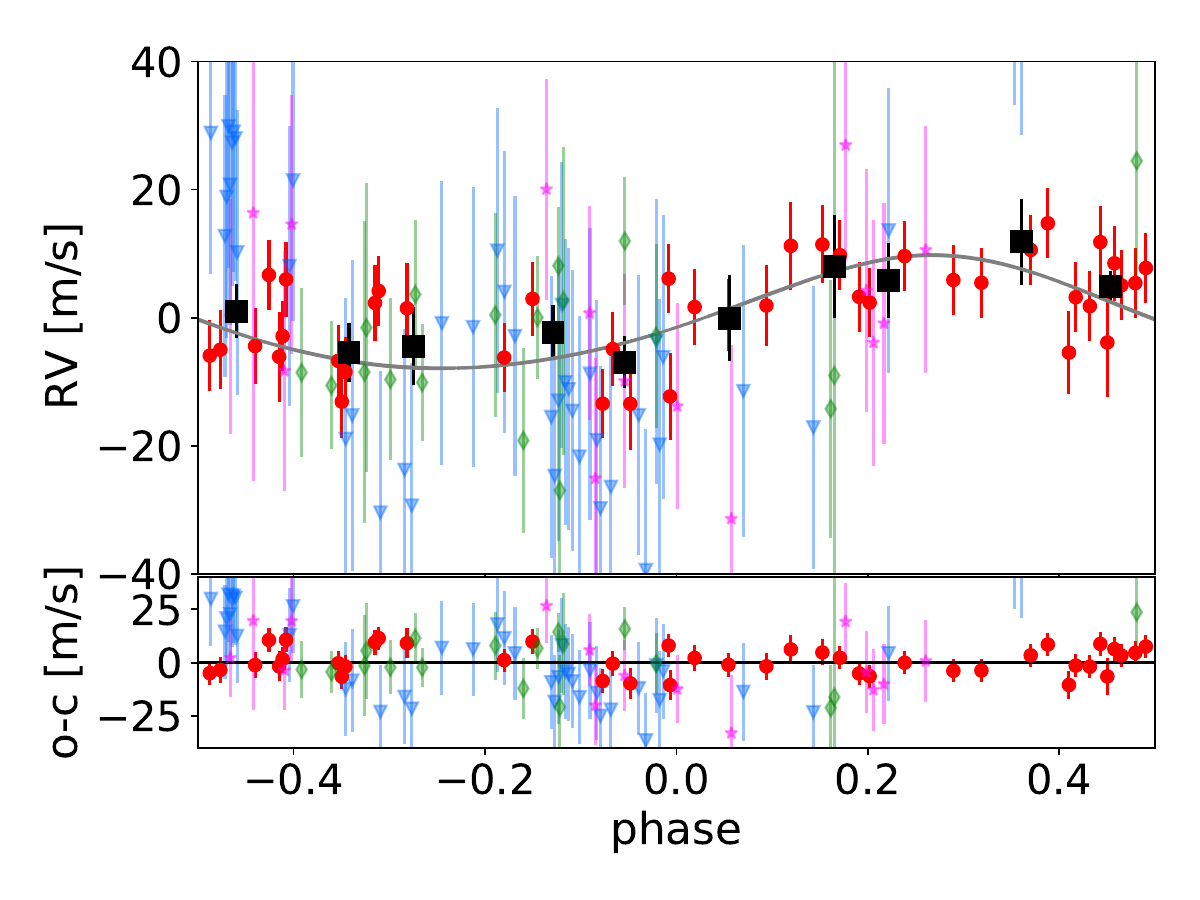} \put(-140,50){TOI\,199\,c}
 
    \caption{{\em Top}: Radial velocity measurements (FEROS: blue, HARPS: red, CHIRON: purple, CORALIE: green) and best-fit two-planet model (grey line) for the combined data. {\em Bottom left}: TTVs (blue circles) and fitted model with \texttt{Exo-Striker} (grey line) for TOI-199 b (top panel) and residuals to the model (bottom panel). The model has been smoothed with a quadratic spline. {\em Bottom centre and right}:  phase-folded representation of the two planetary signals after the RV signal of the other companion was subtracted. The respective RV residuals are shown under each panel, accordingly. The more precise HARPS RVs, which are the main driver of the fit, are highlighted. The black squares show the binned phase-folded RVs.} 
\label{fig:RVs_TTVs} 
\end{figure*}

The phase-folded RVs for TOI-199 b and c are shown in Fig. \ref{fig:RVs_TTVs} (bottom centre and right panels respectively), and the TTVs and fitted model in Fig. \ref{fig:RVs_TTVs}, bottom left panel. The full list of posterior parameters is given in Table \ref{tab:exos-priors-posteriors}, and the posterior probability distribution plots are shown in Figs. \ref{fig:cornerplot_pl1_fit}, \ref{fig:cornerplot_pl2_fit}, \ref{fig:cornerplot_pl_derived}, and \ref{fig:cornerplot_rvs}. The osculating orbital parameters, which evolve dynamically, are given for epoch $\mathrm{BJD=2458256.14}$.

\begin{table*}[ht]
\centering
\caption{Prior and posterior parameter distributions for the joint TTV+RV fit for the TOI-199 system obtained with a self-consistent N-body model with \texttt{Exo-Striker} at epoch BJD=2458256.14. The given values and error bars correspond to the median and 1$\sigma$ of the posterior distributions, respectively. \textit{Top}: Planetary parameters. \textit{Bottom}: RV offsets and jitters.}
\label{tab:exos-priors-posteriors}
\begin{tabular}{lrrrrrr}     
\hline\hline  \noalign{\vskip 0.7mm}
 & \multicolumn{2}{c}{Prior} & \multicolumn{2}{c}{Median \& $1\sigma$} & \multicolumn{2}{c}{Max $-\ln{\mathcal{L}}$} \\
Parameter  \hspace{0.0 mm} & Planet b       & Planet c & Planet b       & Planet c       & Planet b   & Planet c   \\
\hline \noalign{\vskip 0.7mm}
$K$ [m\,s$^{-1}$] & $\mathcal{U}$\tablenotemark{a}(5,20)& $\mathcal{U}$(3,20) & \Kplanetb & \Kplanetc & \KplanetbmaxLn & \KplanetcmaxLn \\
$P$ [day]  & $\mathcal{U}$(    104.60,    105.00)& $\mathcal{U}$(    265.00,    290.00) & \periodplanetb & \periodplanetc & \periodplanetbmaxLn & \periodplanetcmaxLn \\
$e$ & $\mathcal{U}$(      0,      0.4)& $\mathcal{U}$(      0.0,      0.4) & \eccplanetb & \eccplanetc & \eccplanetbmaxLn & \eccplanetcmaxLn \\
$\omega$ [deg] & $\mathcal{U}$(    -30,    30)& $\mathcal{U}$(     270,    450) & \omegaplanetb & \omegaplanetc\tablenotemark{b} & \omegaplanetbmaxLn & \omegaplanetcmaxLn\tablenotemark{b} \\
$M_{\rm 0}$ [deg]& $\mathcal{U}$(     45,    135)& $\mathcal{U}$(     90,    270)  & \Maplanetb & \Maplanetc & \MaplanetbmaxLn & \MaplanetcmaxLn \\
$a$ [au] &  \multicolumn{1}{c}{--} & \multicolumn{1}{c}{--} & \axisplanetb & \axisplanetc & \axisplanetbmaxLn & \axisplanetcmaxLn \\
$m$\tablenotemark{c} [$M_{\rm jup}$] & \multicolumn{1}{c}{--} & \multicolumn{1}{c}{--} & \massplanetb & \massplanetc & \massplanetbmaxLn & \massplanetcmaxLn \\
\hline
RV$_{\rm off\, FEROS}$ [m\,s$^{-1}$] & \multicolumn{2}{c}{$\mathcal{U}$(  51270,  51370)} & \multicolumn{2}{c}{\RVoffFEROS} & \multicolumn{2}{c}{\RVoffFEROSmaxLn} \\ 
RV$_{\rm off\, HARPS}$ [m\,s$^{-1}$] &  \multicolumn{2}{c}{$\mathcal{U}$(  51270,  51370)} & \multicolumn{2}{c}{\RVoffHARPS} & \multicolumn{2}{c}{\RVoffHARPSmaxLn} \\
RV$_{\rm off\, CHIRON}$ [m\,s$^{-1}$]&  \multicolumn{2}{c}{$\mathcal{U}$(    -50,     10)} & \multicolumn{2}{c}{\RVoffCHIRON} & \multicolumn{2}{c}{\RVoffCHIRONmaxLn} \\
RV$_{\rm off\, CORALIE}$ [m\,s$^{-1}$] &  \multicolumn{2}{c}{$\mathcal{U}$(  51270,  51370)} & \multicolumn{2}{c}{\RVoffCORALIE} & \multicolumn{2}{c}{\RVoffCORALIEmaxLn} \\
RV$_{\rm jit\, FEROS}$ [m\,s$^{-1}$] &  \multicolumn{2}{c}{$\mathcal{U}$(      0.01,     40.00)} & \multicolumn{2}{c}{\RVjittFEROS} & \multicolumn{2}{c}{\RVjittFEROSmaxLn} \\
RV$_{\rm jit\, HARPS}$ [m\,s$^{-1}$] &  \multicolumn{2}{c}{$\mathcal{U}$(      0.01,     40.00)} & \multicolumn{2}{c}{\RVjittHARPS} & \multicolumn{2}{c}{\RVjittHARPSmaxLn} \\
RV$_{\rm jit\, CHIRON}$ [m\,s$^{-1}$] &  \multicolumn{2}{c}{$\mathcal{U}$(      0.01,     40.00)} & \multicolumn{2}{c}{\RVjittCHIRON} & \multicolumn{2}{c}{\RVjittCHIRONmaxLn} \\
RV$_{\rm jit\, CORALIE}$ [m\,s$^{-1}$] &  \multicolumn{2}{c}{$\mathcal{U}$(      0.01,     40.00)} & \multicolumn{2}{c}{\RVjittCORALIE} & \multicolumn{2}{c}{\RVjittCORALIEmaxLn} \\
\hline \noalign{\vskip 0.7mm} 
\end{tabular}
\tablenotetext{a}{$\mathcal{U}(a,b)$ indicates a uniform distribution between $a$ and $b$.
\tablenotetext{b}{$360\degr$ has been added to/subtracted from the median and max $-\ln{\mathcal{L}}$ values of $\omega_b$ and $\omega_c$ respectively, to better show the orbital alignment between the two planets.}
\tablenotetext{c}{Mass corresponding to an edge-on coplanar model.}
}
\end{table*}   

 %

\subsection{Dynamical stability}

To examine the dynamical stability of the TOI-199 system, we first considered the classical Hill stability criterion \citep{Gladman93} and the Angular Momentum Deficiency criterion \citep[AMD,][]{Laskar17}. In terms of Hill stability, we found that the two planets are separated by $\approx 7.7 R_H$, well above the $\approx 2.4 R_H$ limit given by \cite{Gladman93} (for initially circular orbits; however, the eccentricities of these planets are small). Likewise, the system is AMD-stable, with values of $\beta = 0.3$ and $\beta_s = 0.02$ respectively (following \citealt{Laskar17}, Sect. 5), well below the $\beta = 1$ limit for collisions between the planets or the innermost planet and the star, respectively. Therefore, we do expect the TOI-199 Saturn-mass pair to be stable at the estimated orbits.

\begin{figure*}[htb]
    \centering
            \includegraphics[width=5.9cm]{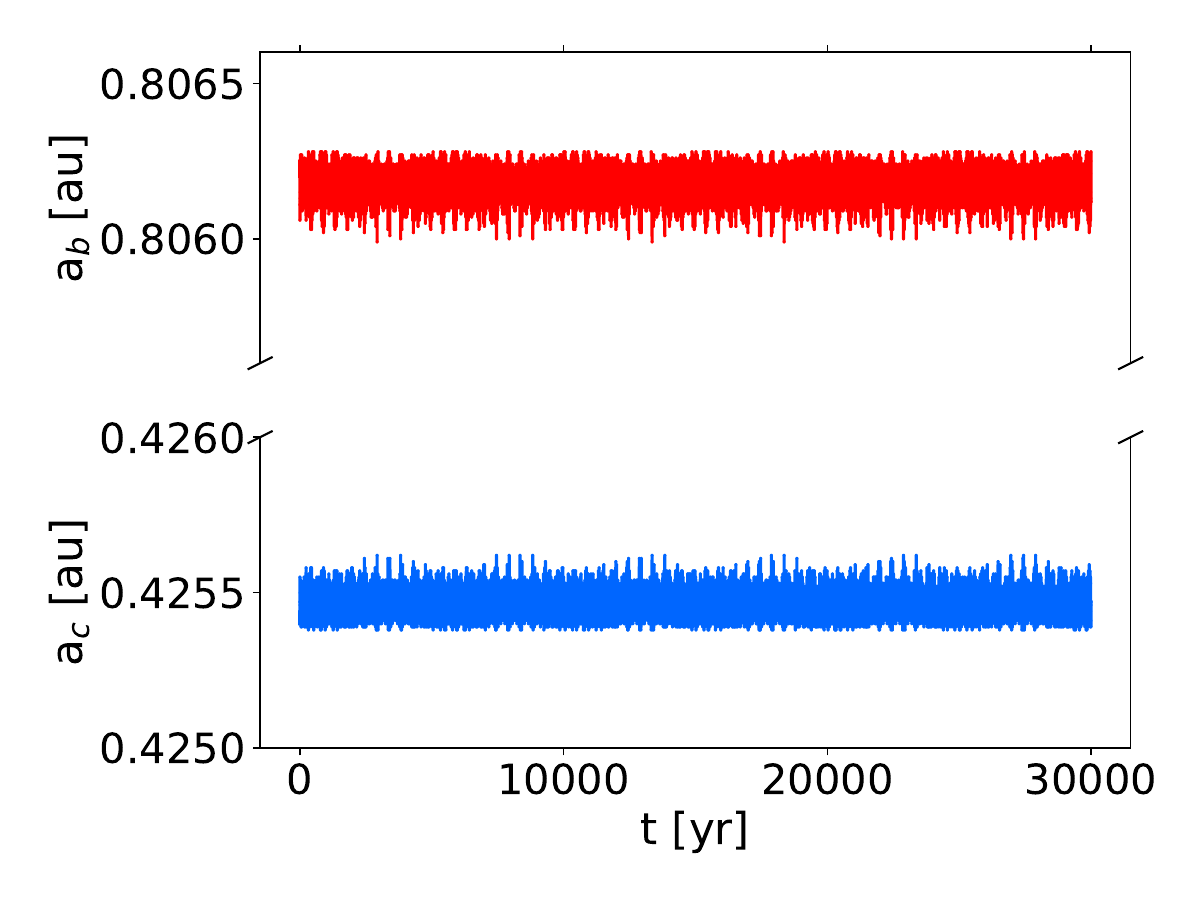}
        \includegraphics[width=5.9cm]{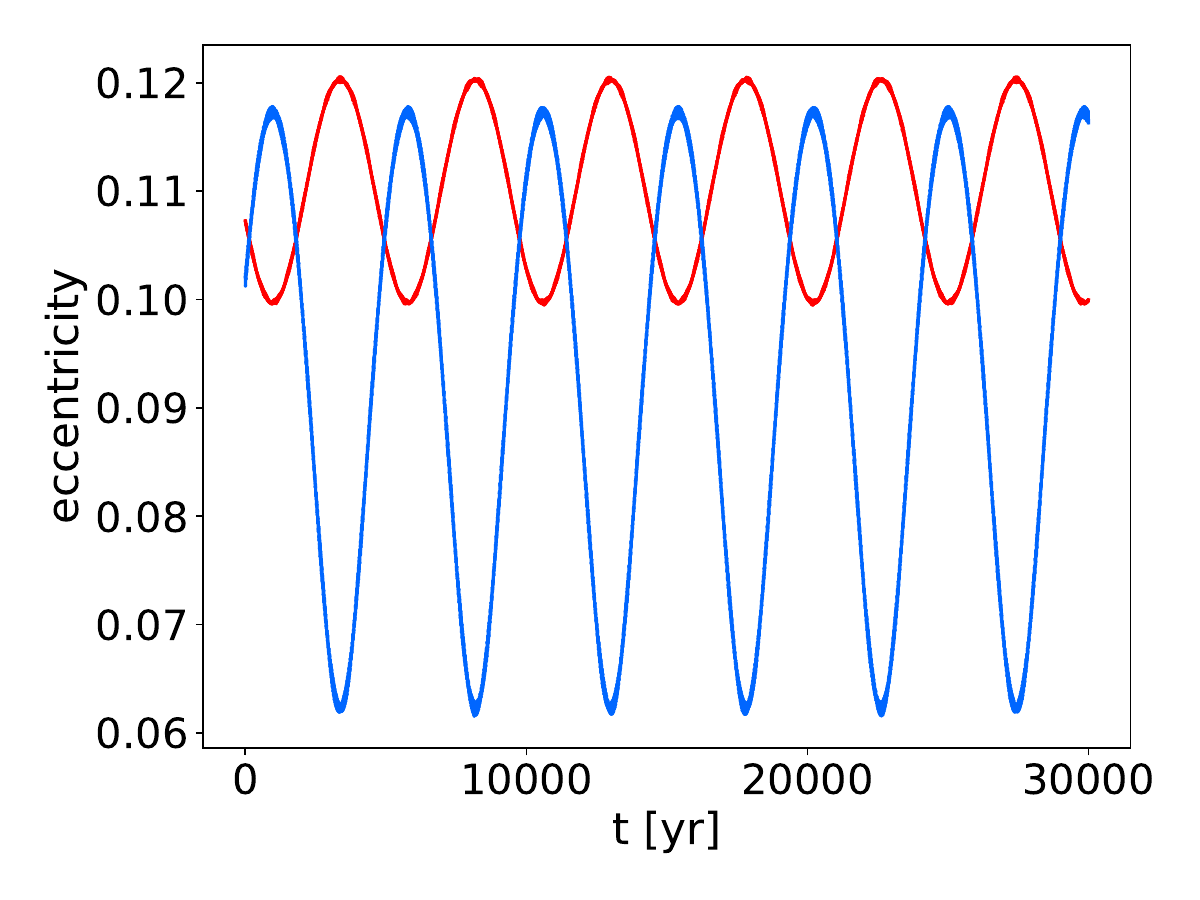}
        \includegraphics[width=5.9cm]{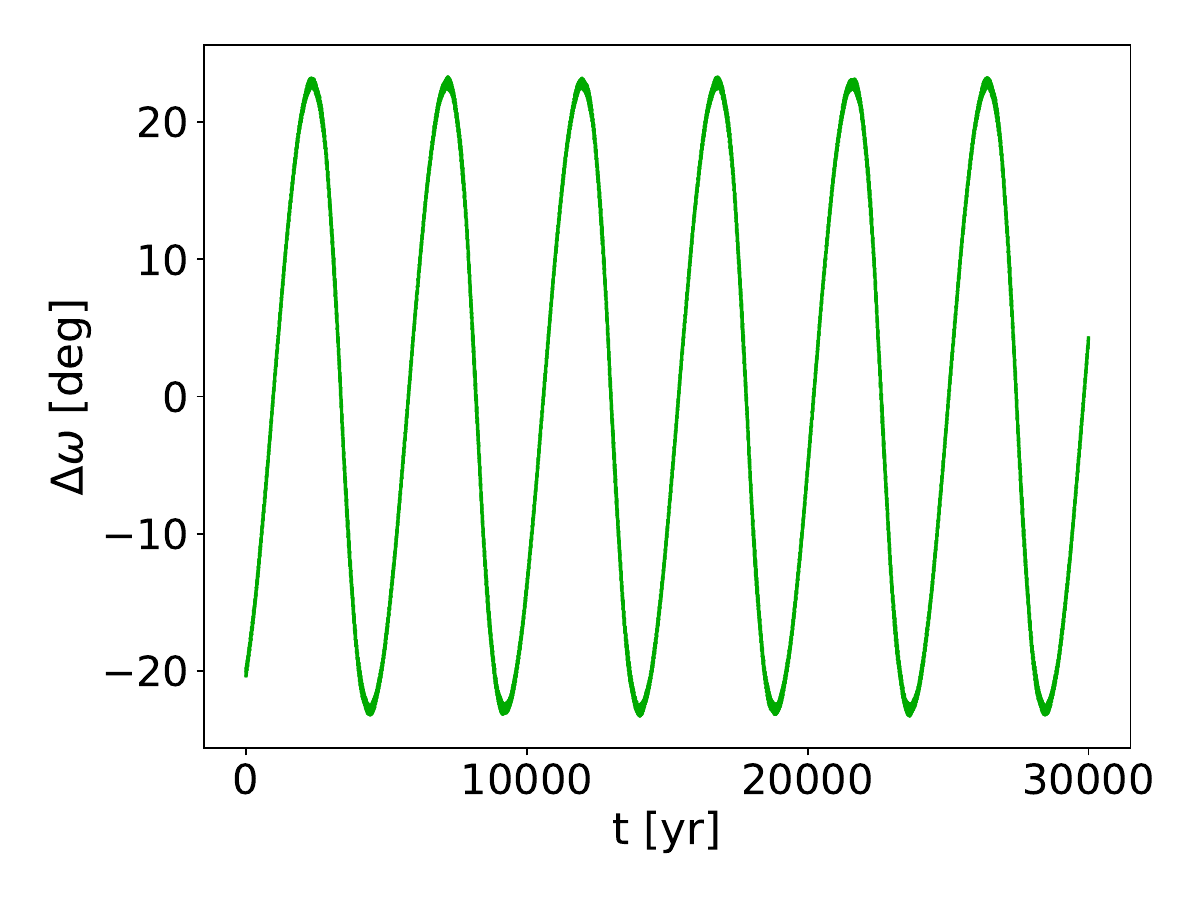} \\

    \caption{N-body orbital evolution of the TOI-199 system for an extent of 30\,000 yr. The initial orbit is taken from our TTV+RV best fit (see Table \ref{tab:exos-priors-posteriors} and is stable for at least 
    10\,Myr. Left to right: evolution of the semi-major axes $a_b$ (blue) and $a_c$ (red), the eccentricities $e_b$ (blue) and $e_c$ (red), and the evolution of $\Delta\omega$ = $\omega_b$ -$\omega_c$, respectively. See text for details.}
    \label{fig:dynamical-stab}
\end{figure*}

Nonetheless we aimed to analyse the dynamical evolution of the the best-fit, and study its properties. 
We performed a numerical N-body integration of the best-fit model with the Wisdom-Holman N-body algorithm \citep[][]{Wisdom1991} for 10\,Myr, adopting a small time step of $dt$ =0.5\,d. Fig. \ref{fig:dynamical-stab} shows the resulting evolution of the orbital semi-major axes, eccentricities, and $\Delta\omega$
for an extent of 30\,000 yr. We find the system to be well separated with no significant changes in the semi-major axes during the 10\,Myr N-body simulation. The eccentricities, however, osculate with notable amplitudes of $\approx 0.06$ for $e_b$ and $\approx 0.02$ for $e_c$. We find that the planetary pair osculates in aligned geometry, with the secular apsidal angle $\Delta\omega = \omega_b - \omega_c$ librating around 0 deg, with a semi-amplitude of $\sim$ 20 deg. 
The libration of the angle between the periapses $\Delta\omega$ is interesting, since it suggests that the planetary orbits must have been locked in secular apsidal alignment during the planet migration. Therefore, this libration of $\Delta\omega$ is an important remnant evidence of the planets' formation and orbital evolution during the disk phase.

\section{Discussion}\label{sec:disc}

\subsection{Possible exomoons in the system} \label{sec:exomoons}

The possibility of stable exomoons orbiting the planets of the TOI-199 system is interesting for two reasons. 
On the one hand, exomoons could trigger TTVs variations in TOI-199\,b, which might be misinterpreted as a TTV signal caused by a second planetary companion \citep{Kipping2022}. 
The relatively small Hill-radius of TOI-199\,b (R$_{\rm Hill}$ $\sim$ 0.014 au), and the orbital analysis of the RV and TTV data we performed in Sec. \ref{sec3.3}, strongly point to the existence of the outer Saturn-mass planet TOI-199\,c, rather than an exomoon. Nonetheless, eliminating the exomoon hypothesis would strengthen the evidence for the existence of TOI-199\,c even further. On the other hand, given the observational evidence of the existence of the Saturn-mass giant TOI-199\,c, it is worth testing the possibility of exomoons around this companion. This planet orbits further out, so the Hill-radii of dynamical influence is somewhat larger (R$_{\rm Hill}$ $\sim$ 0.03 au), positively affecting potential exomoons' stability. Further, TOI-199\,c resides in the Habitable Zone (HZ), which is intriguing for the existence of stable exomoon bodies. The gas giant was likely formed even further out beyond the ice line around TOI-199. Thus, initially icy exomoons, similar to those of Saturn, are possibly in the current warm orbit, and could become potentially habitable, ocean-like, or early Mars-like worlds \citep[see][for further discussion]{Trifonov2020a}.

We study the possibility of stable exomoons in the system following the same numerical setup as in \citet{Trifonov2020a}, who tested the exomoon stability of the Saturn-mass pair of planets around the M-dwarf GJ\,1148. The N-body test was done using a Wisdom-Holman algorithm \citep[][]{Wisdom1991} modified to handle the evolution of test particles in the Jacobi coordinate system \citep{Lee2003}, which allows for invertibility of the system's hierarchy (i.e., making either of the planets a central body in the system, enabling the testing of stable orbits around each planet). Our N-body simulations were performed with a very small time step of $dt$=0.01\,d and for a maximum of 1\,Myr, which we find to be a good balance between CPU resources and the numerical accuracy of the run. We randomly seed an arbitrary number of 4\,000 ``exomoon" mass-less test particles on circular prograde orbits around TOI-199\,b \& c. The semi-major axes of the test exomoons ranged randomly between the planetary Roche limit ($\sim$ 0.0003 au, for both planets), and the planetary Hill-radius, where exomoons are expected to be stable. 
In addition, we studied the tidal interactions between the planets and the exomoons. We adopt Mars-like mass and radius for the exomoons ($m$ = 0.107 $M_\oplus$, $R$ = 0.53 $R_\oplus$). In contrast, for the massive planets TOI-199\,b \& c, we adopt masses and radii from our best-fit analysis (the radius of TOI-199 c is unknown as it does not transit; thus, we assumed the same radius as for TOI-199 b). We integrated their orbits with the \textsc{EqTide}\footnote{\url{https://github.com/RoryBarnes/EqTide}} code \citep{Barnes2017}, which calculates the tidal evolution of two bodies based on the models by \citet{Ferraz-Mello2008}. 
For the rest of the numerical setup, we strictly follow the same \textsc{EqTide} prescription as in \citet{Trifonov2020a}. 

The results from the test particle simulations are illustrated in \autoref{e_a_inner} for TOI-199\,b, and \autoref{e_a_outer} for TOI-199\,c, respectively. In the case of TOI-199\,b, we found that only  $\sim$ 48 \% of the test exomoons remained long-term stable with small eccentricities, usually those near the stability limit, which we found to be at $\sim$ 0.0055 au ($\sim 13$ planet radii, $\sim 11.5\, \mathrm{R_J}$). This stability limit is much smaller than the Hill-radius of the planet. As was shown in \citet{Trifonov2020a} the expected stability limit for prograde exomoons is $\sim$ $0.5\,{\rm R_{Hill}}$, due to the dynamical perturbations of the second planet \citep[see also,][]{Grishin2017}. Similarly, for TOI-199\,c the stability region was found to be $<$ $0.5\,{\rm R_{Hill}}$, where $\sim$ 60 \% of the test particles survived for 1\,M\,yr. As expected, the stable region around TOI-199\,c is larger, close to 0.013 au ($\sim 27\, \mathrm{R_J}$) from the planet. The tidal interactions, however, further limit the possible region where the exomoons could exist around both planets. The Saturn-mass planets likely have comparable rotational periods to Saturn ($\sim$ 10.5 h), which will not be affected significantly over the age of the system due to tidal interaction with the star. Therefore, closer Mars-like exomoons will drift away to longer orbits over time. Finally, we concluded that TOI-199\,b is unlikely to host large exomoons because of the very limited range of semi-major axes of 0.0045-0.0055 au where such bodies can exist. For TOI-199\,c habitable exomoons could exist in the range 0.0045-0.0125 au, comparable to the semi-major axes of the Galilean moons. 

\subsection{Interior models}\label{interior-models}

We modelled the interior evolution of both planets in the system using CEPAM \citep{guillot1995} and a non-grey atmosphere \citep{parmentier2015}. We assumed simple structures consisting of a central dense core,  composed of 50$\%$ ices and 50$\%$ rocks, surrounded by a hydrogen and helium envelope of solar composition. We use the P-$\rho$ relationships from \cite{Hubbard89} to calculate the density in the core. The equation of state used for hydrogen and helium accounts for non-ideal mixing effects \citep{howard2023, chabrier2019}.

\begin{figure}
    \centering
    \includegraphics[width=1\hsize]{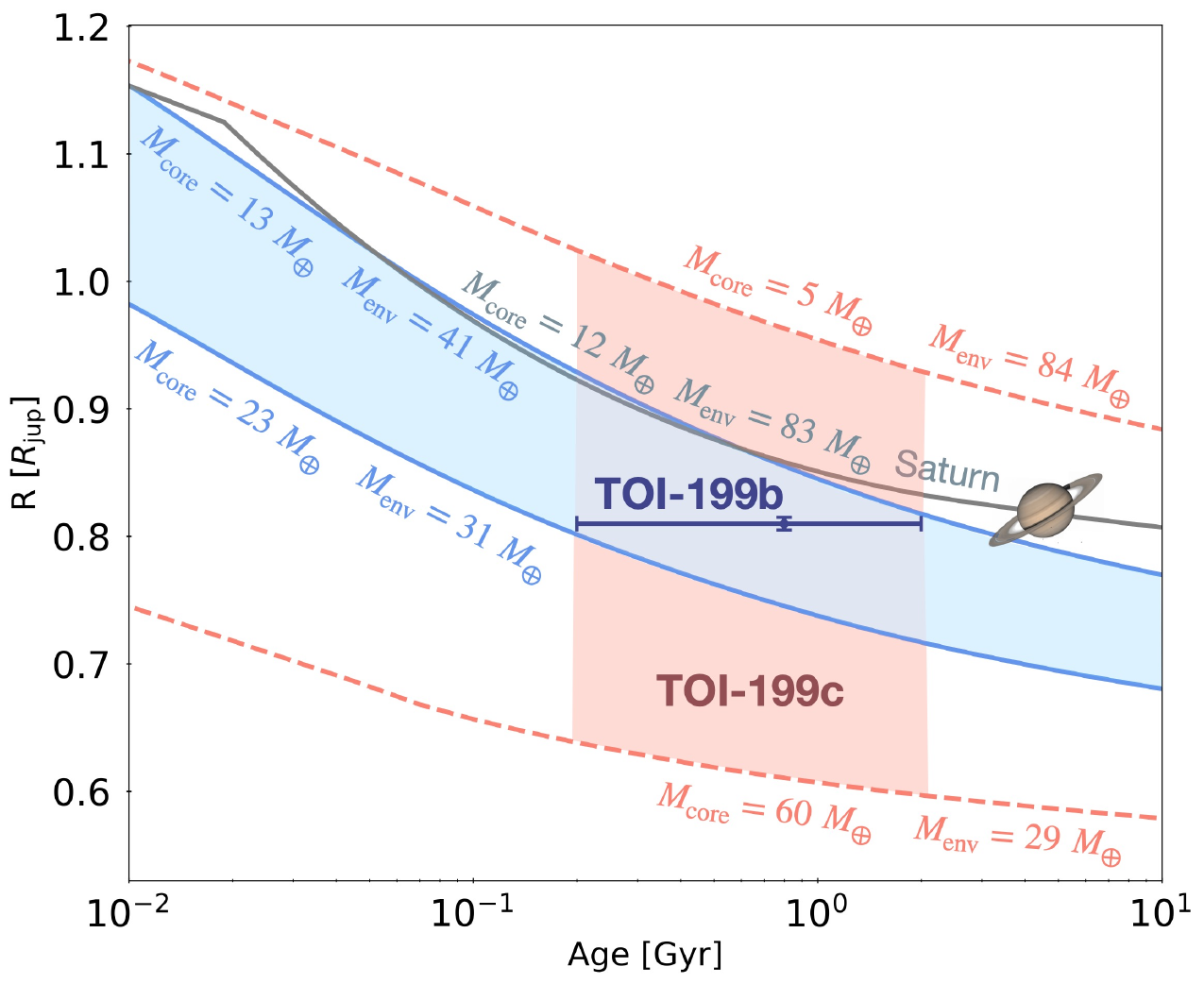}
    \caption{Evolution models of TOI-199\,b and TOI-199\,c compared to a simple model for Saturn. All models assume a central ice-rock core surrounded by a hydrogen and helium envelope of solar composition. $M_{\rm core}$ corresponds to the mass of the ice/rock core while $M_{\rm env}$ corresponds to the mass of the solar-composition envelope. The range of $M_{\rm core}$ and $M_{\rm env}$ compatible with the observational constraints is shown for TOI-199\,b. The blue error bar corresponds to observational constraints on the age and the radius of TOI-199\,b. We also show the range of radii expected for likely extreme compositions of TOI-199\,c in red.}
    \label{fig:interior}
\end{figure}

Figure~\ref{fig:interior} shows the resulting evolution models and the observational constraints. Defining an approximate bulk metallicity as $Z\approx M_{\rm core}/M_{\rm tot}$ and assuming $Z_\odot=0.015$, we find $Z/Z_\odot=16$ to 28 for TOI-199b. This is somewhat larger than what we obtain for a similarly simple model of Saturn although we stress that Saturn's enrichment corresponds to a narrower range of possibilities \citep[between 12 and 15 according to][]{Mankovich21}. However, most of the uncertainty comes from the poor constraint on the age of the system. An accurate determination of its age would thus greatly help to constrain the bulk metallicity of TOI-199 b \citep[see also][]{muller2023}. We unfortunately do not have a measurement of the radius of TOI-199 c, as it does not transit. Using the same evolution models and a composition with cores between 5 and 60 $M_\oplus$ (bulk metallicity between 4 and 45), we obtain expected radii of TOI-199c between 0.6 and 1.03 $R_{\rm J}$.

\subsection{Atmospheric characterization potential}

\begin{figure*}[htb]
    \centering
    \includegraphics[width=1\hsize]{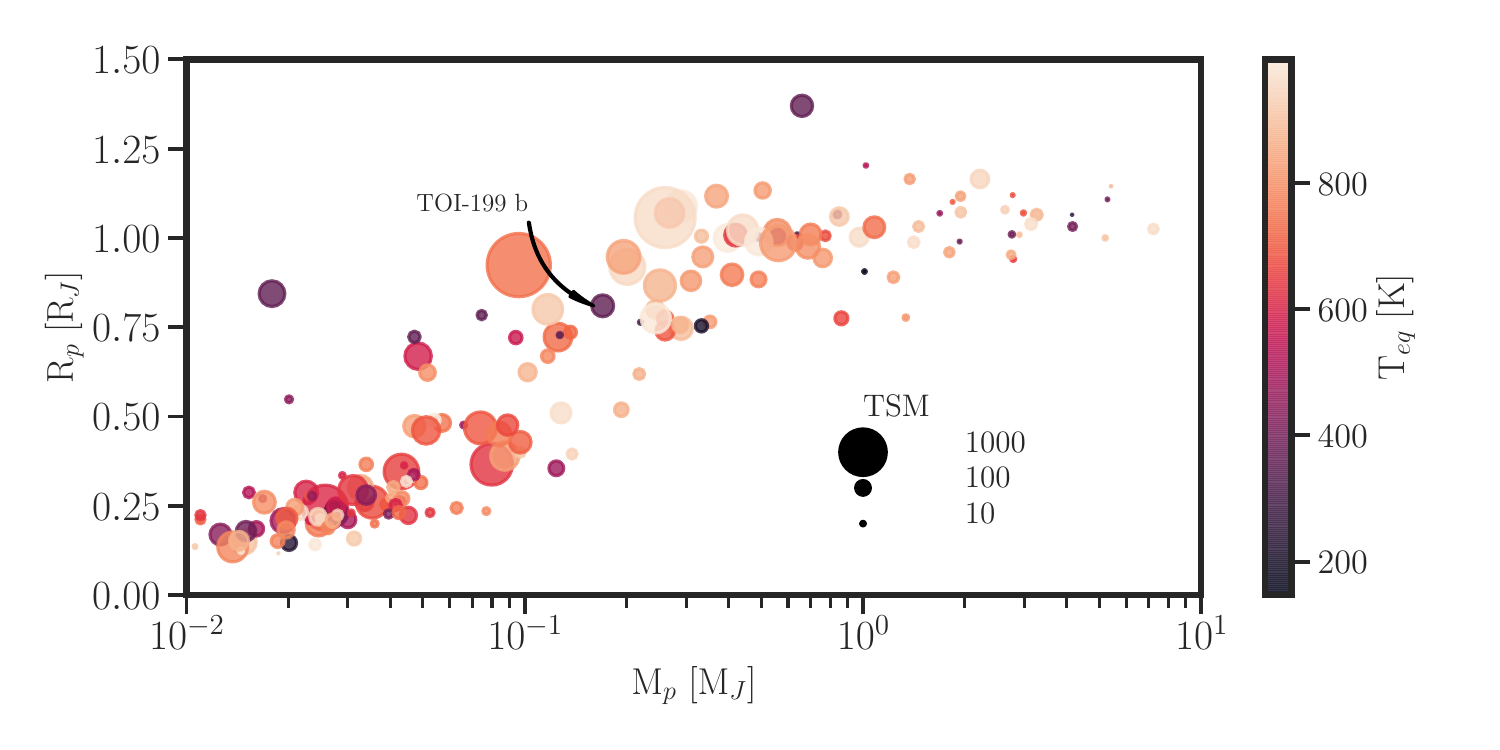}
    \caption{Mass-radius diagram for warm ($\mathrm{T_{eq} < 1000\, K}$) giant planets with masses measured to better than 20\% and radii measured to better than 10\%, similar to our accuracies for TOI-199 b. The markers are colour-coded by equilibrium temperature and scaled by transmission spectroscopy metric. TOI-199 b is significantly cooler than other well-characterized planets with similar masses and radii, and has a high TSM.}
    \label{fig:m-r-tsm}
\end{figure*}

TOI-199 b is a very interesting target for atmospheric characterization through transmission spectroscopy. We computed the transmission spectroscopy metric \citep[TSM,][]{Kempton18} for TOI-199 b, finding a value of 107, above the $\mathrm{TSM=90}$ threshold recommended by \cite{Kempton18}. Compared to other well-characterised warm giants with similar masses and radii, it has a comparable TSM but is notably cooler (Fig. \ref{fig:m-r-tsm}), making it a unique target. 

The intrinsic luminosities of both TOI-199 b and c are between a third to one times the present-day luminosity of Jupiter, implying that their atmospheric structure is mostly governed by the irradiation that they receive \citep[see][]{parmentier2015}. With a photospheric temperature expected to range between 250 and 350\,K, TOI-199 b is an ideal candidate for the observation of the consequences of condensation of water in giant planet atmospheres \citep[see][]{Guillot+2022}.

\subsection{The TOI-199 system in a population context}

\begin{figure*}[htb]
    \centering
    \includegraphics[width=1\hsize]{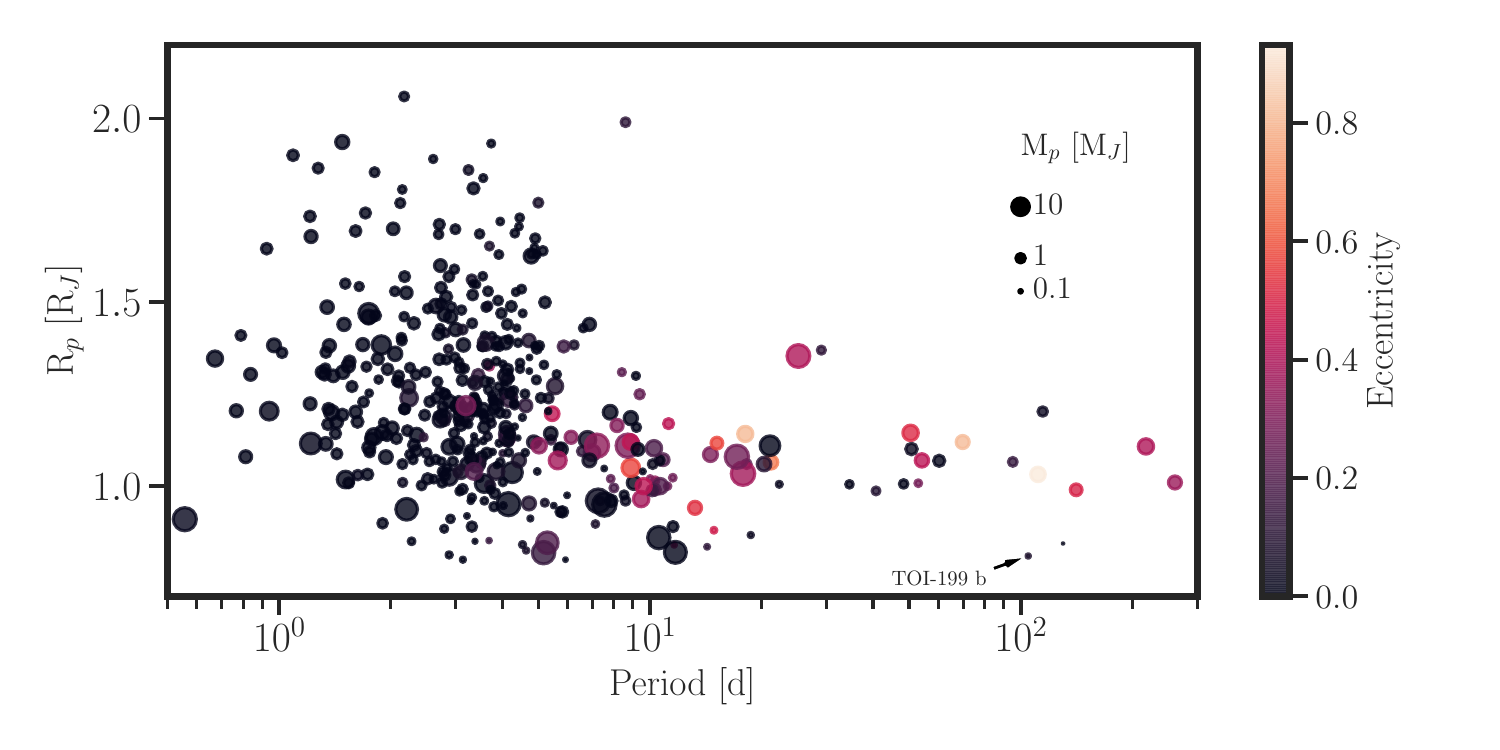}
    \caption{Radius-period diagram for giant planets with masses measured to better than 20\% and radii measured to better than 10\%, similar to our accuracies for TOI-199 b. The markers are colour-coded by eccentricity and scaled by planet mass. TOI-199 occupies a scarcely-populated region of this space.}
    \label{fig:period-r-ecc}
\end{figure*}

The TOI-199 system is composed of two giant planets. The inner planet, TOI-199 b, is a transiting warm giant at a $\mathrm{\periodplanetb \, d}$ period with a $\mathrm{\julietRp\, R_J}$ radius, and a $\mathrm{\massplanetb \, M_J}$ mass, comparable to those of Saturn ($\mathrm{R_S = 0.83 \, R_J}$, $\mathrm{M_S = 0.30 \, M_J}$). To place TOI-199 b in context within the exoplanet populations, we plot the radius-period diagram (Fig. \ref{fig:period-r-ecc}) for giant planets ($\mathrm{R_p \geq 0.8\, R_J}$) with well-constrained masses and radii from the TEPCAT catalogue \citep{Southworth11}\footnote{Available at \url{https://www.astro.keele.ac.uk/jkt/tepcat/}, accessed 6th October 2022.}. TOI-199 b helps populate a so far very sparse region in the period-radius space. It is noticeably less massive, less dense and less eccentric than its closest neighbours, and is the first warm exo-Saturn (i.e., a warm giant with mass and radius similar to those of Saturn) with mass measured to better than 20\% and radius measured to better than 10\%.

\begin{figure}[htb]
    \centering
    \includegraphics[width=1\hsize]{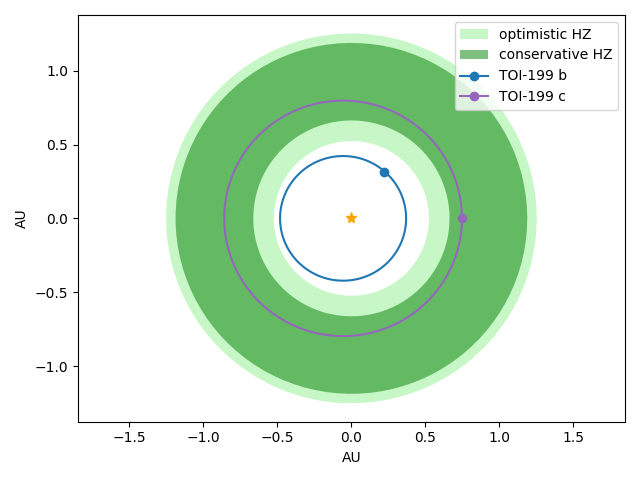}
    \caption{Orbital configuration of the TOI-199 system. The conservative and optimistic Habitable Zones are shown in dark and light green respectively. TOI-199 c is in the Habitable Zone.}
    \label{fig:HZ-orbits}
\end{figure}

The outer planet, TOI-199 c, also falls within the warm giant parameter space at a period of $\mathrm{\periodplanetc \, d}$. With a minimum mass of $\mathrm{m \sin i = \massplanetc \, M_J}$, it is significantly more massive than the inner planet; as it does not transit, we cannot measure its radius. Given its orbital period, TOI-199 c falls into the conservative HZ, as defined by \cite{Kopparapu14}. The orbits of the two planets, and the conservative and optimistic HZ regions for TOI-199, are shown in Fig. \ref{fig:HZ-orbits}. However, given its minimum mass, TOI-199 c is presumably a gas giant, and as such does not possess a surface on which water could be found.

We use the stellar effective temperatures and luminosities listed in the NASA Exoplanet Archive\footnote{Retrieved 2nd June 2022} to compute the Habitable Zones defined by \cite{Kopparapu14} for all stars with temperatures in the $\mathrm{2600-7200\,K}$ range and catalogued luminosities. While most exoplanet host stars (4092 out of the 5035 catalogued) fall in this temperature range, relatively few of those (897) have catalogued luminosities. In these, we find 159 planets in the conservative or optimistic Habitable Zones of their host stars, of which 138 have measured masses. Out of the planets with measured masses, the vast majority (111) have $\mathrm{M_p>10M_\oplus}$, placing them in the Neptune-or-larger regime. True potentially habitable planets, with surfaces on which water could exist in a liquid state, are still rare. However, these giant planets, among which TOI-199 c falls, could still host potentially habitable exomoons.

\section{Conclusions}\label{sec:conc}

We have presented the discovery and characterization of the TOI-199 system, composed of two warm giant planets. The inner planet, TOI-199 b, was first identified as a single transiter in \textit{TESS} photometry, and confirmed using ASTEP, LCO, PEST, Hazelwood, and NEOSSat follow-up photometry, and HARPS, FEROS, CORALIE, and CHIRON RVs. The transits of TOI-199 b showed strong TTVs, pointing to the existence of a second planet in the system. From the joint analysis of the TTVs and RVs, we found that TOI-199 b has a $\mathrm{\periodplanetb \, d}$ period, a mass of $\mathrm{\massplanetb \, M_J}$, and a radius of $\mathrm{\julietRp \, R_J}$, making it the first precisely characterized warm exo-Saturn. Meanwhile, the outer non-transiting planet TOI-199 c has a period of $\mathrm{\periodplanetc \, d}$, placing it in the Habitable Zone, and a minimum mass of $\mathrm{\massplanetc \, M_J}$.

We studied the dynamical stability and the potential for these planets to host exomoons. Our N-body simulations show that the system is stable, with no significant changes in the semi-major axes. The secular apsidal angle $\Delta\omega$, however, librates, suggesting the planets' orbits were locked in secular apsidal alignment during their migration. TOI-199 b is extremely unlikely to host large exomoons. TOI-199 c, on the other hand, could potentially host habitable exomoons, though as it does not transit these would be exceedingly difficult to detect.

TOI-199 b is a unique target for atmospheric characterization. It has a high TSM, and is cooler than other well-characterized giants with similar masses and radii, making it ideal for the study of water condensation in giant planet atmospheres.

TESS will observe one further transit of TOI-199 during Cycle 5, in Sector 67 (transit predicted for 2460143.85625384, 18 July 2023). We also predict transits on 2460248.7352902 (31 October 2023), 2460353.58791697 (13 February 2024) and 2460458.46233884 (27 May 2024), which will not be observed by TESS. Ground-based monitoring will thus be vital to the continued study of this TTV system.

\acknowledgments
This paper includes data collected by the \textit{TESS} mission, which are publicly available from the Mikulski Archive for Space Telescopes (MAST). Funding for the \textit{TESS} mission is provided by NASA's Science Mission directorate. The specific observations analyzed can be accessed via DOI \dataset[10.17909/cbph-wd44]{https://doi.org/10.17909/cbph-wd44}.

This research has made use of the Exoplanet Follow-up Observation Program website, which is operated by the California Institute of Technology, under contract with the National Aeronautics and Space Administration under the Exoplanet Exploration Program.
We acknowledge the use of public \textit{TESS} Alert data from the pipelines at the \textit{TESS} Science Office and at the \textit{TESS} Science Processing Operations Center.

Resources supporting this work were provided by the NASA High-End Computing (HEC) Program through the NASA Advanced Supercomputing (NAS) Division at Ames Research Center for the production of the SPOC data products. 

Based on observations collected at the European Organisation for Astronomical Research in the Southern Hemisphere under ESO programmes 0101.C-0510, 0102.C-0451, 0104.C-0413, 106.21ER.001, 0102.A-9003, 0102.A-9006, 0102.A-9011, 0102.A-9029, 0103.A-9008, and 0104.A-9007.

This research has made use of the Spanish Virtual Observatory (https://svo.cab.inta-csic.es) project funded by MCIN/AEI/10.13039/501100011033/ through grant PID2020-112949GB-I00.

T.T. acknowledges support by the DFG Research Unit FOR 2544 "Blue Planets around Red Stars" project No. KU 3625/2-1.
T.T. further acknowledges support by the BNSF program "VIHREN-2021" project No. KP-06-DV/5.

A.J., R.B. and M.H. acknowledge support from ANID - Millennium Science Initiative - ICN12\_009. A.J. acknowledges additional support from FONDECYT project 1210718. R.B. acknowledges support from FONDECYT Project 1120075 and from project IC120009 “Millennium Institute of Astrophysics (MAS)” of the Millenium Science Initiative. This work was funded by the Data Observatory Foundation.

The results reported herein benefited from collaborations and/or information exchange within the program “Alien Earths” (supported by the National Aeronautics and Space Administration under agreement No. 80NSSC21K0593) for NASA’s Nexus for Exoplanet System Science (NExSS) research coordination network sponsored by NASA’s Science Mission Directorate.

This work makes use of observations from the ASTEP telescope. ASTEP benefited from the support of the French and Italian polar agencies IPEV and PNRA in the framework of the Concordia station program, from OCA, INSU, Idex UCAJEDI (ANR- 15-IDEX-01) and ESA through the Science Faculty of the European Space Research and Technology Centre (ESTEC).

This work makes use of observations from the LCOGT network. Part of the LCOGT telescope time was granted by NOIRLab through the Mid-Scale Innovations Program (MSIP). MSIP is funded by NSF. K.A.C. acknowledges support from the TESS mission via subaward s3449 from MIT.

The postdoctoral fellowship of K.B. is funded by F.R.S.-FNRS grant T.0109.20 and by the Francqui Foundation.

The contributions of M.L., S.U. and S.S. have been carried out within the framework of the NCCR PlanetS supported by the Swiss National Science Foundation under grants 51NF40\_182901 and 51NF40\_205606. M.L. acknowledges support of the Swiss National Science Foundation under grant number PCEFP2\_194576.

This research received funding from the European Research Council (ERC) under the European Union's Horizon 2020 research and innovation programme (grant agreement n$^\circ$ 803193/BEBOP), and from the Science and Technology Facilities Council (STFC; grant n$^\circ$ ST/S00193X/1).

D. D. acknowledges support from the NASA Exoplanet Research Program grant 18-2XRP18\_2-0136.

This work has made use of data from the European Space Agency (ESA) mission
{\it Gaia} (\url{https://www.cosmos.esa.int/gaia}), processed by the {\it Gaia}
Data Processing and Analysis Consortium (DPAC,
\url{https://www.cosmos.esa.int/web/gaia/dpac/consortium}). Funding for the DPAC
has been provided by national institutions, in particular the institutions
participating in the {\it Gaia} Multilateral Agreement.

J.~K. gratefully acknowledges the support of the Swedish National Space Agency (SNSA; DNR 2020-00104) and of the Swedish Research Council  (VR: Etableringsbidrag 2017-04945).

M. T. P. acknowledges support of the ANID-Fondecyt Post-doctoral fellowship no. 3210253.

%

\facilities{\textit{TESS}, ASTEP, LCOGT, PEST, NEOSSat, Hazelwood, FEROS/MPG2.2m, HARPS/ESO3.6m, CORALIE, CHIRON, HRCam/SOAR}


\software{\texttt{tesseract} (Rojas et al. in prep),
          \texttt{CERES} \citep{Brahm17CERES}, 
          \texttt{juliet} \citep{Espinoza19juliet},
          \texttt{ZASPE}~\citep{Brahm17ZASPE},
          \texttt{radvel}~\citep{Fulton18},
          \texttt{emcee}~\citep{Foreman-Mackey13},
          \texttt{MultiNest}~\citep{Feroz09},
          \texttt{PyMultiNest}~\citep{Buchner14},
          \texttt{batman}~\citep{Kreidberg15},
          \texttt{celerite}~\citep{Foreman-Mackey17},
          AstroImageJ \citep{Collins:2017}, 
          TAPIR \citep{Jensen:2013},
          \texttt{Exo-Striker} \citep{TrifonovExoStriker},
          \textsc{EqTide} \citep{Barnes2017}, 
          limb-darkening \citep{Espinoza15}
          }


\newpage

\appendix

\section{Radial velocity and activity indices}
\restartappendixnumbering

In this appendix, we show the RVs and activity indicators (where applicable) obtained from the spectroscopical data. Tables \ref{tab:harps-data} and \ref{tab:feros-data} show the HARPS and FEROS results respectively, both obtained from the \texttt{ceres} pipeline. Table \ref{tab:coralie-data} shows the CORALIE results, obtained using the standard CORALIE DRS. Table \ref{tab:chiron-data} shows the CHIRON results, obtained following \cite{Jones19}.

\begin{table*}[pht]
\begin{center}
\caption{RV and activity indices obtained from the HARPS spectra.}
\label{tab:harps-data}
\centering
\resizebox{\textwidth}{!}{%
\begin{tabular}{llllllllll}
\hline  \hline
BJD - 2457000 [d] & RV [km/s] &  Bisector &   FWHM & SNR &  $\mathrm{H_\alpha}$ & $\mathrm{\log(R'_{HK})}$  &  Na~II  &  He~I  \\
\hline
1464.79399978 &  $51.3356 \pm 0.0052$	 &  $-0.021 \pm 0.007$	 & 7.7540	& 40 &	$0.1205 \pm 0.0016$ & $-4.8668 \pm 0.0356$	&  $0.1918 \pm 0.0025$	&   $0.5018 \pm 0.0037$ \\
1464.78179568 &  $51.3470 \pm 0.0046$	 &  $-0.003 \pm 0.006$	 & 7.7509	& 43 &	$0.1237 \pm 0.0014$ & $-4.8177 \pm 0.0296$	&  $0.1923 \pm 0.0022$	&   $0.5066 \pm 0.0034$ \\
1465.78521663 &  $51.3439 \pm 0.0022$	 &  $-0.004 \pm 0.003$	 & 7.7473	& 64 &	$0.1254 \pm 0.0010$ & $-4.7620 \pm 0.0187$	&  $0.1922 \pm 0.0014$	&   $0.5031 \pm 0.0023$ \\
1465.77425849 &  $51.3450 \pm 0.0021$	 &  $ 0.002 \pm 0.003$	 & 7.7630	& 66 &	$0.1285 \pm 0.0009$ & $-4.7613 \pm 0.0177$	&  $0.1907 \pm 0.0013$	&   $0.5046 \pm 0.0022$ \\
1466.77843100 &  $51.3505 \pm 0.0030$	 &  $-0.008 \pm 0.004$	 & 7.7446	& 55 &	$0.1278 \pm 0.0011$ & $-4.7720 \pm 0.0206$	&  $0.1921 \pm 0.0016$	&   $0.4959 \pm 0.0026$ \\
1466.76744544 &  $51.3554 \pm 0.0030$	 &  $-0.003 \pm 0.004$	 & 7.7638	& 55 &	$0.1282 \pm 0.0011$ & $-4.7677 \pm 0.0205$	&  $0.1908 \pm 0.0016$	&   $0.4999 \pm 0.0026$ \\
1481.68018901 &  $51.3350 \pm 0.0024$	 &  $-0.001 \pm 0.003$	 & 7.7732	& 62 &	$0.1292 \pm 0.0011$ & $-4.6661 \pm 0.0143$	&  $0.1979 \pm 0.0016$	&   $0.5001 \pm 0.0025$ \\
1481.69624159 &  $51.3375 \pm 0.0022$	 &  $ 0.005 \pm 0.003$	 & 7.7611	& 64 &	$0.1294 \pm 0.0011$ & $-4.6882 \pm 0.0149$	&  $0.1921 \pm 0.0015$	&   $0.5017 \pm 0.0024$ \\
1482.71116935 &  $51.3298 \pm 0.0027$	 &  $ 0.009 \pm 0.003$	 & 7.7900	& 59 &	$0.1268 \pm 0.0010$ & $-4.7413 \pm 0.0206$	&  $0.1938 \pm 0.0015$	&   $0.5029 \pm 0.0024$ \\
1483.75203132 &  $51.3331 \pm 0.0022$	 &  $ 0.001 \pm 0.003$	 & 7.7623	& 64 &	$0.1276 \pm 0.0010$ & $-4.7612 \pm 0.0188$	&  $0.1937 \pm 0.0014$	&   $0.5058 \pm 0.0023$ \\
1483.74069812 &  $51.3355 \pm 0.0021$	 &  $ 0.004 \pm 0.003$	 & 7.7595	& 66 &	$0.1271 \pm 0.0010$ & $-4.7476 \pm 0.0179$	&  $0.1940 \pm 0.0014$	&   $0.5037 \pm 0.0022$ \\
1576.52948882 &  $51.3330 \pm 0.0046$	 &  $-0.014 \pm 0.006$	 & 7.7801	& 43 &	$0.1265 \pm 0.0016$ & $-5.4237 \pm 0.1398$	&  $0.1907 \pm 0.0022$	&   $0.4997 \pm 0.0035$ \\
1583.54058377 &  $51.3452 \pm 0.0030$	 &  $-0.002 \pm 0.004$	 & 7.7792	& 55 &	$0.1259 \pm 0.0012$ & $-5.0498 \pm 0.0507$	&  $0.1980 \pm 0.0017$	&   $0.4979 \pm 0.0026$ \\
1765.74216151 &  $51.3561 \pm 0.0032$	 &  $-0.003 \pm 0.004$	 & 7.8046	& 54 &	$0.1302 \pm 0.0013$ & $-4.7421 \pm 0.0195$	&  $0.1988 \pm 0.0018$	&   $0.5000 \pm 0.0027$ \\
1766.80813997 &  $51.3575 \pm 0.0020$	 &  $ 0.003 \pm 0.002$	 & 7.7842	& 78 &	$0.1294 \pm 0.0009$ & $-4.6460 \pm 0.0131$	&  $0.2015 \pm 0.0012$	&   $0.4983 \pm 0.0020$ \\
1774.87929440 &  $51.3546 \pm 0.0049$	 &  $ 0.003 \pm 0.006$	 & 7.8038	& 42 &	$0.1294 \pm 0.0017$ & $-4.7973 \pm 0.0351$	&  $0.1970 \pm 0.0024$	&   $0.4962 \pm 0.0035$ \\
1774.88874566 &  $51.3474 \pm 0.0052$	 &  $ 0.008 \pm 0.007$	 & 7.8170	& 40 &	$0.1321 \pm 0.0018$ & $-4.7826 \pm 0.0346$	&  $0.1998 \pm 0.0024$	&   $0.4983 \pm 0.0036$ \\
1802.66127481 &  $51.3363 \pm 0.0020$	 &  $ 0.002 \pm 0.002$	 & 7.7880	& 95 &	$0.1274 \pm 0.0007$ & $-4.6571 \pm 0.0127$	&  $0.1954 \pm 0.0010$	&   $0.5001 \pm 0.0016$ \\
1810.75114647 &  $51.3459 \pm 0.0028$	 &  $-0.005 \pm 0.004$	 & 7.7686	& 57 &	$0.1292 \pm 0.0013$ & $-4.7578 \pm 0.0180$	&  $0.1945 \pm 0.0016$	&   $0.4978 \pm 0.0026$ \\
1830.82116768 &  $51.3358 \pm 0.0020$	 &  $-0.004 \pm 0.002$	 & 7.7822	& 85 &	$0.1238 \pm 0.0008$ & $-4.7114 \pm 0.0151$	&  $0.1925 \pm 0.0010$	&   $0.5029 \pm 0.0017$ \\
1833.70474847 &  $51.3457 \pm 0.0028$	 &  $-0.006 \pm 0.004$	 & 7.7964	& 57 &	$0.1288 \pm 0.0012$ & $-4.6735 \pm 0.0165$	&  $0.1925 \pm 0.0016$	&   $0.5042 \pm 0.0026$ \\
1838.71542850 &  $51.3396 \pm 0.0052$	 &  $-0.012 \pm 0.007$	 & 7.7865	& 40 &	$0.1266 \pm 0.0018$ & $-4.7433 \pm 0.0274$	&  $0.1962 \pm 0.0025$	&   $0.5072 \pm 0.0037$ \\
1849.67005564 &  $51.3633 \pm 0.0020$	 &  $ 0.008 \pm 0.002$	 & 7.7852	& 73 &	$0.1263 \pm 0.0010$ & $-4.6783 \pm 0.0142$	&  $0.1933 \pm 0.0012$	&   $0.5039 \pm 0.0020$ \\
1866.75015160 &  $51.3558 \pm 0.0025$	 &  $-0.000 \pm 0.003$	 & 7.8227	& 61 &	$0.1223 \pm 0.0010$ & $-4.7484 \pm 0.0206$	&  $0.1849 \pm 0.0014$	&   $0.5093 \pm 0.0023$ \\
1877.60993471 &  $51.3525 \pm 0.0038$	 &  $ 0.002 \pm 0.005$	 & 7.7812	& 48 &	$0.1234 \pm 0.0013$ & $-4.7426 \pm 0.0208$	&  $0.1882 \pm 0.0017$	&   $0.5054 \pm 0.0028$ \\
1884.59857785 &  $51.3588 \pm 0.0046$	 &  $ 0.008 \pm 0.006$	 & 7.8175	& 43 &	$0.1279 \pm 0.0016$ & $-4.6631 \pm 0.0191$	&  $0.1892 \pm 0.0022$	&   $0.4970 \pm 0.0033$ \\
1893.60218111 &  $51.3560 \pm 0.0034$	 &  $-0.004 \pm 0.004$	 & 7.7699	& 52 &	$0.1254 \pm 0.0012$ & $-4.7414 \pm 0.0190$	&  $0.1917 \pm 0.0017$	&   $0.5006 \pm 0.0027$ \\
1898.61246619 &  $51.3532 \pm 0.0020$	 &  $ 0.001 \pm 0.002$	 & 7.7709	& 73 &	$0.1247 \pm 0.0008$ & $-4.7345 \pm 0.0163$	&  $0.1911 \pm 0.0011$	&   $0.5048 \pm 0.0019$ \\
2177.64271415 &  $51.3599 \pm 0.0020$	 &  $ 0.004 \pm 0.003$	 & 7.7965	& 68 &	$0.1365 \pm 0.0011$ & $-4.6513 \pm 0.0145$	&  $0.1975 \pm 0.0015$	&   $0.4980 \pm 0.0024$ \\
2180.65538975 &  $51.3580 \pm 0.0020$	 &  $ 0.002 \pm 0.002$	 & 7.7665	& 80 &	$0.1306 \pm 0.0009$ & $-4.7086 \pm 0.0143$	&  $0.1975 \pm 0.0012$	&   $0.5002 \pm 0.0021$ \\
2190.65829783 &  $51.3609 \pm 0.0020$	 &  $-0.001 \pm 0.002$	 & 7.7770	& 73 &	$0.1272 \pm 0.0011$ & $-4.7286 \pm 0.0158$	&  $0.1927 \pm 0.0014$	&   $0.5037 \pm 0.0023$ \\
2204.58471260 &  $51.3515 \pm 0.0020$	 &  $-0.002 \pm 0.002$	 & 7.7795	& 83 &	$0.1304 \pm 0.0009$ & $-4.6921 \pm 0.0139$	&  $0.1990 \pm 0.0012$	&   $0.5045 \pm 0.0020$ \\
2212.60431636 &  $51.3490 \pm 0.0020$	 &  $ 0.002 \pm 0.002$	 & 7.7460	& 87 &	$0.1243 \pm 0.0008$ & $-4.7019 \pm 0.0138$	&  $0.1964 \pm 0.0012$	&   $0.5074 \pm 0.0019$ \\
2226.68671558 &  $51.3532 \pm 0.0020$	 &  $ 0.007 \pm 0.002$	 & 7.8213	& 69 &	$0.1322 \pm 0.0010$ & $-4.7181 \pm 0.0173$	&  $0.1964 \pm 0.0014$	&   $0.4970 \pm 0.0023$ \\
2231.56880343 &  $51.3579 \pm 0.0021$	 &  $ 0.004 \pm 0.003$	 & 7.7757	& 66 &	$0.1279 \pm 0.0011$ & $-4.7062 \pm 0.0161$	&  $0.1948 \pm 0.0016$	&   $0.5016 \pm 0.0025$ \\
2237.61010265 &  $51.3390 \pm 0.0040$	 &  $-0.006 \pm 0.005$	 & 7.7522	& 47 &	$0.1220 \pm 0.0015$ & $-4.8071 \pm 0.0257$	&  $0.1939 \pm 0.0022$	&   $0.5079 \pm 0.0034$ \\
2239.55155959 &  $51.3482 \pm 0.0020$	 &  $-0.005 \pm 0.002$	 & 7.7232	& 69 &	$0.1220 \pm 0.0010$ & $-4.7566 \pm 0.0166$	&  $0.1914 \pm 0.0014$	&   $0.5086 \pm 0.0023$ \\
2243.58440735 &  $51.3482 \pm 0.0020$	 &  $-0.017 \pm 0.002$	 & 7.7637	& 81 &	$0.1262 \pm 0.0009$ & $-4.7494 \pm 0.0154$	&  $0.1984 \pm 0.0012$	&   $0.4996 \pm 0.0020$ \\
2246.58730329 &  $51.3594 \pm 0.0024$	 &  $-0.012 \pm 0.003$	 & 7.7880	& 62 &	$0.1284 \pm 0.0012$ & $-4.6911 \pm 0.0170$	&  $0.1985 \pm 0.0016$	&   $0.4994 \pm 0.0026$ \\
2248.58606747 &  $51.3446 \pm 0.0069$	 &  $ 0.005 \pm 0.009$	 & 7.8496	& 33 &	$0.1346 \pm 0.0023$ & $-5.1308 \pm 0.0994$	&  $0.2088 \pm 0.0034$	&   $0.5027 \pm 0.0048$ \\
2250.61259800 &  $51.3579 \pm 0.0030$	 &  $ 0.004 \pm 0.004$	 & 7.8126	& 55 &	$0.1316 \pm 0.0013$ & $-4.7977 \pm 0.0232$	&  $0.1978 \pm 0.0019$	&   $0.4986 \pm 0.0029$ \\
2252.58098883 &  $51.3554 \pm 0.0020$	 &  $ 0.003 \pm 0.002$	 & 7.7913	& 78 &	$0.1330 \pm 0.0010$ & $-4.6907 \pm 0.0146$	&  $0.1983 \pm 0.0013$	&   $0.5048 \pm 0.0021$ \\
2256.58379243 &  $51.3577 \pm 0.0020$	 &  $ 0.008 \pm 0.002$	 & 7.7786	& 78 &	$0.1249 \pm 0.0009$ & $-4.7789 \pm 0.0167$	&  $0.1926 \pm 0.0012$	&   $0.5055 \pm 0.0020$ \\
2259.59184291 &  $51.3615 \pm 0.0020$	 &  $-0.001 \pm 0.002$	 & 7.7748	& 73 &	$0.1245 \pm 0.0010$ & $-4.7708 \pm 0.0174$	&  $0.1926 \pm 0.0013$	&   $0.5065 \pm 0.0022$ \\
2265.62666161 &  $51.3503 \pm 0.0024$	 &  $-0.002 \pm 0.003$	 & 7.7465	& 62 &	$0.1239 \pm 0.0011$ & $-4.8104 \pm 0.0231$	&  $0.1965 \pm 0.0016$	&   $0.4979 \pm 0.0025$ \\
2268.63412182 &  $51.3521 \pm 0.0036$	 &  $-0.015 \pm 0.005$	 & 7.8106	& 50 &	$0.1287 \pm 0.0014$ & $-4.8145 \pm 0.0280$	&  $0.1980 \pm 0.0020$	&   $0.5015 \pm 0.0031$ \\
2278.56738606 &  $51.3531 \pm 0.0032$	 &  $-0.002 \pm 0.004$	 & 7.7978	& 54 &	$0.1265 \pm 0.0013$ & $-4.7587 \pm 0.0234$	&  $0.1954 \pm 0.0017$	&   $0.5059 \pm 0.0027$ \\
2282.52973675 &  $51.3633 \pm 0.0020$	 &  $ 0.001 \pm 0.002$	 & 7.7765	& 78 &	$0.1263 \pm 0.0009$ & $-4.8303 \pm 0.0193$	&  $0.2002 \pm 0.0011$	&   $0.5020 \pm 0.0018$ \\
\hline
\end{tabular}
}
\end{center}
\end{table*}

\begin{table*}[pht]
\begin{center}
\caption{RV and activity indices obtained from the FEROS spectra.}
\label{tab:feros-data}
\centering
\resizebox{\textwidth}{!}{%
\begin{tabular}{lllllllll}
\hline  \hline
BJD - 2457000 [d] & RV [km/s] & Bisector & FWHM & SNR & $\mathrm{H_\alpha}$ & $\mathrm{\log(R'_{HK})}$ & Na~II & He~I \\
\hline
1449.83201319 & $51.3599 \pm 0.0069$ & $-0.004 \pm 0.011$ & 10.0274 &	  79 &	   $0.1291 \pm 0.0022$ & $-4.6161 \pm 0.0289$ & $0.1834 \pm 0.0033$ & $0.5030 \pm 0.0057$ \\
1449.71864651 & $51.3394 \pm 0.0054$ & $-0.013 \pm 0.009$ &  9.9826 &	 111 &	   $0.1313 \pm 0.0015$ & $-4.6156 \pm 0.0210$ & $0.1815 \pm 0.0023$ & $0.5026 \pm 0.0039$ \\
1450.76285679 & $51.3461 \pm 0.0055$ & $-0.013 \pm 0.009$ & 10.0078 &	 108 &	   $0.1299 \pm 0.0016$ & $-4.6040 \pm 0.0212$ & $0.1817 \pm 0.0024$ & $0.5006 \pm 0.0041$ \\
1450.76988749 & $51.3538 \pm 0.0057$ & $-0.012 \pm 0.009$ & 10.0048 &	 102 &	   $0.1323 \pm 0.0017$ & $-4.6358 \pm 0.0235$ & $0.1854 \pm 0.0025$ & $0.5021 \pm 0.0042$ \\
1451.79336283 & $51.3548 \pm 0.0057$ & $-0.028 \pm 0.009$ & 10.0136 &	 103 &	   $0.1295 \pm 0.0017$ & $-4.6330 \pm 0.0239$ & $0.1890 \pm 0.0025$ & $0.5023 \pm 0.0044$ \\
1451.78645985 & $51.3606 \pm 0.0057$ & $-0.002 \pm 0.009$ &  9.9988 &	 103 &	   $0.1298 \pm 0.0017$ & $-4.6299 \pm 0.0236$ & $0.1903 \pm 0.0025$ & $0.5039 \pm 0.0043$ \\
1452.80242598 & $51.3384 \pm 0.0067$ & $-0.035 \pm 0.011$ & 10.0351 &	  83 &	   $0.1251 \pm 0.0020$ & $-4.6715 \pm 0.0313$ & $0.1854 \pm 0.0031$ & $0.5002 \pm 0.0054$ \\
1467.76807806 & $51.3298 \pm 0.0056$ & $-0.017 \pm 0.009$ & 10.0192 &	 106 &	   $0.1345 \pm 0.0016$ & $-4.6849 \pm 0.0249$ & $0.1846 \pm 0.0025$ & $0.5017 \pm 0.0041$ \\
1468.76772262 & $51.3428 \pm 0.0060$ & $-0.019 \pm 0.010$ & 10.0365 &	  96 &	   $0.1360 \pm 0.0018$ & $-4.6456 \pm 0.0251$ & $0.1817 \pm 0.0028$ & $0.5055 \pm 0.0046$ \\
1469.76959746 & $51.1795 \pm 0.0058$ & $-0.018 \pm 0.009$ & 10.0202 &	 100 &	   $0.1291 \pm 0.0018$ & $-4.6321 \pm 0.0234$ & $0.1865 \pm 0.0027$ & $0.5037 \pm 0.0045$ \\
1483.81130478 & $51.2990 \pm 0.0061$ & $-0.003 \pm 0.010$ & 10.1302 &	  94 &	   $0.1345 \pm 0.0017$ & $-4.7680 \pm 0.0660$ & $0.1849 \pm 0.0026$ & $0.5054 \pm 0.0045$ \\
1485.69646887 & $51.3025 \pm 0.0119$ & $-0.025 \pm 0.017$ & 10.0762 &	  41 &	   $0.2045 \pm 0.0058$ & $-4.7093 \pm 0.0825$ & $0.1729 \pm 0.0081$ & $0.4653 \pm 0.0117$ \\
1493.73812753 & $51.2874 \pm 0.0067$ & $-0.027 \pm 0.011$ & 10.0567 &	  83 &	   $0.1448 \pm 0.0022$ & $-4.6505 \pm 0.0326$ & $0.1873 \pm 0.0032$ & $0.5004 \pm 0.0056$ \\
1500.63798240 & $51.2951 \pm 0.0062$ & $-0.021 \pm 0.010$ & 10.0234 &	  91 &	   $0.1430 \pm 0.0020$ & $-4.6571 \pm 0.0269$ & $0.1919 \pm 0.0029$ & $0.5043 \pm 0.0051$ \\
1502.70459267 & $51.2900 \pm 0.0069$ & $-0.028 \pm 0.011$ & 10.0955 &	  80 &	   $0.1559 \pm 0.0024$ & $-4.5820 \pm 0.0281$ & $0.1890 \pm 0.0033$ & $0.4965 \pm 0.0054$ \\
1521.58004932 & $51.2470 \pm 0.0056$ & $-0.024 \pm 0.009$ & 10.0306 &	 106 &	   $0.1433 \pm 0.0017$ & $-4.6306 \pm 0.0233$ & $0.1931 \pm 0.0025$ & $0.5044 \pm 0.0041$ \\
1542.57046626 & $51.3174 \pm 0.0060$ & $-0.035 \pm 0.010$ & 10.1118 &	  96 &	   $0.1319 \pm 0.0018$ & $-4.6890 \pm 0.0292$ & $0.1905 \pm 0.0027$ & $0.5047 \pm 0.0045$ \\
1543.55517435 & $51.3081 \pm 0.0056$ & $-0.039 \pm 0.009$ & 10.1183 &	 106 &	   $0.1340 \pm 0.0017$ & $-4.6422 \pm 0.0244$ & $0.1849 \pm 0.0024$ & $0.5099 \pm 0.0041$ \\
1544.63920077 & $51.3197 \pm 0.0058$ & $-0.019 \pm 0.009$ & 10.1491 &	 100 &	   $0.1426 \pm 0.0019$ & $-4.7059 \pm 0.0375$ & $0.1933 \pm 0.0027$ & $0.5075 \pm 0.0047$ \\
1545.61458508 & $51.3345 \pm 0.0072$ & $-0.029 \pm 0.011$ & 10.1500 &	  76 &	   $0.1321 \pm 0.0024$ & $-4.7325 \pm 0.0479$ & $0.1960 \pm 0.0037$ & $0.5021 \pm 0.0062$ \\
1546.60962993 & $51.3222 \pm 0.0073$ & $-0.045 \pm 0.011$ & 10.2209 &	  74 &	   $0.1492 \pm 0.0025$ & $-4.5902 \pm 0.0365$ & $0.1898 \pm 0.0036$ & $0.5011 \pm 0.0060$ \\
1547.54838034 & $51.3209 \pm 0.0063$ & $-0.036 \pm 0.010$ & 10.1565 &	  91 &	   $0.1409 \pm 0.0020$ & $-4.6714 \pm 0.0286$ & $0.1952 \pm 0.0029$ & $0.4914 \pm 0.0048$ \\
1548.60710448 & $51.3172 \pm 0.0059$ & $-0.013 \pm 0.010$ & 10.1383 &	  99 &	   $0.1421 \pm 0.0018$ & $-4.6428 \pm 0.0300$ & $0.1980 \pm 0.0027$ & $0.5019 \pm 0.0047$ \\
1550.63138466 & $51.3094 \pm 0.0058$ & $-0.003 \pm 0.009$ & 10.1810 &	 102 &	   $0.1408 \pm 0.0018$ & $-4.6807 \pm 0.0367$ & $0.1934 \pm 0.0026$ & $0.5031 \pm 0.0046$ \\
1553.63083558 & $51.3212 \pm 0.0086$ & $-0.010 \pm 0.013$ & 10.2209 &	  61 &	   $0.1461 \pm 0.0031$ & $-4.5733 \pm 0.0522$ & $0.1894 \pm 0.0047$ & $0.5110 \pm 0.0077$ \\
1555.61367223 & $51.3100 \pm 0.0058$ & $-0.029 \pm 0.009$ & 10.1904 &	 100 &	   $0.1331 \pm 0.0018$ & $-4.7084 \pm 0.0363$ & $0.1892 \pm 0.0027$ & $0.5092 \pm 0.0047$ \\
1556.63579254 & $51.2989 \pm 0.0071$ & $-0.021 \pm 0.011$ & 10.2023 &	  77 &	   $0.1389 \pm 0.0024$ & $-4.6131 \pm 0.0425$ & $0.1980 \pm 0.0036$ & $0.5017 \pm 0.0060$ \\
1559.61969760 & $51.3009 \pm 0.0064$ & $-0.020 \pm 0.010$ & 10.1919 &	  88 &	   $0.1434 \pm 0.0022$ & $-4.6768 \pm 0.0421$ & $0.1912 \pm 0.0031$ & $0.5054 \pm 0.0053$ \\
1567.57829965 & $51.3085 \pm 0.0056$ & $-0.017 \pm 0.009$ & 10.1049 &	 106 &	   $0.1301 \pm 0.0018$ & $-4.6912 \pm 0.0307$ & $0.1914 \pm 0.0026$ & $0.5057 \pm 0.0045$ \\
1569.60503227 & $51.2835 \pm 0.0064$ & $-0.023 \pm 0.010$ & 10.1956 &	  89 &	   $0.1329 \pm 0.0020$ & $-4.7453 \pm 0.0601$ & $0.1900 \pm 0.0029$ & $0.5023 \pm 0.0051$ \\
1572.58523826 & $51.3181 \pm 0.0069$ & $-0.032 \pm 0.011$ & 10.2139 &	  80 &	   $0.1330 \pm 0.0023$ & $-4.6869 \pm 0.0467$ & $0.1893 \pm 0.0033$ & $0.4985 \pm 0.0057$ \\
1573.54735541 & $51.3016 \pm 0.0084$ & $-0.031 \pm 0.013$ & 10.1954 &	  62 &	   $0.1544 \pm 0.0031$ & $-4.5939 \pm 0.0391$ & $0.1793 \pm 0.0045$ & $0.5042 \pm 0.0075$ \\
1574.54805848 & $51.3149 \pm 0.0068$ & $-0.025 \pm 0.011$ & 10.1603 &	  82 &	   $0.1415 \pm 0.0022$ & $-4.6917 \pm 0.0391$ & $0.1847 \pm 0.0031$ & $0.4994 \pm 0.0055$ \\
1597.47834342 & $51.3063 \pm 0.0086$ & $-0.013 \pm 0.013$ & 10.1328 &	  61 &	   $0.1431 \pm 0.0030$ & $-4.6326 \pm 0.0483$ & $0.1712 \pm 0.0046$ & $0.5005 \pm 0.0074$ \\
1617.47697353 & $51.3056 \pm 0.0062$ & $-0.022 \pm 0.010$ & 10.1409 &	  93 &	   $0.1354 \pm 0.0019$ & $-4.5700 \pm 0.0482$ & $0.1773 \pm 0.0028$ & $0.5070 \pm 0.0048$ \\
1674.93453293 & $51.3783 \pm 0.0075$ & $-0.025 \pm 0.011$ & 10.0186 &	  71 &	   $0.1312 \pm 0.0024$ & $-4.7272 \pm 0.0608$ & $0.1862 \pm 0.0038$ & $0.5134 \pm 0.0060$ \\
1676.93586168 & $51.3733 \pm 0.0087$ & $-0.048 \pm 0.013$ &  9.9677 &	  59 &	   $0.1390 \pm 0.0032$ & $-4.6440 \pm 0.0566$ & $0.1904 \pm 0.0051$ & $0.5115 \pm 0.0076$ \\
1718.84166145 & $51.3502 \pm 0.0062$ & $-0.018 \pm 0.010$ & 10.0267 &	  91 &	   $0.1362 \pm 0.0020$ & $-4.6830 \pm 0.0346$ & $0.2024 \pm 0.0030$ & $0.5057 \pm 0.0051$ \\
1722.85657508 & $51.3357 \pm 0.0061$ & $-0.029 \pm 0.010$ & 10.0151 &	  94 &	   $0.1450 \pm 0.0021$ & $-4.6067 \pm 0.0246$ & $0.1899 \pm 0.0030$ & $0.4967 \pm 0.0048$ \\
1723.87675972 & $51.3533 \pm 0.0066$ & $-0.035 \pm 0.010$ & 10.0086 &	  85 &	   $0.1394 \pm 0.0022$ & $-4.6187 \pm 0.0290$ & $0.2095 \pm 0.0034$ & $0.5060 \pm 0.0054$ \\
1724.89283013 & $51.3512 \pm 0.0072$ & $-0.023 \pm 0.011$ & 10.0126 &	  76 &	   $0.1385 \pm 0.0024$ & $-4.5978 \pm 0.0298$ & $0.1991 \pm 0.0037$ & $0.4994 \pm 0.0061$ \\
1725.85732443 & $51.3523 \pm 0.0070$ & $-0.026 \pm 0.011$ & 10.0140 &	  77 &	   $0.1332 \pm 0.0023$ & $-4.5306 \pm 0.0281$ & $0.1999 \pm 0.0035$ & $0.5056 \pm 0.0058$ \\
1784.84308988 & $51.3200 \pm 0.0066$ & $-0.014 \pm 0.011$ &  9.9842 &	  83 &     $0.1262 \pm 0.0021$ & $-4.6930 \pm 0.0291$ & $0.1838 \pm 0.0032$ & $0.5004 \pm 0.0053$ \\
1793.78206387 & $51.3171 \pm 0.0057$ & $-0.017 \pm 0.009$ & 10.0122 &	 103 &     $0.1311 \pm 0.0017$ & $-4.6046 \pm 0.0219$ & $0.1887 \pm 0.0025$ & $0.5059 \pm 0.0043$ \\
1800.70371165 & $51.3282 \pm 0.0069$ & $ 0.002 \pm 0.011$ & 10.0512 &	  80 &     $0.1361 \pm 0.0023$ & $-4.6515 \pm 0.0289$ & $0.1813 \pm 0.0034$ & $0.5125 \pm 0.0058$ \\
1802.71106430 & $51.3217 \pm 0.0063$ & $-0.013 \pm 0.010$ &  9.9732 &	  89 &     $0.1281 \pm 0.0018$ & $-4.7028 \pm 0.0376$ & $0.1800 \pm 0.0028$ & $0.5017 \pm 0.0048$ \\
1805.73619653 & $51.3148 \pm 0.0057$ & $-0.005 \pm 0.009$ & 10.0003 &	 102 &     $0.1349 \pm 0.0017$ & $-4.6241 \pm 0.0249$ & $0.1847 \pm 0.0025$ & $0.5060 \pm 0.0044$ \\
1912.53943145 & $51.3314 \pm 0.0070$ & $-0.015 \pm 0.011$ & 10.2001 &	  79 &     $0.1375 \pm 0.0023$ & $-4.6775 \pm 0.0315$ & $0.1976 \pm 0.0035$ & $0.5020 \pm 0.0059$ \\
\hline
\end{tabular}
}
\end{center}
\end{table*}

\begin{table*}[pht]
\begin{center}
\caption{RV and activity indices obtained from the CORALIE spectra.}
\label{tab:coralie-data}
\centering
\begin{tabular}{lllll}
\hline  \hline
BJD - 2457000 [d] & RV [km/s] & FWHM & Bisector & H$_\alpha$   \\
\hline
1479.75545134023	& $51310.503 \pm 0.009$ & $9088.42 \pm 12.85$ &	$-40.45  \pm 12.72$ & $0.2397 \pm 0.0038$ \\
1489.74439708004	& $51318.693 \pm 0.022$ & $9006.89 \pm 12.74$ &	$ 13.24  \pm 31.27$ & $0.2398 \pm 0.0068$ \\
1496.593841759954	& $51311.109 \pm 0.012$ & $9055.17 \pm 12.81$ &	$-44.33  \pm 16.77$ & $0.2341 \pm 0.0045$ \\
1503.72372861998	& $51325.855 \pm 0.011$ & $9110.66 \pm 12.88$ &	$-72.14  \pm 15.29$ & $0.2358 \pm 0.0041$ \\
1505.67841371009	& $51312.657 \pm 0.008$ & $9104.72 \pm 12.88$ &	$-22.37  \pm 11.33$ & $0.2356 \pm 0.0035$ \\
1526.5760778198 	& $51332.347 \pm 0.015$ & $9173.53 \pm 12.97$ &	$-47.52  \pm 21.63$ & $0.2270 \pm 0.0050$ \\
1534.58861656999	& $51315.643 \pm 0.014$ & $9091.04 \pm 12.86$ &	$-94.11  \pm 19.42$ & $0.2294 \pm 0.0049$ \\
1538.639787740074	& $51335.519 \pm 0.009$ & $9098.83 \pm 12.87$ &	$-49.80  \pm 12.03$ & $0.2261 \pm 0.0035$ \\
1544.60949906986	& $51343.382 \pm 0.008$ & $9082.49 \pm 12.84$ &	$-66.87  \pm 11.31$ & $0.2337 \pm 0.0034$ \\
1563.598259389866	& $51340.052 \pm 0.009$ & $9098.37 \pm 12.87$ &	$-55.33  \pm 12.79$ & $0.2343 \pm 0.0037$ \\
1572.54610658996	& $51321.501 \pm 0.014$ & $9144.88 \pm 12.93$ &	$-69.62  \pm 19.26$ & $0.2394 \pm 0.0049$ \\
1622.46025983011	& $51313.385 \pm 0.020$ & $9159.30 \pm 12.95$ &	$-37.24  \pm 27.81$ & $0.2236 \pm 0.0058$ \\
1623.46099031018	& $51319.046 \pm 0.060$ & $9304.44 \pm 13.16$ &	$-149.08 \pm 84.31$ & $0.2113 \pm 0.0134$ \\
1709.894646670204	& $51345.850 \pm 0.017$ & $9101.04 \pm 12.87$ &	$-103.87 \pm 24.21$ & $0.2419 \pm 0.0059$ \\
1744.779055220075	& $51326.377 \pm 0.013$ & $9150.97 \pm 12.94$ &	$-61.98  \pm 17.69$ & $0.2552 \pm 0.0047$ \\
1762.736076259986	& $51324.292 \pm 0.023$ & $9144.08 \pm 12.93$ &	$-116.84 \pm 32.75$ & $0.2499 \pm 0.0070$ \\
1818.58744085999	& $51295.316 \pm 0.024$ & $9150.28 \pm 12.94$ &	$-44.14  \pm 33.62$ & $0.2352 \pm 0.0068$ \\
1819.59875601018	& $51325.253 \pm 0.024$ & $9186.31 \pm 12.99$ &	$-42.85  \pm 33.45$ & $0.2201 \pm 0.0067$ \\
\hline
\end{tabular}
\end{center}
\end{table*}

\begin{table*}[pht]
\begin{center}
\caption{RV and activity indices obtained from the CHIRON spectra.}
\label{tab:chiron-data}
\centering
\begin{tabular}{llll}
\hline  \hline
BJD - 2457000 [d] & RV [m/s] & Bisector & FWHM  \\
\hline
1553.54058 & $  6.0 \pm  7.6$ & $  14.1 \pm 27.4$ & $12.090 \pm 0.304$ \\
1563.54900 & $ -9.1 \pm  7.5$ & $  72.0 \pm 29.0$ & $12.088 \pm 0.302$ \\
1578.55488 & $-18.7 \pm  6.0$ & $   4.0 \pm 36.5$ & $12.033 \pm 0.293$ \\
1723.90308 & $ -1.8 \pm 10.6$ & $  32.5 \pm 23.4$ & $12.254 \pm 0.309$ \\
1724.89900 & $  0.0 \pm 10.8$ & $ -14.1 \pm 22.8$ & $12.219 \pm 0.305$ \\
1730.93316 & $ 18.4 \pm 39.1$ & $-135.7 \pm 98.1$ & $12.188 \pm 0.285$ \\
1739.87310 & $ -2.2 \pm 11.3$ & $ -76.0 \pm 39.2$ & $12.248 \pm 0.306$ \\
1741.91746 & $ 21.4 \pm 13.6$ & $   2.7 \pm 42.8$ & $12.205 \pm 0.297$ \\
1814.73821 & $ 14.1 \pm  8.6$ & $ -39.6 \pm 31.2$ & $12.135 \pm 0.302$ \\
1828.71190 & $-26.4 \pm 11.3$ & $ -12.0 \pm 35.0$ & $12.025 \pm 0.309$ \\
1867.57241 & $-25.9 \pm 22.7$ & $ -74.5 \pm 55.4$ & $12.116 \pm 0.290$ \\
2173.80175 & $ 18.7 \pm  7.3$ & $ -12.5 \pm 35.1$ & $12.134 \pm 0.302$ \\
2173.80890 & $ 41.6 \pm 11.7$ & $   5.0 \pm 25.2$ & $12.115 \pm 0.307$ \\
2173.81606 & $ 53.4 \pm 11.3$ & $ 128.5 \pm 43.5$ & $12.163 \pm 0.308$ \\
2179.73480 & $ -3.1 \pm 12.3$ & $  26.9 \pm 35.4$ & $12.139 \pm 0.309$ \\
2179.74196 & $ 29.2 \pm 11.1$ & $  40.8 \pm 32.7$ & $12.105 \pm 0.314$ \\
2179.74911 & $  4.0 \pm 11.8$ & $   5.9 \pm 39.6$ & $12.081 \pm 0.300$ \\
2181.72309 & $-11.2 \pm 12.5$ & $ -19.6 \pm 38.7$ & $12.164 \pm 0.312$ \\
2181.73025 & $ -4.5 \pm 11.3$ & $   3.8 \pm 43.9$ & $12.130 \pm 0.311$ \\
2181.73740 & $ 21.5 \pm 12.4$ & $   8.0 \pm 42.3$ & $12.137 \pm 0.307$ \\
2184.73563 & $ -6.7 \pm 12.1$ & $ -99.8 \pm 40.0$ & $12.108 \pm 0.298$ \\
2184.74278 & $ 20.4 \pm 10.0$ & $   3.7 \pm 39.9$ & $12.094 \pm 0.302$ \\
2184.74992 & $ -6.9 \pm 12.3$ & $ -90.0 \pm 42.2$ & $12.172 \pm 0.300$ \\
2196.68466 & $ 34.3 \pm 11.2$ & $  -9.6 \pm 50.2$ & $12.096 \pm 0.315$ \\
2196.69181 & $ -6.3 \pm 12.8$ & $  -9.9 \pm 36.7$ & $12.116 \pm 0.305$ \\
2196.69895 & $ -2.6 \pm 12.7$ & $ -41.2 \pm 32.7$ & $12.101 \pm 0.307$ \\
\hline
\end{tabular}
\end{center}
\end{table*}

\section{Instrumental priors and posteriors for the TTV extraction}
\restartappendixnumbering

In this appendix, we list the instrumental and GP prior and posterior distributions for the TTV extraction with \texttt{juliet}.

\startlongtable
\begin{deluxetable*}{lll}
\tablecaption{Prior and posterior instrumental parameter distributions for the TTV extraction with juliet. \label{tab:juliet-inst-pp}}
\tablehead{\colhead{Parameter} & \colhead{Prior\tablenotemark{a}} & \colhead{Posterior}}
\startdata
q$_\mathrm{{1,TESS}}$ \dotfill & $\mathcal{TN}(0.39,0.1,0,1)$ & \julietqoneTESS \\
q$_\mathrm{{2,TESS}}$ \dotfill & $\mathcal{TN}(0.31,0.1,0,1)$ & \julietqtwoTESS \\
q$_\mathrm{{1,ASTEP}}$ \dotfill & $\mathcal{TN}(0.5,0.1,0,1)$ & \julietqoneASTEP \\
q$_\mathrm{{2,ASTEP}}$ \dotfill & $\mathcal{TN}(0.35,0.1,0,1)$ & \julietqtwoASTEP \\
q$_\mathrm{{1,PEST}}$ \dotfill & $\mathcal{TN}(0.5,0.1,0,1)$ & \julietqonePEST \\
q$_\mathrm{{2,PEST}}$ \dotfill & $\mathcal{TN}(0.5,0.1,0,1)$ & \julietqtwoPEST \\
q$_\mathrm{{1,NEOSsat}}$ \dotfill & $\mathcal{TN}(0.59,0.1,0,1)$ & \julietqoneNEOS \\ 
q$_\mathrm{{2,NEOSSat}}$ \dotfill & $\mathcal{TN}(0.39,0.1,0,1)$ & \julietqtwoNEOS \\ 
q$_\mathrm{{1,Hwd}}$ \dotfill & $\mathcal{TN}(0.5,0.1,0,1)$ & \julietqoneHwd \\
q$_\mathrm{{2,Hwd}}$ \dotfill & $\mathcal{TN}(0.35,0.1,0,1)$ & \julietqtwoHwd \\
q$_\mathrm{{1,LCO,1}}$ \dotfill & $\mathcal{TN}(0.33,0.1,0,1)$ & \julietqoneLCOone \\ 
q$_\mathrm{{2,LCO,1}}$ \dotfill & $\mathcal{TN}(0.29,0.1,0,1)$ & \julietqtwoLCOone \\ 
q$_\mathrm{{1,LCO,2}}$ \dotfill & $\mathcal{TN}(0.68,0.1,0,1)$ & \julietqoneLCOtwo \\ 
q$_\mathrm{{2,LCO,2}}$ \dotfill & $\mathcal{TN}(0.45,0.1,0,1)$ & \julietqtwoLCOtwo \\ 
q$_\mathrm{{1,LCO,3}}$ \dotfill & $\mathcal{TN}(0.41,0.1,0,1)$ & \julietqoneLCOthr \\ 
q$_\mathrm{{2,LCO,3}}$ \dotfill & $\mathcal{TN}(0.32,0.1,0,1)$ & \julietqtwoLCOthr \\ 
q$_\mathrm{{1,LCO,4}}$ \dotfill & $\mathcal{TN}(0.41,0.1,0,1)$ & \julietqoneLCOfour \\ 
q$_\mathrm{{2,LCO,4}}$ \dotfill & $\mathcal{TN}(0.32,0.1,0,1)$ & \julietqtwoLCOfour \\ 
q$_\mathrm{{1,LCO,5}}$ \dotfill & $\mathcal{TN}(0.68,0.1,0,1)$ & \julietqoneLCOfive \\ 
q$_\mathrm{{2,LCO,5}}$ \dotfill & $\mathcal{TN}(0.45,0.1,0,1)$ & \julietqtwoLCOfive \\ 
q$_\mathrm{{1,LCO,6}}$ \dotfill & $\mathcal{TN}(0.41,0.1,0,1)$ & \julietqoneLCOsix \\ 
q$_\mathrm{{2,LCO,6}}$ \dotfill & $\mathcal{TN}(0.32,0.1,0,1)$ & \julietqtwoLCOsix \\ 
$m_\mathrm{{d,TESS,2}}$ \dotfill & 1.0 (fixed) & 1.0 (fixed) \\
$m_\mathrm{{flux,TESS,2}}$ \dotfill & $\mathcal{N}(0,0.1)$ & \julietmfluxTESStwo \\ 
$\sigma_\mathrm{{TESS,2}}$ \dotfill & $\mathcal{J}(0.1,1000)$ & \julietsigmaTESStwo \\ 
$m_\mathrm{{d,TESS,10}}$ \dotfill & 1.0 (fixed) & 1.0 (fixed) \\
$m_\mathrm{{flux,TESS,10}}$ \dotfill & $\mathcal{N}(0,0.1)$ & \julietmfluxTESSten\\ 
$\sigma_\mathrm{{TESS,10}}$ \dotfill & $\mathcal{J}(0.1,1000)$ & \julietsigmawTESSten \\
$m_\mathrm{{d,TESS,13}}$ \dotfill & 1.0 (fixed) & 1.0 (fixed) \\
$m_\mathrm{{flux,TESS,13}}$ \dotfill & $\mathcal{N}(0,0.1)$ & \julietmfluxTESSthir \\ 
$\sigma_\mathrm{{TESS,13}}$ \dotfill & $\mathcal{J}(0.1,1000)$ & \julietsigmawTESSthir \\ 
$m_\mathrm{{d,TESS,29}}$ \dotfill & 1.0 (fixed) & 1.0 (fixed) \\
$m_\mathrm{{flux,TESS,29}}$ \dotfill & $\mathcal{N}(0,0.1)$ & \julietmfluxTESStn \\ 
$\sigma_\mathrm{{TESS,29}}$ \dotfill & $\mathcal{J}(0.1,1000)$ & \julietsigmawTESStn\\ 
$m_\mathrm{{d,TESS,32}}$ \dotfill & 1.0 (fixed) & 1.0 (fixed) \\
$m_\mathrm{{flux,TESS,32}}$ \dotfill & $\mathcal{N}(0,0.1)$ & \julietmfluxTESStt \\ 
$\sigma_\mathrm{{TESS,32}}$ \dotfill & $\mathcal{J}(0.1,1000)$ & \julietsigmawTESStt \\ 
$m_\mathrm{{d,TESS,36}}$ \dotfill & 1.0 (fixed) & 1.0 (fixed) \\
$m_\mathrm{{flux,TESS,36}}$ \dotfill & $\mathcal{N}(0,0.1)$ & \julietmfluxTESSts \\ 
$\sigma_\mathrm{{TESS,36}}$ \dotfill & $\mathcal{J}(0.1,1000)$ & \julietsigmawTESSts \\ 
$m_\mathrm{{flux,TESS,63}}$ \dotfill & $\mathcal{N}(0,0.1)$ & \julietmfluxTESSst \\ 
$\sigma_\mathrm{{TESS,63}}$ \dotfill & $\mathcal{J}(0.1,1000)$ & \julietsigmawTESSst \\ 
$m_\mathrm{{d,ASTEP,1}}$ \dotfill & 1.0 (fixed) & 1.0 (fixed) \\
$m_\mathrm{{flux,ASTEP,1}}$ \dotfill & $\mathcal{N}(0,0.1)$ & \julietmfluxASTEPone \\ 
$\sigma_\mathrm{{ASTEP,1}}$ \dotfill & $\mathcal{J}(0.1,1000)$ & \julietsigmawASTEPone \\ 
$m_\mathrm{{d,ASTEP,2}}$ \dotfill & 1.0 (fixed) & 1.0 (fixed) \\
$m_\mathrm{{flux,ASTEP,2}}$ \dotfill & $\mathcal{N}(0,0.1)$ & \julietmfluxASTEPtwo \\ 
$\sigma_\mathrm{{ASTEP,2}}$ \dotfill & $\mathcal{J}(0.1,1000)$ & \julietsigmawASTEPtwo \\ 
$m_\mathrm{{d,ASTEP,3}}$ \dotfill & 1.0 (fixed) & 1.0 (fixed) \\
$m_\mathrm{{flux,ASTEP,3}}$ \dotfill & $\mathcal{N}(0,0.1)$ & \julietmfluxASTEPthr \\ 
$\sigma_\mathrm{{ASTEP,3}}$ \dotfill & $\mathcal{J}(0.1,1000)$ & \julietsigmawASTEPthr \\ 
$m_\mathrm{{d,PEST}}$ \dotfill & 1.0 (fixed) & 1.0 (fixed) \\
$m_\mathrm{{flux,PEST}}$ \dotfill & $\mathcal{N}(0,0.1)$ & \julietmfluxPEST \\ 
$\sigma_\mathrm{{PEST}}$ \dotfill & $\mathcal{J}(0.1,1000)$ & \julietsigmawPEST \\ 
$m_\mathrm{{d,NEOSSat}}$ \dotfill & 1.0 (fixed) & 1.0 (fixed) \\
$m_\mathrm{{flux,NEOSSat}}$ \dotfill & $\mathcal{N}(0,0.1)$ & \julietmfluxNEOS \\ 
$\sigma_\mathrm{{NEOSSat}}$ \dotfill & $\mathcal{J}(0.1,1000)$ & \julietsigmawNEOS \\ 
$m_\mathrm{{d,Hwd}}$ \dotfill & 1.0 (fixed) & 1.0 (fixed) \\
$m_\mathrm{{flux,Hwd}}$ \dotfill & $\mathcal{N}(0,0.1)$ & \julietmfluxHwd \\ 
$\sigma_\mathrm{{Hwd}}$ \dotfill & $\mathcal{J}(0.1,1000)$ & \julietsigmawHwd \\ 
$m_\mathrm{{d,LCO,1}}$ \dotfill & 1.0 (fixed) & 1.0 (fixed) \\
$m_\mathrm{{flux,LCO,1}}$ \dotfill & $\mathcal{N}(0,0.1)$ & \julietmfluxLCOone \\ 
$\sigma_\mathrm{{LCO,1}}$ \dotfill & $\mathcal{J}(0.1,1000)$ & \julietsigmawLCOone \\ 
$m_\mathrm{{d,LCO,2}}$ \dotfill & 1.0 (fixed) & 1.0 (fixed) \\
$m_\mathrm{{flux,LCO,2}}$ \dotfill & $\mathcal{N}(0,0.1)$ & \julietmfluxLCOtwo \\ 
$\sigma_\mathrm{{LCO,2}}$ \dotfill & $\mathcal{J}(0.1,1000)$ & \julietsigmawLCOtwo \\ 
$m_\mathrm{{d,LCO,3}}$ \dotfill & 1.0 (fixed) & 1.0 (fixed) \\
$m_\mathrm{{flux,LCO,3}}$ \dotfill & $\mathcal{N}(0,0.1)$ & \julietmfluxLCOthr \\ 
$\sigma_\mathrm{{LCO,3}}$ \dotfill & $\mathcal{J}(0.1,1000)$ & \julietsigmawLCOthr \\ 
$m_\mathrm{{d,LCO,4}}$ \dotfill & 1.0 (fixed)  & 1.0 (fixed) \\
$m_\mathrm{{flux,LCO,4}}$ \dotfill & $\mathcal{N}(0,0.1)$ & \julietmfluxLCOfo\\ 
$\sigma_\mathrm{{LCO,4}}$ \dotfill & $\mathcal{J}(0.1,1000)$ & \julietsigmawLCOfo \\ 
$m_\mathrm{{d,LCO,5}}$ \dotfill & 1.0 (fixed) & 1.0 (fixed) \\
$m_\mathrm{{flux,LCO,5}}$ \dotfill & $\mathcal{N}(0,0.1)$ & \julietmfluxLCOfi \\ 
$\sigma_\mathrm{{LCO,5}}$ \dotfill & $\mathcal{J}(0.1,1000)$ & \julietsigmawLCOfi \\ 
$m_\mathrm{{d,LCO,6}}$ \dotfill & 1.0 (fixed) & 1.0 (fixed) \\
$m_\mathrm{{flux,LCO,6}}$ \dotfill & $\mathcal{N}(0,0.1)$ & \julietmfluxLCOsix \\ 
$\sigma_\mathrm{{LCO,6}}$ \dotfill & $\mathcal{J}(0.1,1000)$ & \julietsigmawLCOsix \\ 
$\sigma_\mathrm{{GP, TESS, 2}}$  \dotfill & $\mathcal{J}(0.0023,0.0052)$  & \julietGPsigmaTESStwo \\
$\rho_\mathrm{{GP, TESS, 2}}$   \dotfill & $\mathcal{J}(3.0,5.9)$ & \julietGPrhoTESStwo \\
$\sigma_\mathrm{{GP, TESS, 10}}$   \dotfill & $\mathcal{J}(0.007,0.010)$  & \julietGPsigmaTESSten\\
$\rho_\mathrm{{GP, TESS, 10}}$  \dotfill & $\mathcal{J}(0.63,0.79)$ & \julietGPrhoTESSten \\
$\sigma_\mathrm{{GP, TESS, 13}}$  \dotfill & $\mathcal{J}(0.0028,0.0036)$ & \julietGPsigmaTESSthir \\
$\rho_\mathrm{{GP, TESS, 13}}$  \dotfill & $\mathcal{J}(0.64,0.81)$  & \julietGPrhoTESSthir \\
$\sigma_\mathrm{{GP, TESS, 29}}$  \dotfill & $\mathcal{J}(0.0032,0.0040)$ & \julietGPsigmaTESStn \\
$\rho_\mathrm{{GP, TESS, 29}}$   \dotfill & $\mathcal{J}(0.39,0.47)$ & \julietGPrhoTESStn \\
$\sigma_\mathrm{{GP, TESS, 32}}$  \dotfill & $\mathcal{J}(0.0053,0.0265)$ & \julietGPsigmaTESStt \\
$\rho_\mathrm{{GP, TESS, 32}}$   \dotfill & $\mathcal{J}(6.1,20.0)$ & \julietGPrhoTESStt  \\
$\sigma_\mathrm{{GP, TESS, 36}}$   \dotfill & $\mathcal{J}(0.0049,0.0061)$ & \julietGPsigmaTESSts \\
$\rho_\mathrm{{GP, TESS, 36}}$   \dotfill & $\mathcal{J}(0.39,0.47)$ & \julietGPrhoTESSts \\
$\sigma_\mathrm{{GP, TESS, 63}}$   \dotfill & $\mathcal{J}(0.0049,0.0141)$ & \julietGPsigmaTESSst \\
$\rho_\mathrm{{GP, TESS, 63}}$   \dotfill & $\mathcal{J}(3.2,7.0)$ & \julietGPrhoTESSst \\
\enddata
\tablenotetext{a}{$\mathcal{U}(a,b)$ indicates a uniform distribution between $a$ and $b$; $\mathcal{TN}(a, b)$ a normal distribution with mean $a$ and standard deviation $b$ $\mathcal{N}(a, b, c, d)$ a normal distribution with mean $a$, standard deviation $b$, limited to the $[c,d]$ range; $\mathcal{J}(a,b)$ a Jeffreys or log-uniform distribution between $a$ and $b$.}
\end{deluxetable*}

\section{Posterior probability distributions}
\restartappendixnumbering

\begin{figure*}[htb]
    \centering
    \includegraphics[width=1\hsize]{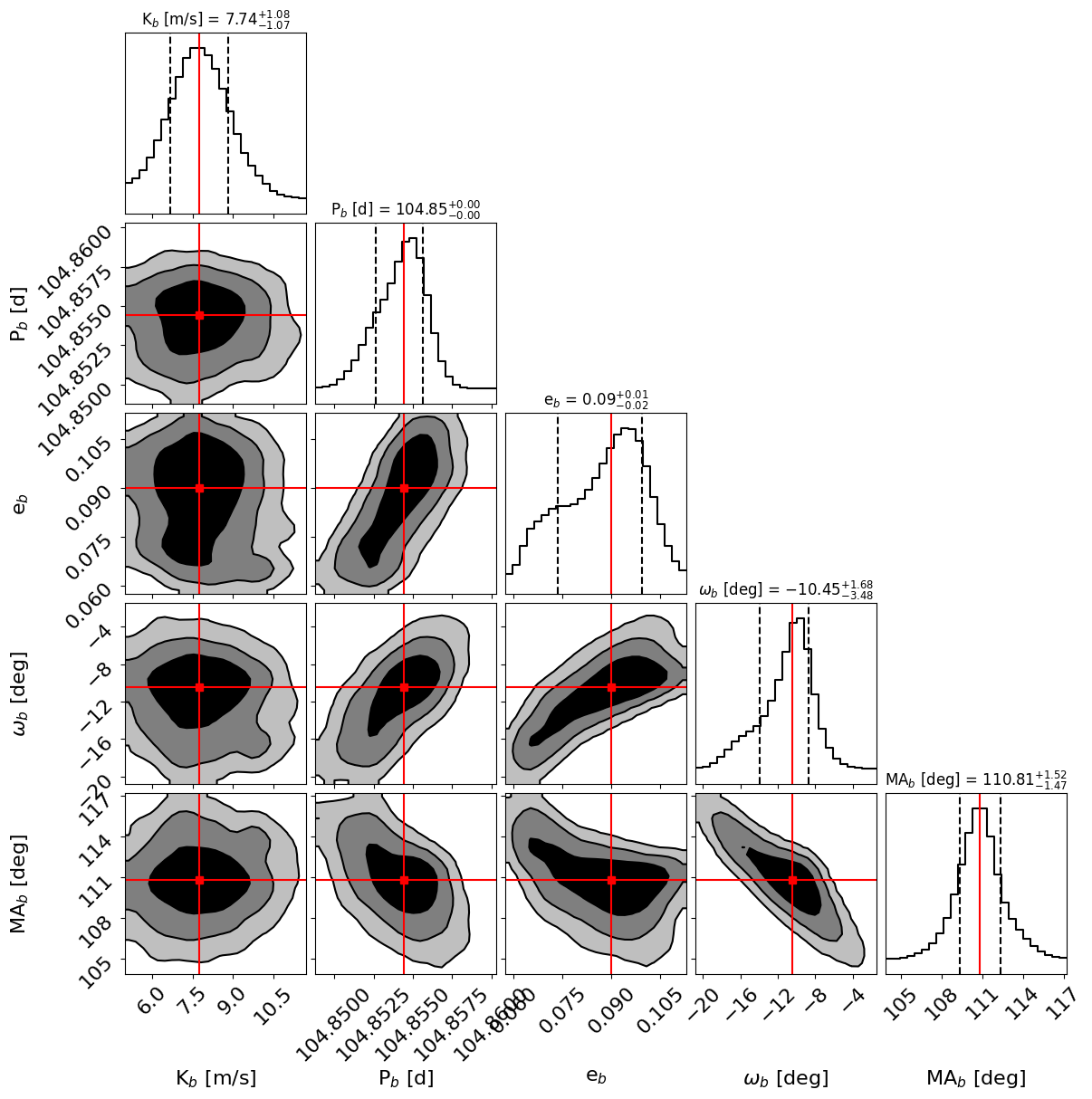}
    \caption{Cornerplot of the posterior distributions of the fitted planet parameters for TOI-199 b for the joint modelling of the \textit{TESS} and ASTEP TTVs, and the HARPS, FEROS, CHIRON, and CORALIE RVs. The distributions are explored using nested sampling. The red crosses indicate the median values, and the black contour lines the 1$\sigma$, 2$\sigma$, and 3$\sigma$ confidence levels. For $\omega_b$, note that it is a circular argument and thus, e.g., $-10\degr = 350\degr$.}
    \label{fig:cornerplot_pl1_fit}
\end{figure*}

\begin{figure*}[htb]
    \centering
    \includegraphics[width=1\hsize]{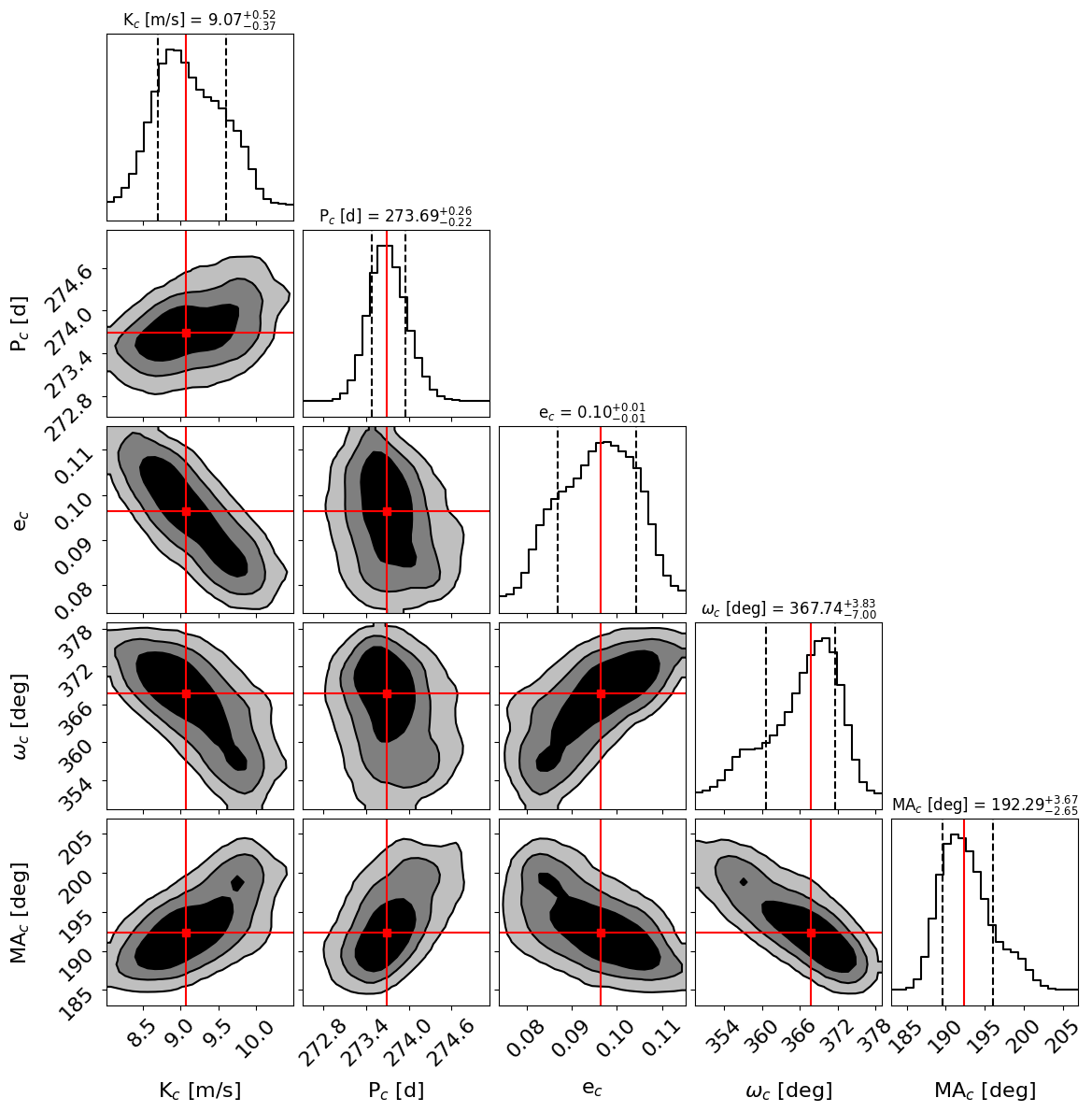}
    \caption{Cornerplot of the posterior distributions of the planet parameters for TOI-199 c the joint modelling of the \textit{TESS} and ASTEP TTVs, and the HARPS, FEROS, CHIRON, and CORALIE RVs. The distributions are explored using nested sampling. The red crosses indicate the median values, and the black contour lines the 1$\sigma$, 2$\sigma$, and 3$\sigma$ confidence levels. For $\omega_c$, note that it is a circular argument and thus, e.g., $370\degr = 10\degr$.}
    \label{fig:cornerplot_pl2_fit}
\end{figure*}

\begin{figure*}[htb]
    \centering
    \includegraphics[width=1\hsize]{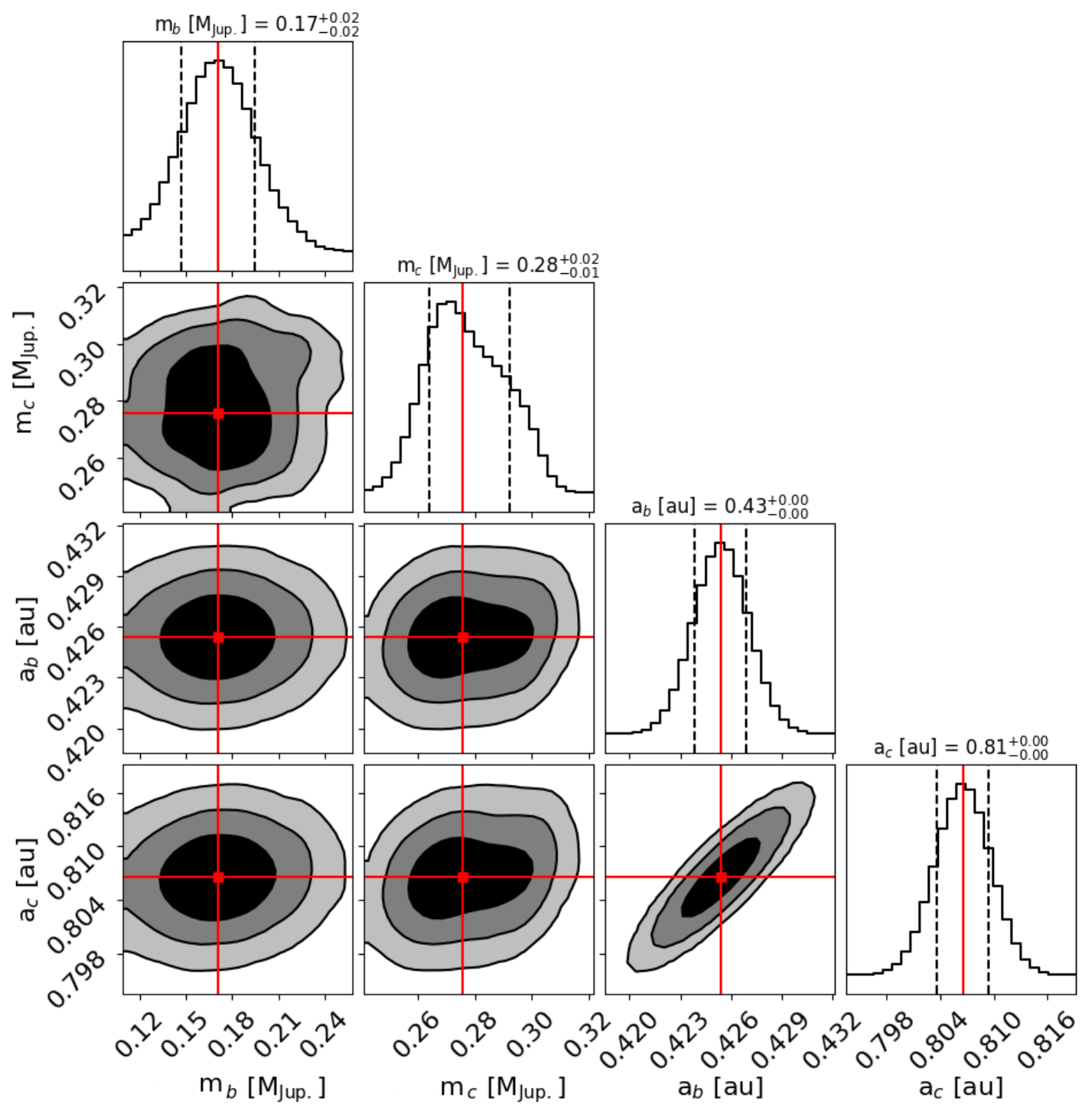}
    \caption{Cornerplot of the posterior distributions of the derived planet parameters for the joint modelling of the \textit{TESS} and ASTEP TTVs, and the HARPS, FEROS, CHIRON, and CORALIE RVs. The distributions are explored using nested sampling. The red crosses indicate the median values, and the black contour lines the 1$\sigma$, 2$\sigma$, and 3$\sigma$ confidence levels.}
    \label{fig:cornerplot_pl_derived}
\end{figure*}

\begin{figure*}[htb]
    \centering
    \includegraphics[width=1\hsize]{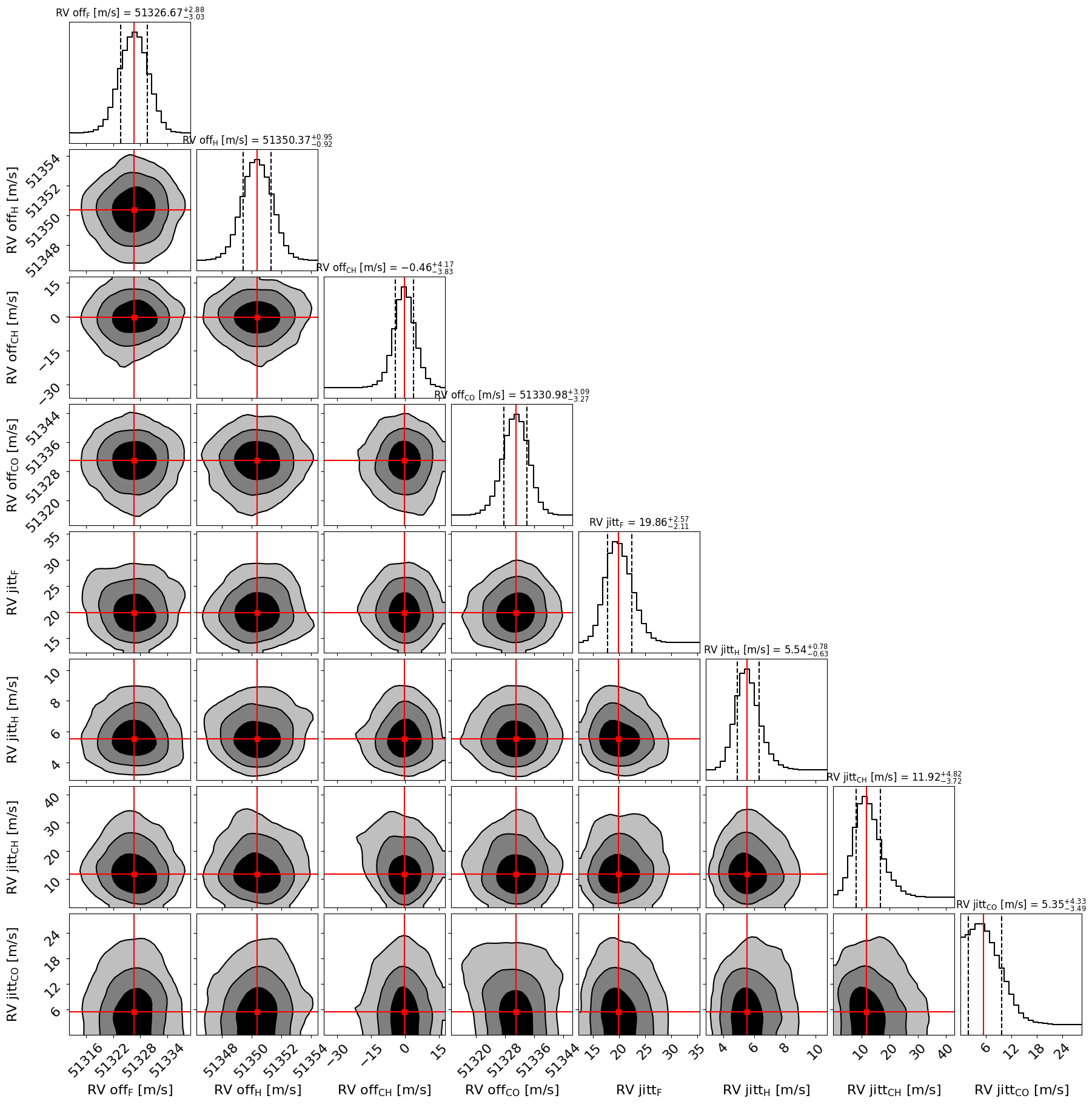}
    \caption{Cornerplot of the posterior distributions of the instrumental RV parameters for the joint modelling of the \textit{TESS} and ASTEP TTVs, and the HARPS, FEROS, CHIRON, and CORALIE RVs. The distributions are explored using nested sampling. The red crosses indicate the median values, and the black contour lines the 1$\sigma$, 2$\sigma$, and 3$\sigma$ confidence levels. Subindex H refers to HARPS, F to FEROS, CH to CHIRON, and CO to CORALIE.}
    \label{fig:cornerplot_rvs}
\end{figure*}

In this appendix, we show the posterior probability distribution of the joint TTV+RV modelling with \texttt{Exo-Striker}. For ease of viewing, we have separated the cornerplots into fitted planetary parameters (Figs. \ref{fig:cornerplot_pl1_fit} and \ref{fig:cornerplot_pl2_fit}), derived planetary parameters (Fig. \ref{fig:cornerplot_pl_derived}) and instrumental RV parameters (Fig. \ref{fig:cornerplot_rvs}).


\section{Exomoon dynamical evolution}
\restartappendixnumbering

In this appendix, we show the dynamical evolution of the test particles used in the exomoon analysis described in Sec. \ref{sec:exomoons}. 

\begin{figure*}[tp]
    \centering
    \includegraphics[width=18cm]{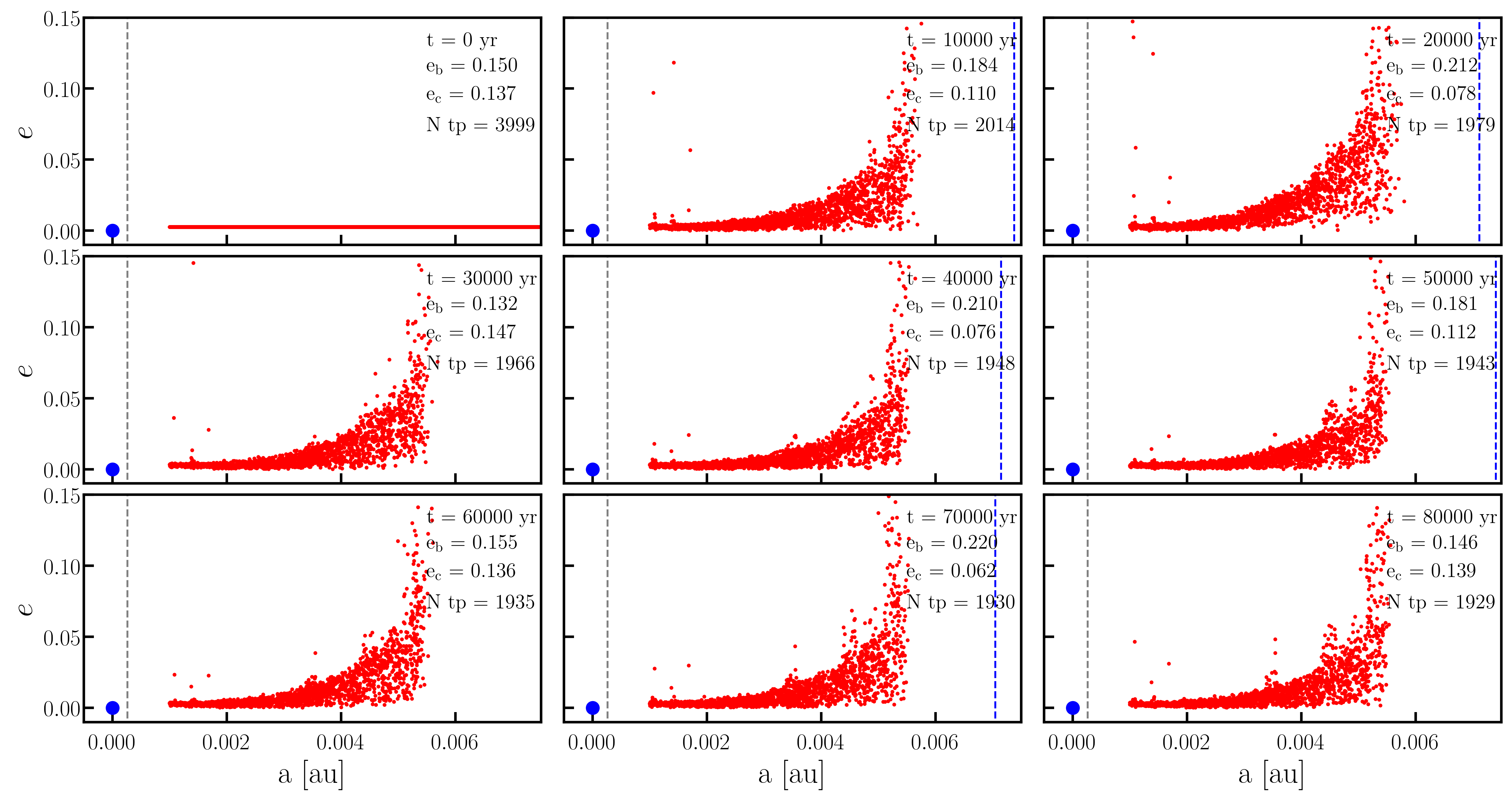} 

    \caption{
Evolution of mass-less test particles (i.e., exomoons) around TOI-199\,b, under the gravitational perturbation of the outer planet. Shown are the position of TOI-199\,b (blue dot), the planetary Roche limit (gray dashed line), and the planetary $0.5R_{\rm Hill}$ (blue dashed line), which scales with (1-$e_b$) due to the dynamical perturbations of TOI-199\,c. 
    }
    \label{e_a_inner} 
\end{figure*}

\begin{figure*}[tp]
    \centering
    \includegraphics[width=18cm]{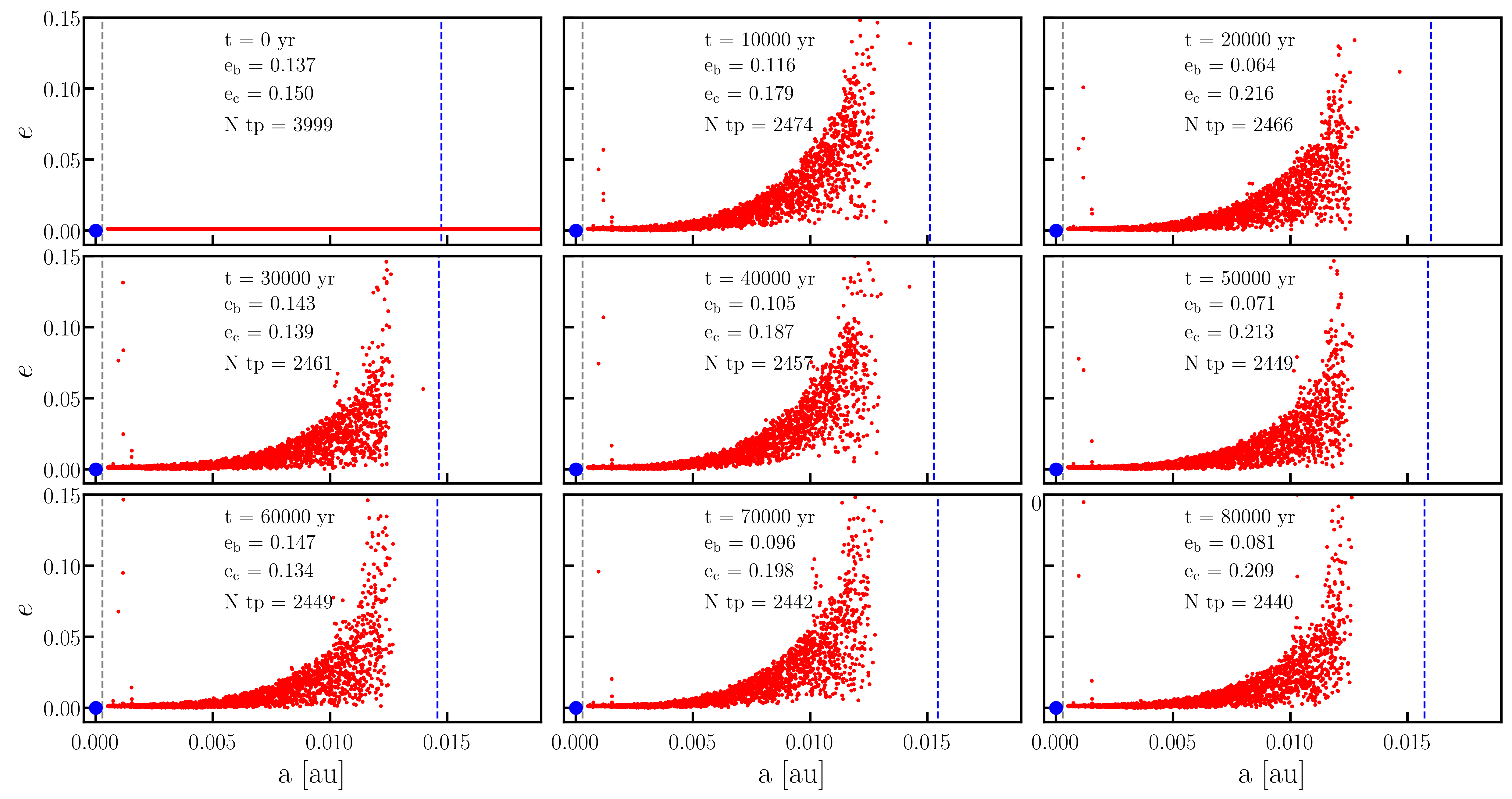} 

    \caption{
Same as in \autoref{e_a_inner}, but for TOI-199\,c.}
    \label{e_a_outer} 
\end{figure*}

\bibliography{sample63}{}
\bibliographystyle{aasjournal}



\end{document}